\documentclass[a4paper,11pt]{article}
\pdfoutput=1

\usepackage{jheppub} 

\usepackage[T1]{fontenc} 

\usepackage{amssymb}\usepackage[normalem]{ulem} 
\usepackage{ulem}

\usepackage{graphicx} 
\usepackage{graphics}
\usepackage{epsfig}
\usepackage{color}
\usepackage{amssymb}
\usepackage{amsmath}
\usepackage{doi}
\usepackage{dsfont}
\usepackage{setspace}
\usepackage{float}
\usepackage{caption}
\usepackage{subcaption}

\usepackage{slashed}
\usepackage{booktabs}
\usepackage{multirow}
\usepackage{soul}
\title{Four top final states with NLO accuracy in perturbative QCD: 4 lepton  channel}

\author{Nikolaos Dimitrakopoulos}
\author{and Malgorzata Worek\,}

\affiliation{Institute for Theoretical Particle Physics
and Cosmology, RWTH Aachen University, \\D-52056 Aachen, Germany}

\emailAdd{ndimitrak@physik.rwth-aachen.de}
\emailAdd{worek@physik.rwth-aachen.de}

\abstract{Triggered by the observation of four top-quark production at the LHC by the ATLAS and CMS collaboration we report on the calculation of 
the next-to-leading order QCD corrections to the Standard Model process $pp \to t\bar{t}t\bar{t}$ in the $4\ell$ top-quark decay channel. We take into account higher-order QCD effects in both the production and decays of the four top quarks. The latter effects are treated in the narrow width approximation, which preserves top-quark spin correlations  throughout the calculation. We present results for two selected renormalisation and factorisation scale settings and three different PDF sets. Furthermore, we study the main theoretical uncertainties that are associated with the neglected higher-order terms in the perturbative expansion and with the parameterisation of the PDF sets. The results at the integrated and differential fiducial cross-section level are shown for the LHC Run III center-of-mass energy of $\sqrt{s}=13.6$ TeV. Our findings  are particularly relevant for precise measurements of the four top-quark fiducial cross sections and for the modelling of top-quark decays at the LHC.}

\dedicated{\rm TTK-24-02, P3H-24-001}

\keywords{Higher-Order Perturbative Calculations, Specific QCD Phenomenology, Top Quark}

\begin{document} 

\maketitle
\flushbottom

%
\section{Introduction}
\label{sec:introduction}
%

Being the heaviest known elementary particle of the Standard Model (SM), the top quark has a special relation with the SM Higgs boson due to the magnitude of its Yukawa coupling $(Y_t)$ which is of the order of one. The top quark is also predicted to have large couplings to hypothetical new particles that might appear in many models beyond the SM (BSM). In this respect, rare processes involving the top quark are particularly important to be studied at the Large Hadron Collider (LHC). Not only can they explore the parameters of the SM in regions hitherto inaccessible, but also, if the BSM couplings to the top quark are indeed significant, serve as golden candidates to reveal signs of new physics. Production of four top-quarks in $pp$ collisions, that requires a partonic center-of-mass energy of at least $4m_t$,  is one of the rarest and heaviest processes accessible at the LHC. Although the $pp \to t\bar{t}t\bar{t}$ process occurs predominantly through the strong interaction, non-negligible contributions arise also from electroweak (EW) processes occurring through the exchange of the electroweak gauge and/or SM Higgs boson. Due to EW contributions involving the Higgs boson exchange,  the  $pp \to t\bar{t}H(H\to  t\bar{t})$  process can be used to constrain the Higgs boson Yukawa coupling \cite{Cao:2016wib,Cao:2019ygh}, complementary to 
loop sensitivity through electroweak  corrections in $t\bar{t}$ production \cite{Schmidt:1992et,Kuhn:2013zoa,Martini:2021uey} and tree-level sensitivity in the $pp \to t\bar{t}H$ and/or $pp \to tH$ process \cite{Artoisenet:2013puc,Demartin:2014fia,Demartin:2015uha,Demartin:2016axk,Bahl:2020wee,Bortolato:2020zcg,Hermann:2022vit}. Moreover, in various new physics scenarios, the small SM cross section for the production of four top-quarks can be significantly enhanced. These BSM scenarios include among others models with a composite right-handed top quark \cite{Lillie:2007hd,Pomarol:2008bh,Kumar:2009vs,Banelli:2020iau}, 
heavy Kaluza-Klein gluons and quarks from extra dimensions \cite{Cacciapaglia:2011kz}, scalar gluons occurring in supersymmetric models and produced in pairs  \cite{Plehn:2008ae,Calvet:2012rk}, singly produced top-philic resonances \cite{Kim:2016plm,Darme:2021gtt}  or a heavy scalar or pseudoscalar boson from extended electroweak symmetry breaking sectors produced in association with a top-quark pair  \cite{BhupalDev:2014bir,Craig:2015jba,Gori:2016zto,Craig:2016ygr}. In addition, the $pp\to t\bar{t}t\bar{t}$ process provides a direct way to constrain the SM Effective Field Theory  Wilson coefficients sensitive to the quartic couplings between top-quarks \cite{Zhang:2017mls,Ethier:2021bye,Aoude:2022deh}. If the scale of new physics is too high to be observed directly at the LHC, it can manifest itself as a modification of the SM $t\bar{t}t\bar{t}$ cross section and cause modifications in various differential cross-section distributions. Given such rich phenomenology of the SM $pp\to t\bar{t}t\bar{t}$ process at the LHC, the measurements of four-top integrated and differential cross sections must be accompanied by very precise and accurate theoretical predictions for the process. Such theoretical predictions should be a prerequisite for shedding light on possible signals of new physics that may emerge in this rare channel. 

With its short lifetime, however, the top quark decays before 
top-flavoured hadrons or $t\bar{t}$-bound states can
be formed. We therefore only have access to the properties of the top quark through its decay products. Given that the top quark decays
almost $100\%$ of the time into $t \to W b$, its final states are
classified via  the $W$ gauge boson decays into $W\to \ell
\nu_\ell$ and $W \to q\bar{q}^{\, \prime}$. Consequently, the SM $pp 
\to t\bar{t}t\bar{t} \to W^+ W^- W^+ W^- b\bar{b} b\bar{b}$ process produces a plethora of final states, all of which yield interesting and distinctive signatures at the LHC.  The dominant $4t$  decay mode is the mono-leptonic channel with a branching ratio $({\cal BR})$ of $40\%$.   It is followed by the fully hadronic and opposite-sign  dilepton modes with $ {\cal BR} \approx 20\%$ each, the 
same-sign dilepton and the trilepton modes with ${\cal BR}
\approx 10\%$ each  and finally by the fully leptonic mode
with the smallest contribution of the order of $1.2\%$ 
\cite{Alvarez:2016nrz}.  The last branching
ratio is further reduced to $0.2\%$ when $\ell^\pm=e^\pm,\mu^\pm$  are only considered and the $\tau$ lepton is not taken
into account. Both ATLAS and CMS have recently observed the
production of four top-quarks with the LHC Run II energy of
$13$ TeV using the integrated luminosity of about $140$ fb${}^{-1}$ 
\cite{ATLAS:2023ajo,CMS:2023ftu}. Events with two same-sign, three, or
four charged leptons (electrons and muons) and additional light- and/or 
$b$-jets have been analysed to achieve the required significance of
more than $5 \sigma$  that is needed to claim observation.  Independent of the experiment and taking  into  account current uncertainties of the
measurements, the obtained  $t\bar{t}t\bar{t}$ cross sections are  in 
agreement with the SM  prediction. At present, however, both ATLAS and CMS measurements are limited by statistical uncertainties. The largest systematic uncertainties, on the other hand, come from the modelling of the additional jet activity in $t\bar{t}W^\pm$ production, the normalisation of the  $pp \to t\bar{t}Z$ background process and the modelling of the $pp \to t\bar{t}t\bar{t}$ signal process. Therefore, an accurate theoretical description of $t\bar{t}t\bar{t}$ production at the LHC comprising all the features of the process is much needed and more relevant than ever.  

On the theory side, NLO QCD corrections for the $pp \to t\bar{t}t\bar{t}$ process with stable top quarks have been calculated for the first time 
 a while ago \cite{Bevilacqua:2012em} and a few years later recomputed in Refs. \cite{Alwall:2014hca,Maltoni:2015ena}. Besides NLO QCD corrections, a further step towards more precise predictions for this process has been achieved by evaluating the complete-NLO corrections \cite{Frederix:2017wme} that additionally include NLO EW higher-order effects and subleading contributions at perturbative orders from ${\cal O}(\alpha_s^5)$ to ${\cal O}(\alpha^5)$. However, it has been shown that there are cancellations between different terms in $\alpha_s$ and $\alpha$, so in general the 
higher-order effects for the $pp \to t\bar{t}t\bar{t}$ process are
dominated by $\alpha_s$ corrections to the leading QCD
process at ${\cal O}(\alpha_s^4)$. Finally, very recently the calculation for $pp
\to t\bar{t}t\bar{t}$ at the next-to-leading logarithmic accuracy
has been carried out \cite{vanBeekveld:2022hty} including
also constant ${\cal O}(\alpha_s)$ non-logarithmic contributions that
do not vanish at the absolute production threshold defined as $M^2/\hat{s}\rightarrow 1$, where $M=4t$. In all these studies only the magnitude of the higher-order effects in the total production rate of four top quarks has been investigated, while the reliable description of the fiducial cross sections is still lacking. For realistic studies, the final state with $12$ particles must be considered that consists of four $b$-jets, additional light jets and/or charged leptons together with the (significant) missing transverse momentum. Here, too, reliable theoretical predictions are needed. The first attempt in this direction has already been accomplished in Ref. \cite{Jezo:2021smh}  by matching theoretical 
predictions for the $pp \to t\bar{t}t\bar{t}$ process with the 
dipole-style $p_T$-ordered parton shower from the \textsc{Pythia8} Monte Carlo (MC)
program \cite{Sjostrand:2006za,Sjostrand:2014zea} using  the 
\textsc{Powheg} framework \cite{Nason:2004rx,Alioli:2010xd}.  In this study, top-quark decays in the $\ell  + jets$ channel have been modelled at LO. Consequently, spin correlations are also described at the LO level in the perturbative expansion. For the first time, not only the impact of NLO QCD corrections and $t\bar{t}$ spin correlation effects, but also the importance of the subleading EW production modes, that have been included with the LO accuracy only, has been investigated at the differential fiducial cross-section level. In addition, a comparison with the results obtained within the \textsc{MG5}${}_{-}$\textsc{aMC@NLO} framework \cite{Alwall:2014hca}, that employs the \textsc{MC@NLO} matching to parton shower programs \cite{Frixione:2002ik,Frixione:2003ei}, has been performed. Basically, parton showers represent higher-order corrections to the hard process. However, in this approach an approximation scheme is used in which only collinear parton splitting or soft gluon emission contributions are included. Thus, to cover the entire fiducial phase space it is important to incorporate higher-order effects in top-quark decays already at the matrix element level. This would represent a further step towards the complete description of the jet radiation pattern for this process. 

Unfortunately,  NLO QCD predictions for $pp \to t\bar{t}t\bar{t}$ that are valid at the level of observable particles such as leptons, $b$-jets and light-jets originating either from the production of top quarks or in their decays would require the calculation of higher order corrections to $2\to 12$ process, which is a formidable task at present. Fortunately, the problem can be simplified by examining only the (quadruple) resonance contributions for top-quarks and $W$ gauge bosons while including spin degrees of freedom exactly. Indeed,  all QCD corrections for the  $pp \to t\bar{t}t\bar{t}$ process can be decomposed into factorizable and non-factorizable contributions. The latter corrections imply a cross-talk between $t\bar{t}t\bar{t}$ production and $t/\bar{t}$ decays as well as between the $t$ and $\bar{t}$ decays. In the limit when $\Gamma/m \to 0$, these corrections must vanish. The exact way in which they vanish has already been described ample times in literature,  see e.g. \cite{Fadin:1993kt,Melnikov:1993np,Melnikov:1995fx,Denner:1997ia,Beenakker:1999ya,Melnikov:2009dn,Campbell:2012uf,Bevilacqua:2019quz,Czakon:2020qbd}, and goes under the name of the narrow-width-approximation (NWA). The aim of this paper is to provide theoretical predictions for the $pp \to t\bar{t}t\bar{t}$ process that include NLO QCD corrections to $t\bar{t}t\bar{t}$ production and four top-quark decays in the NWA. In our study, we consider the $4\ell$ top-quark decay channel  comprising the following decay chain $pp \to t\bar{t}t\bar{t} \to W^+W^-W^+W^- b\bar{b}\, b\bar{b}\to \ell^+ \nu_\ell \, \ell^- \bar{\nu}_\ell  \, \ell^+ \nu_\ell \, \ell^- \bar{\nu}_\ell  \, b\bar{b} \, b\bar{b}$, where $\ell^\pm=e^\pm,\mu^\pm$. This is the first time that such a complete study for this process is conducted at the NLO level in QCD. With a full set of higher-order QCD effects included, we can investigate their effect at the integrated fiducial cross-section level as well as for various infrared-safe observables. In addition, such theoretical predictions allow us to qualitatively increase the level of precision and accuracy for the SM $pp\to t\bar{t}t\bar{t}$ process, making future comparison with LHC data more feasible.

The paper is organised as follows. In Section \ref{sec:calculation} we describe our calculation and the \textsc{Helac-Nlo} MC program that we use in our studies. Input parameters and fiducial phase-space cuts for simulating the ATLAS and/or CMS detector response are given in Section \ref{sec:setup}. We briefly review the NWA framework in Section \ref{sec:nwa}, where we also provide the explicit expressions for the NLO QCD cross section.  Numerical results for the integrated fiducial cross sections for the LHC Run III energy of $\sqrt{s}=13.6$ TeV are presented in Section \ref{sec:integrated}. In particular, we show the results 
for two renormalisation and factorisation scale choices and three different PDF sets. On the other hand, differential 
cross-section distributions, which are generated only for the dynamic scale setting, are shown in Section \ref{sec:diff}.  The theoretical uncertainties of the integrated and differential fiducial cross sections, that are associated with neglected higher order terms in the perturbative expansion and with different parametrisations of the parton distribution functions (PDFs), are also given in Section \ref{sec:integrated} and Section \ref{sec:diff}. In addition, we study there the differences and similarities between the expanded and unexpanded NLO QCD results and investigate the magnitude and impact of higher-order QCD effects in top-quark decays. Finally, in Section \ref{summary} our conclusions are given.

%
\section{Description of the calculation}
\label{sec:calculation}
%

\begin{figure}[t!]
        \centering        
        \captionsetup{skip=10pt}
        \includegraphics[height=1in,width=1\textwidth,angle=0]{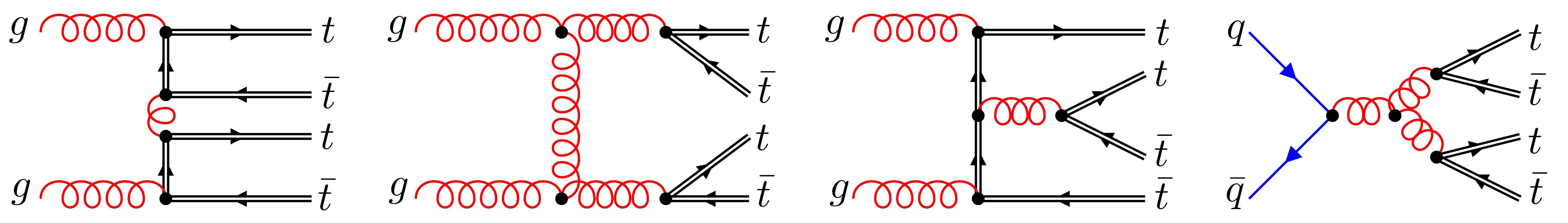}       
    \caption{\textit{\textit{Representative Feynman diagrams for the process $pp \to t\bar{t}t\bar{t} \to W^+W^-W^+W^-b\bar{b} \,b\bar{b} \to\ell^+ \nu_\ell \, \ell^- \bar{\nu}_\ell \,  \ell^+ \nu_\ell \, \ell^- \bar{\nu}_\ell \,   b\bar{b} \, b\bar{b}$ at the order of $\mathcal{O}(\alpha_s^4 \alpha^8)$.  For simplicity, the decays of top quarks are not shown in the diagrams.  All graphs were generated with FeynGame \cite{Harlander:2020cyh}.}}}
             \label{fig:feynman}
\end{figure}

In high-energy $pp$ collisions at the LHC, the production of four top quarks can take place, followed by their decay into various final states. This process is initiated either by the interaction of two gluons ($gg$) or by a pair of a quark and an anti-quark ($q\bar{q}$). For the fully leptonic top-quark decay channel that we are focusing on, it can be described as follows: $gg/q\bar{q} \to t\bar{t}t\bar{t} \to W^+W^-W^+W^-b\bar{b} \,b\bar{b}\to\ell^+ \nu_\ell \, \ell^- \bar{\nu}_\ell \,  \ell^+ \nu_\ell \, \ell^- \bar{\nu}_\ell \,   b\bar{b} \, b\bar{b}$, where $q = u,d,s,c,b$ and $\ell^{\pm}$ stands for $e^{\pm}, \mu^{\pm}$. The leading order (LO) contributions considered in this paper are of the order of $\mathcal{O}(\alpha_s^4 \alpha^8)$ and they consist of Feynman diagrams containing four intermediate top quarks and $W$ bosons along with their corresponding decay products. This implies that for the production of four top quarks, we are only taking into account the dominant QCD contributions, while the electroweak coupling becomes relevant only during the decay phase. At LO the $gg$ subprocess involves a total of 72 Feynman diagrams, while for each $q\bar{q}$ subprocess, the corresponding number is 14. Representative Feynman diagrams are depicted in Figure \ref{fig:feynman}, where, for simplicity, the decays of top quarks are not shown. 

In addition, we calculate next-to-leading-order (NLO) QCD corrections of the order of $\mathcal{O}(\alpha_s^5 \alpha^8)$. These higher-order corrections comprise both virtual and real emission contributions. The virtual corrections are derived by calculating the interference between the sum of all one-loop diagrams with the Born amplitude. For the real emission contributions, an additional QCD parton is emitted, leading to the following subprocesses:
\begin{itemize}
    \item $gg\to t t \bar{t} \bar{t} \,  (g) \to W^+W^-W^+W^-b\bar{b} \,b\bar{b} \,(g) \to \ell^+ \nu_\ell \, \ell^- \bar{\nu}_\ell \,  \ell^+ \nu_\ell \, \ell^- \bar{\nu}_\ell \,   b\bar{b} \, b\bar{b} \, (g) \,,$ 
    \item $\bar{q}q/q\bar{q}\to t t \bar{t} \bar{t} \, (g) \to W^+W^-W^+W^-b\bar{b} \,b\bar{b} \, (g) \to \ell^+ \nu_\ell \, \ell^- \bar{\nu}_\ell \,  \ell^+ \nu_\ell \, \ell^- \bar{\nu}_\ell \,   b\bar{b} \, b\bar{b} \,(g)\,,$ 
    \item  $gq / qg \to t t \bar{t} \bar{t} \, q \to W^+W^-W^+W^-b\bar{b} \,b\bar{b} \, q \to \ell^+ \nu_\ell \, \ell^- \bar{\nu}_\ell \,  \ell^+ \nu_\ell \, \ell^- \bar{\nu}_\ell \,   b\bar{b} \, b\bar{b} \,q\,,$ 
    \item $g\bar{q} / \bar{q} g \to t t \bar{t} \bar{t} \, \bar{q} \to W^+W^-W^+W^-b\bar{b} \,b\bar{b} \, \bar{q} \to \ell^+ \nu_\ell \, \ell^- \bar{\nu}_\ell \,  \ell^+ \nu_\ell \, \ell^- \bar{\nu}_\ell \,   b\bar{b} \, b\bar{b} \, \bar{q}\,,$ 
\end{itemize}
where in the case of the $gg/q\bar{q}/\bar{q}q$ initiated subprocesses $(g)$ indicates the fact that the additional gluon can be emitted during both the production and decay stages. Throughout our study, we adopt the NWA for the treatment of the top quark and the $W$ gauge boson. This approximation is valid when a width $(\Gamma)$ of an unstable particle is much smaller than its mass $(m)$, allowing us to safely take the limit $\Gamma/m \to 0$ in the Breit-Wigner propagator
\begin{equation} \label{nwa_formula}
    \lim_{\Gamma/m \to 0} \dfrac{1}{(p^2 - m^2)^2 + m^2 \Gamma^2} = \dfrac{\pi}{m \Gamma} \delta(p^2 - m^2)\,. \\ \vspace{0.2cm}
\end{equation}
Indeed, this is the case in our process where
$\Gamma_t / m_t \approx 0.008$ and $\Gamma_W /m_W \approx 0.026$, justifying our treatment. The NWA ensures that the considered particles are kept on-shell, which allows us to separately analyse the production and decay stages of the top quarks and investigate the impact of higher-order corrections at each of these stages. Hence, in this study, we examine various scenarios involving NLO QCD corrections applied either in both the production and decay stages or solely during the production of four top quarks. We label the former approach as $\sigma^{\rm NLO}_{\rm full}$ and the latter as $\sigma^{\rm NLO}_{\rm LO_{dec}}$. Additionally, we are going to assess the effects of expanding the top-quark width in our calculations, referred to as $\sigma^{\rm NLO}_{\rm exp}$. More details about these three approaches are going to be provided in Section \ref{sec:nwa}.

Our NLO QCD calculations are performed with the help of the \textsc{Helac-NLO} program \cite{Bevilacqua:2011xh}. In this framework the calculation of scattering amplitudes is based on the Dyson-Schwinger recursive algorithm  \cite{Draggiotis:1998gr,Kanaki:2000ey,Draggiotis:2002hm,Cafarella:2007pc}. The  \textsc{Helac-NLO} program includes \textsc{Helac-1loop} \cite{vanHameren:2009dr} for the evaluation of the numerators of the loop integrals and the rational terms, \textsc{CutTools} \cite{Ossola:2007ax}, which implements the OPP method for the numerical evaluation of one-loop amplitudes, based on a decomposition at the integrand level \cite{Ossola:2006us,Ossola:2008xq,Draggiotis:2009yb}, and \textsc{OneLoop} \cite{vanHameren:2010cp} for the evaluation of the scalar integrals. The preservation of gauge symmetries by this approach is explicitly checked by studying Ward identities up to the one-loop level. For the $gg$ subprocess we perform this test for every phase-space point. Quadruple precision is used to recompute events which fail the gauge-invariance check. For the $q\bar{q}$ subprocess, on the other hand, we use the so-called scale test \cite{Badger:2010nx}. The latter test, that is based on momentum rescaling, is also performed for each phase-space point. Also in this case, for the failed points the amplitude is recomputed using higher precision. In order to optimise the performance of the \textsc{Helac-1loop} system reweighting techniques, helicity and colour sampling methods are used. We employ the Catani-Seymour dipole subtraction formalism  \cite{Catani:1996vz} to extract the soft and collinear infrared singularities and to combine them with the virtual corrections. Specifically, the formulation presented in Ref. \cite{Catani:2002hc} for massive quarks as implemented in the \textsc{Helac-Dipoles} MC program \cite{Czakon:2009ss} is employed. It has been further extended by us to work for arbitrary helicity eigenstates and colour configurations of the external partons. Furthermore, \textsc{Helac-Dipoles} comprises another subtraction scheme \cite{Bevilacqua:2013iha} that uses the momentum mapping and the splitting functions derived in the context of an improved parton shower formulation by Nagy and Soper \cite{Nagy:2007ty}. Also in this case random polarisation and colour sampling of the external partons are employed to optimise the calculations. Compared to standard dipole subtraction, this scheme features a significantly smaller number of subtraction terms and facilitates the matching of NLO calculations with parton showers including quantum interference \cite{Czakon:2015cla}.  For both subtraction schemes a phase-space restriction on the contribution of the subtraction terms, called $\alpha_{\rm max}$, is implemented  \cite{Nagy:1998bb,Nagy:2003tz,Bevilacqua:2009zn,Czakon:2015cla}.  After combining virtual and real corrections, singularities connected to collinear configurations in the final state as well as soft divergences in the initial and final states cancel for collinear-safe observables after applying a jet algorithm. Singularities connected to collinear initial-state splittings are removed via factorisation by parton distribution function redefinitions. The phase space integration is performed with the help of the Monte Carlo generator \textsc{Kaleu} \cite{vanHameren:2010gg}, including \textsc{Parni} \cite{vanHameren:2007pt} for the importance sampling. For the real corrections  \textsc{Kaleu} is equipped with additional, special channels that proved to be important for phase-space optimisation in the subtracted real emission part of the calculation \cite{Bevilacqua:2010qb}. 

We have conducted multiple cross-checks to ensure the accuracy of our results. To begin with, we checked the coefficients of the poles in $\epsilon$, i.e. $1/\epsilon$ and $1/\epsilon^2$, of the virtual amplitudes for $gg$ and $q\bar{q}$ subprocesses with \textsc{MadGraph}${}_{-}$\textsc{aMC@NLO} \cite{Alwall:2014hca} for a few random phase-space points for stable $t\bar{t}t\bar{t}$ production and for the $t\to W^+ b \to \ell^+ \nu_\ell \,b$ decay.  In the next step, we repeated this procedure using the \textsc{Recola} MC program \cite{Actis:2016mpe}. This time, however, also the finite parts along with the coefficients of the poles in $\epsilon$ were cross-checked without separating the production stage from the top-quark decays. Because \textsc{Recola} is able to provide matrix elements in the so-called double-pole approximation \cite{Denner:2000bj,Accomando:2004de,Denner:2016jyo}, it was rather straightforward to prepare this program for the use in the NWA by interfacing it to the \textsc{Helac-NLO} system. This approach has already been tested before in the case of NLO QCD calculations for  $pp \to t\bar{t}jj$ \cite{Bevilacqua:2022ozv} and $pp \to t\bar{t}\gamma\gamma$ \cite{Stremmer:2023kcd}. For the real corrections, we have checked for several phase-space points that the poles in $\epsilon$ that appear in the ${\cal I}$-operator and those appearing in the virtual part of our calculations cancel. We have confirmed that the real emission part of the calculation is independent of any phase-space restriction imposed by the unphysical $\alpha_{\rm max}$ parameter. This check has been applied to both the Catani-Seymour and Nagy-Soper subtraction schemes. Finally, having two independent subtraction schemes allowed us to cross-check the correctness of the real emission part of the calculation in an even more robust way.

All our LO and NLO results are stored in the modified \textsc{Les Houches Files} \cite{Alwall:2006yp}, which are subsequently converted into the \textsc{Root Ntuple Files} \cite{Antcheva:2009zz,Bern:2013zja}. This approach offers a significant advantage as it allows us to efficiently reweight these events. As a consequence, we can produce differential distributions for new observables, perform their rebinning or generate theoretical predictions for various scenarios, including different PDF sets, other scale settings, and various (more exclusive) cuts on the final state particles, all without the need to rerun the entire process from scratch. All the above tasks are carried out using the  \textsc{HEPlot} program \cite{Bevilacqua:HEPlot}.

%
\section{Computational setup}
\label{sec:setup}
%

We consider the process
\begin{equation}
   pp \to t\bar{t}t\bar{t} +X \to W^+W^-W^+W^- b\bar{b}b\bar{b} +X \to \ell^+ \nu_\ell\, \ell^- \bar{\nu}_\ell \,  \,\ell^+ \nu_\ell\, \ell^- \bar{\nu}_\ell  \, b\bar{b}b\bar{b}+X \\ \vspace{0.2cm}
\end{equation}
at NLO in QCD, where $\ell^\pm$ stands for $\ell^\pm= e^\pm, \mu^\pm$. If instead $\ell = e^\pm, \mu^\pm, \tau^\pm$, then all the results presented in this paper should be multiplied by the following factor $81/16 \approx 5.0625$ to obtain the correct normalisation. To perform this calculation we use the NWA. For simplicity, we refer to this process as $pp \to t\bar{t}t\bar{t}$ in the $4\ell$ top-quark decay channel. We do not consider $\tau$ leptons as they are often analysed separately at the LHC due to their rather rich and complex decay pattern. Our results are obtained for LHC Run III center-of-mass energy of $\sqrt{s} = 13.6$ TeV. The numerical values we use for the SM input parameters are given below:
\begin{equation}
\begin{array}{lll}
 G_{ \mu}= 1.166 3787 \cdot 10^{-5} ~{\rm GeV}^{-2}\,,  
& \quad \quad \quad &  m_{t}=172.5 ~{\rm GeV} \,,
\vspace{0.2cm}\\
 m_{W}= 80.379 ~{\rm GeV} \,, 
&&\Gamma_{W}^{\rm NLO} = 2.0972 ~{\rm GeV}\,, 
\vspace{0.2cm}\\
  m_{Z}=91.1876  ~{\rm GeV} \,, 
&& m_b = 0 ~{\rm GeV}\,.
\end{array}
\end{equation}
Apart from the top quark and the $W$ boson, all the other particles appearing in our process are treated as massless. Also, we keep the Cabibbo-Kobayashi-Maskawa mixing matrix diagonal for all our calculations. Since the leptonic $W$-boson decays do not receive NLO QCD corrections, to account for some higher-order effects in their decays,  we use the NLO QCD value for the $W$  gauge boson width as calculated for $\mu_R = m_W$. We utilise the $\Gamma_W^{\rm NLO}$ value everywhere, i.e. in the calculation of LO and NLO matrix elements. Regarding the width of the top quark, we adopt its LO (NLO) value in LO (NLO) calculations. The LO and NLO values of $\Gamma_t$ are determined using the formulas provided in Ref. \cite{Denner:2012yc}, where $\alpha_s(\mu_R=m_t)$ is used:
\begin{equation}
\label{top_widths}
\begin{array}{lll}
 \Gamma_{t}^{\rm LO} = 1.4806842 ~{\rm GeV}\,, &
 \quad \quad\quad \quad &
 \Gamma_{t}^{\rm NLO} = 1.3535983  ~{\rm GeV}\,.
\end{array}
\end{equation}
The electromagnetic coupling $\alpha$ is evaluated in the $G_\mu$-scheme as follows:
\begin{equation}
\alpha_{G_\mu}=\frac{\sqrt{2}}{\pi} 
\,G_F \,m_W^2  \,\sin^2\theta_W \,,
~~~~~~~~~~~~~~~~~~
\sin^2\theta_W = 1-\frac{m_W^2}{m_Z^2}\,.\\ \vspace{0.2cm}
\end{equation}
In accordance with the guidance for Standard Model processes provided by the PDF4LHC working group \cite{Butterworth:2015oua}, we utilise three individual, modern PDF sets: MSHT20 \cite{Bailey:2020ooq} (our default PDF set),  NNPDF3.1 \cite{NNPDF:2017mvq} and CT18 \cite{Hou:2019efy}. Since there is no LO PDF set available for CT18, we have substituted it with CT14. In particular, in this case we employ CT14llo with $\alpha_s (m_Z) = 0.130$ for our LO calculations \cite{Dulat:2015mca}. The running of the strong coupling is performed with a two-loop (one-loop) accuracy at NLO (at LO) and provided by the LHAPDF library \cite{Buckley:2014ana} involving five active flavours.

The limitations of fixed-order calculations lead to an inherent dependence on the renormalisation $(\mu_R)$ and factorisation scale $(\mu_F)$. The theoretical uncertainty of the  cross section, associated with neglected higher-order terms in the perturbative expansion, can be estimated by
varying $\mu_R$ and $\mu_F$ in $\alpha_s$ and PDFs, up and down by a factor $2$ around the central value of the scale of the process $(\mu_0)$ in the range
\begin{equation}
    \frac{1}{2} \le \frac{\mu_R}{\mu_0} 
    , \frac{\mu_F}{\mu_0} \le 2\,.\\ \vspace{0.2cm}
\end{equation}
Traditionally, the following additional condition is required
\begin{equation}
    \frac{1}{2} \le \frac{\mu_R}{\mu_F} \le 2\,.\\ \vspace{0.2cm}
\end{equation}
We are looking for the minimum and maximum of the resulting cross sections, taking into account the following pairs
\begin{equation}
\label{scan}
\left(\frac{\mu_R}{\mu_0}\,,\frac{\mu_F}{\mu_0}\right) = \Big\{
\left(2,1\right),\left(0.5,1  
\right),\left(1,2\right), (1,1), (1,0.5), (2,2),(0.5,0.5)
\Big\} \,.\\ \vspace{0.2cm}
\end{equation}
At this point, it is important to note that we use the same numerical value for the top-quark width across all scale choices during the scale variation. We have observed though that the uncertainties stemming from varying the $\mu_R$ scale in the calculation of $\Gamma_t$ are of the order of $4\%$ only. As we will discuss in Section \ref{sec:integrated}, these uncertainties are similar in magnitude to the internal PDF uncertainties.   In our studies, we make use of two scale settings. For the fixed scale choice we use 
\begin{equation}\label{fixedscale}
\mu_0 = 2m_t \,,\\ \vspace{0.2cm}
\end{equation}
which has already been employed in our previous studies for the $pp \to t\bar{t}t\bar{t}$ process with stable top quarks \cite{Bevilacqua:2012em}. On the other hand, for the dynamical scale setting we utilise 
\begin{equation}\label{dynamicalscale}
\mu_0 = E_T/4 \,,\\ \vspace{0.2cm}
\end{equation}
with $E_T$ defined according to
\begin{equation}
E_T=\sum_{i=1,2} m_T(t_i) 
+ \sum_{i=1,2}m_T(\,\bar{t}_i\,) \,, \\ \vspace{0.2cm}
\end{equation}
where the transverse masses $m_T(t)$ and $m_T(\,\bar{t}\,)$ are given by 
\begin{equation}
m_T(t) = \sqrt{m_t^2 + p_T^2(t) }\,, \\ \vspace{0.2cm}
\quad \quad\quad\quad\quad
m_T(\,\bar{t}\,) = \sqrt{m_t^2 + p_T^2(\,\bar{t}\,)}\,. \\ \vspace{0.2cm}
\end{equation}
This particular scale choice has already been explored in other studies for this process, see e.g. Refs. \cite{Bevilacqua:2012em,Jezo:2021smh,Frederix:2017wme}. We note that in our approach, we construct the momenta of the top quarks from their decay products without accounting for the potential presence of an additional light jet, even if it arises in top-quark decays. We have checked, however, that its inclusion in the construction of top quark momenta does not have any significant impact on our results. Jets are constructed from all final-state partons with $|\eta| < 5$ with the help of the infrared-safe anti-$k_{T}$ jet algorithm \cite{Cacciari:2008gp} with a clustering parameter $R = 0.4$. We require at least four $b$-jets as well as four charged leptons. All final states have to fulfil the following selection criteria that closely mimic the ATLAS and CMS detector response \cite{ATLAS:2023ajo,CMS:2023ftu}: 
\begin{equation}
\begin{array}{lll}
 p_{T,\,\ell}>25 ~{\rm GeV}\,,    
 &\quad \quad \quad \quad\quad|y_\ell|<2.5\,,&
\quad \quad \quad \quad \quad
\Delta R_{\ell
 \ell} > 0.4\,,\\[0.2cm]
p_{T,\,b}>25 ~{\rm GeV}\,,  
&\quad \quad\quad\quad\quad |y_b|<2.5 \,, 
 &\quad \quad\quad \quad \quad
\Delta R_{bb}>0.4\,.
\end{array}
\end{equation}
We set no restriction on the kinematics of the extra light jet (if resolved) and employ no cut on the missing transverse momentum.

%
\section{Explicit expressions for the NWA cross section through NLO}
\label{sec:nwa}
%

In the following, we briefly review some of the features of the NWA that will be exploited throughout our study. The fully factorised form of the differential cross section for the $pp\to t\bar{t}t\bar{t}$ process including top-quark decays in the NWA  can be given by, see e.g.   \cite{Melnikov:2009dn, Campbell:2012uf,Bevilacqua:2019quz,Behring:2019iiv,Czakon:2020qbd,Bevilacqua:2022twl}
\begin{equation}
\label{allorders}
    d\sigma = d\sigma_{t\bar{t}t\bar{t}} \times \frac{d\Gamma_t}{\Gamma_t}  \times \frac{d\Gamma_{\bar{t}}}{\Gamma_t} \times \frac{d\Gamma_t}{\Gamma_t}\times \frac{d\Gamma_{\bar{t}}}{\Gamma_t} \,, \\\vspace{0.2cm}
\end{equation}
where $d\sigma_{tt\bar{t}\bar{t}}$ denotes the  cross section for on-shell $pp \to t\bar{t}t\bar{t}$ production and $d\Gamma_t$ ($d\Gamma_{\bar{t}}$) is the top-quark decay rate for $t\to W^+b \to \ell^+ \nu_\ell b$ ($\bar{t}\to W^- \bar{b} \to \ell^- \bar{\nu}_\ell \bar{b}$). Furthermore, the symbol $\times$ represents spin correlations. This formula holds to all orders in $\alpha_s$. However, all the terms on the right-hand side of Eq. \eqref{allorders} can be expanded in terms of $\alpha_s$ to the required order in perturbative expansion. Thus, for example at the NLO level we have 
\begin{equation}
\begin{split}
\label{expansion1}
    d\sigma^{\rm NLO} &= d\sigma_{tt\bar{t}\bar{t}}^{(0)} + \alpha_s d\sigma_{tt\bar{t}\bar{t}}^{(1)} \,, \\[0.2cm]
    d\Gamma_t^{\rm NLO} &= d\Gamma_t^{(0)} + \alpha_s d\Gamma_t^{(1)} \,,
    \end{split}
\end{equation}
where $d\sigma^{(0)}, \, d\Gamma^{(0)}$  and $d\sigma^{(1)}, \, d\Gamma^{(1)}$  denote LO and NLO  contributions to the production and decay processes, respectively. In addition, in Eq. \eqref{allorders} we have the top-quark width, $\Gamma_t$, which enters the NWA calculation as a parameter through the denominator. There are different ways to treat the $1/\Gamma_t$ factor. For example, it is possible to set the value of $\Gamma_t$ to a numerical value corresponding to the perturbative order of the full calculation, $\Gamma_t^{\rm NLO}$. Alternatively, one can formally expand $\Gamma_t$ in powers of $\alpha_s$ and keep only the terms consistent with the order of the performed calculation. Thus, for example at NLO in QCD we would  have 
\begin{equation}
\label{expansion2}
\Gamma_t^{\rm NLO} = \Gamma_t^{(0)} + \alpha_s \Gamma_t^{(1)} \,.\\\vspace{0.2cm}
\end{equation}
By incorporating Eq. \eqref{expansion1}  into Eq. \eqref{allorders} we obtain the following expression for $d\sigma^{\textrm{LO}}$ at the LO level 
\begin{equation}
\label{nwa_lo}
d\sigma^{\textrm{LO}} = 
 ~d\sigma^{\rm LO}_{t\bar{t}t\bar{t}} \times \frac{d\Gamma^{0}_{t}}
  {\Gamma_{t}^{\rm LO}}
  \times \frac{ d\Gamma^{0}_{\bar{t}}}{\Gamma_{t}^{\rm LO}}
 \times \frac{d\Gamma^{0}_{t}}
  {\Gamma_{t}^{\rm LO}}
  \times \frac{d\Gamma^{0}_{\bar{t}}}{\Gamma_{t}^{\rm LO}} \,,\\\vspace{0.2cm}
\end{equation}
whereas, at the NLO QCD level we would have instead $d\sigma^{\textrm{NLO}}_{\rm full}$ given by 
\begin{equation}
\label{nwa_notexp}
\begin{split}
d\sigma^{\textrm{NLO}}_{\rm full} & ~=
 d\sigma^{\rm NLO}_{t\bar{t}t\bar{t}} \times \frac{d\Gamma^{0}_{t}}
  {\Gamma_{t}^{\rm NLO}}
  \times \frac{ d\Gamma^{0}_{\bar{t}}}{\Gamma_{t}^{\rm NLO}}
 \times \frac{d\Gamma^{0}_{t}}
  {\Gamma_{t}^{\rm NLO}}
  \times \frac{d\Gamma^{0}_{\bar{t}}}{\Gamma_{t}^{\rm NLO}} \\[0.2cm]
 & ~+ d\sigma^{\rm LO}_{t\bar{t}t\bar{t}} \times \frac{d\Gamma^{1}_{t}}
  {\Gamma_{t}^{\rm NLO}}
  \times \frac{ d\Gamma^{0}_{\bar{t}}}{\Gamma_{t}^{\rm NLO}}
 \times \frac{d\Gamma^{0}_{t}}
  {\Gamma_{t}^{\rm NLO}}
  \times \frac{d\Gamma^{0}_{\bar{t}}}{\Gamma_{t}^{\rm NLO}} \\[0.2cm]
  & ~+ d\sigma^{\rm LO}_{t\bar{t}t\bar{t}} \times \frac{d\Gamma^{0}_{t}}
  {\Gamma_{t}^{\rm NLO}}
  \times \frac{ d\Gamma^{1}_{\bar{t}}}{\Gamma_{t}^{\rm NLO}}
 \times \frac{d\Gamma^{0}_{t}}
  {\Gamma_{t}^{\rm NLO}}
  \times \frac{d\Gamma^{0}_{\bar{t}}}{\Gamma_{t}^{\rm NLO}} \\[0.2cm]
  & ~+ d\sigma^{\rm LO}_{t\bar{t}t\bar{t}} \times \frac{d\Gamma^{0}_{t}}
  {\Gamma_{t}^{\rm NLO}}
  \times \frac{ d\Gamma^{0}_{\bar{t}}}{\Gamma_{t}^{\rm NLO}}
 \times \frac{d\Gamma^{1}_{t}}
  {\Gamma_{t}^{\rm NLO}}
  \times \frac{d\Gamma^{0}_{\bar{t}}}{\Gamma_{t}^{\rm NLO}} \\[0.2cm]
  & ~+ d\sigma^{\rm LO}_{t\bar{t}t\bar{t}} \times \frac{d\Gamma^{0}_{t}}
  {\Gamma_{t}^{\rm NLO}}
  \times \frac{ d\Gamma^{0}_{\bar{t}}}{\Gamma_{t}^{\rm NLO}}
 \times \frac{d\Gamma^{0}_{t}}
  {\Gamma_{t}^{\rm NLO}}
  \times 
  \frac{d\Gamma^{1}_{\bar{t}}}{\Gamma_{t}^{\rm NLO}} \,.
  \end{split}
\end{equation}
The latter formula describes the full NLO QCD calculation with higher order QCD corrections incorporated in both $t\bar{t}t\bar{t}$ production and top-quark decays. It is worth pointing out that the various terms in Eq. \eqref{nwa_notexp} are separately infrared finite and do not interfere with each other. There is no cross-talk between the production stage and decays, as well as between different decays of the top quarks.  The presence of $\Gamma_t^{\rm NLO}$ in Eq. \eqref{nwa_notexp}, which is a function of the strong coupling constant, spoils the rigorous expansion of the cross section in powers of $\alpha_s$. If also Eq. \eqref{expansion2} is incorporated into Eq. \eqref{allorders} we would rather have the following expression 
\begin{equation}
\label{nwa_exp}
d\sigma^{\rm NLO}_{\textrm{exp}} = d\sigma^{\rm NLO}  = d\sigma^{\textrm{NLO}}_{\rm full} \times \left(\dfrac{\Gamma^{\textrm{NLO}}_{t}}{\Gamma^{\textrm{LO}}_{t}} \right)^4 - d\sigma^{\textrm{LO}} \times \dfrac{4(\Gamma^{\textrm{NLO}}_{t} - \Gamma^{\textrm{LO}}_{t})}{\Gamma^{\textrm{LO}}_{t}} \,,\\\vspace{0.2cm}
\end{equation}
that represents the corresponding expanded NLO QCD result, labelled as $d\sigma^{\rm NLO}_{\textrm{exp}}$. We note that in Eq. \eqref{nwa_exp}  the $d\sigma^{\textrm{LO}}$ part is calculated with the NLO PDF set, thus, with the two-loop running of $\alpha_s$. Moreover, having all the components of the NLO QCD calculation in the NWA at our disposal, we can also study the case where the NLO QCD corrections in top-quark decays are simply neglected. This last case we label as $d\sigma^{\rm NLO}_{\rm LO_{dec}}$. The contribution to this result comes from the first line of Eq. \eqref{nwa_notexp} only, while  $\Gamma_t^{\rm LO}$ is used in the calculation. Thus, we can write for $d\sigma^{\textrm{NLO}}_{\rm LO_{dec}}$
\begin{equation}
d\sigma^{\textrm{NLO}}_{\rm LO_{dec}} = 
 ~d\sigma^{\rm NLO}_{t\bar{t}t\bar{t}} \times \frac{d\Gamma^{0}_{t}}
  {\Gamma_{t}^{\rm LO}}
  \times \frac{ d\Gamma^{0}_{\bar{t}}}{\Gamma_{t}^{\rm LO}}
 \times \frac{d\Gamma^{0}_{t}}
  {\Gamma_{t}^{\rm LO}}
  \times \frac{d\Gamma^{0}_{\bar{t}}}{\Gamma_{t}^{\rm LO}} \,.\\\vspace{0.2cm}
\end{equation}
In this work we consider predictions in the NWA based on the various levels of accuracy, as described above. Specifically, we are going to provide the results  for $\sigma^{\textrm{NLO}}_{\textrm{exp}}$  (our default findings at NLO QCD) and $\sigma^{\textrm{NLO}}_{\rm LO_{dec}}$ at the integrated and differential cross-section level. Furthermore, we show the results for the $\sigma^{\rm NLO}_{\rm full}$ case. The goal is to analyse the similarities and differences between the three approaches and to estimate the size of higher-order corrections in top-quark decays. As mentioned earlier, in the literature these latter effects are approximated only by parton shower programs. 

%
\section{Integrated fiducial cross sections}
\label{sec:integrated}
%

We begin the presentation of our results for the $pp\to t\bar{t}t\bar{t}$ process in the $4\ell$ top-quark decay channel with a discussion of the integrated fiducial cross section for the LHC Run III energy of $\sqrt{s} = 13.6$ TeV. When considering the input parameters and the phase-space cuts as detailed in Section \ref{sec:setup}, our LO and NLO findings as calculated in the NWA for the fixed scale setting, $\mu_R=\mu_F=\mu_0=2m_t$, and the default MSHT20 PDF set, are given by:
\begin{align}
\begin{split}
    \sigma^{\textrm{LO}} ( \textrm{MSHT20}, \mu_0 &= 2m_t ) = 4.3868(3) ^{+73\%}_{-40\%} \textrm{ [scale] ab} \,,\\ 
    \sigma^{\textrm{NLO}} ( \textrm{MSHT20}, \mu_0 &= 2m_t ) = 4.895(2) ^{+13\%}_{-20\%}  
    \textrm{ [scale] }^{+4\%}_{-3\%}\textrm{ [PDF] ab}\,.
\end{split}
\end{align}
We use the expanded version of the NWA at NLO in QCD, which for the sake of brevity is denoted simply by $\sigma^{\rm NLO}$. On the other hand, when the phase-space dependent scale choice is employed, $\mu_R=\mu_F=\mu_0=E_T/4$, we obtain: 
\begin{align}
\begin{split}
    \sigma^{\textrm{LO}} ( \textrm{MSHT20}, \mu_0 &= E_T/4 ) = 4.7479(3) ^{+74\%}_{-40\%} \textrm{ [scale] ab} \,,\\ 
    \sigma^{\textrm{NLO}} ( \textrm{MSHT20}, \mu_0 &= E_T/4 ) = 5.170(3) ^{+12\%}_{-20\%} \textrm{ [scale] }^{+4\%}_{-3\%}\textrm{ [PDF] ab} \,.
\end{split}
\end{align}
At the LO level, the dominant contribution comes from the $gg$ channel, accounting for approximately $88\%$ of the full integrated fiducial cross section, while the remaining contributions that come from the $q\bar{q}/\bar{q}q$ channels collectively make up about $12\%$. The NLO QCD corrections for this process are rather moderate, in the range of $9\%-12\%$, depending on the scale setting. They are much smaller than the LO scale uncertainties, which are of the order of $70\%$ for both scale choices. The incorporation of NLO calculations significantly reduces the latter uncertainties down to $20\%$. We note here, that for both scale settings and independently of the order in perturbative expansion these uncertainties are asymmetric. Apart from the theoretical error arising from the scale dependence, we must also account for another source of uncertainty linked to the parameterisation of PDFs. We use the prescription from the MSHT20 group to provide the $68\%$ confidence level (C.L.) PDF uncertainties. These internal PDF uncertainties are of the order of $4\%$ for both scale settings. Thus, they are well below the theoretical uncertainties due to the scale dependence.  The latter uncertainties remain the dominant source of the systematics.
\begin{table}[t!]
\centering
\scalebox{0.9}{
\begin{tabular}{ccccccc}
\midrule\midrule
PDF & $\sigma^{\textrm{LO}}$ [ab] & $\delta_{scale}$ & $\sigma^{\textrm{NLO}}$ [ab] & $\delta_{scale}$ & $\delta_{PDF}$  &  $\mathcal{K} = \sigma^{\textrm{NLO}}/\sigma^{\textrm{LO}}$\\
\midrule\midrule
\multicolumn{7}{c}{\centering $\mu_R = \mu_F = \mu_0 = 2m_t$}\\
\midrule\midrule
MSHT20 & 4.3868(3) & \begin{tabular}[c]{@{}c@{}}$+3.2237$ (73\%)\\$-1.7332$ (40\%)\end{tabular} & 4.895(2) & \begin{tabular}[c]{@{}c@{}}$+0.624$ (13\%)\\$-1.002$ (20\%)\end{tabular} & \begin{tabular}[c]{@{}c@{}}$+0.211$ (4\%)\\$-0.156$ (3\%) \end{tabular} & 1.12\\
\\
NNPDF3.1 & 3.7389(2) & \begin{tabular}[c]{@{}c@{}}$+2.6811$ (72\%)\\$-1.4545$ (39\%)\end{tabular} & 4.846(2) & \begin{tabular}[c]{@{}c@{}}$+0.632$ (13\%)\\$-1.002$ (21\%)\end{tabular} & \begin{tabular}[c]{@{}c@{}}$+0.105$ (2\%)\\$-0.105$ (2\%) \end{tabular} & 1.30\\
\\
CT18 & 4.6757(3) & \begin{tabular}[c]{@{}c@{}}$+3.3754$ (72\%)\\$-1.8311$ (39\%)\end{tabular} & 4.857(2) & \begin{tabular}[c]{@{}c@{}}$+0.620$ (13\%)\\$-0.992$ (20\%)\end{tabular} & \begin{tabular}[c]{@{}c@{}}$+0.289$ (6\%)\\$-0.236$ (5\%) \end{tabular} & 1.04\\
\midrule\midrule
\multicolumn{7}{c}{\centering $\mu_R = \mu_F = \mu_0 = E_T/4$}\\
\midrule\midrule
MSHT20 & 4.7479(3) & \begin{tabular}[c]{@{}c@{}}$+3.5156$ (74\%)\\$-1.8855$ (40\%)\end{tabular} & 5.170(3) & \begin{tabular}[c]{@{}c@{}}$+0.638$ (12\%)\\$-1.056$ (20\%)\end{tabular} & \begin{tabular}[c]{@{}c@{}}$+0.219$ (4\%)\\$-0.162$ (3\%) \end{tabular} & 1.09\\
\\
NNPDF3.1 & 4.0930(3) & \begin{tabular}[c]{@{}c@{}}$+2.9792$ (73\%)\\$-1.6063$ (39\%)\end{tabular} & 5.126(3) & \begin{tabular}[c]{@{}c@{}}$+0.634$ (12\%)\\$-1.055$ (21\%)\end{tabular} & \begin{tabular}[c]{@{}c@{}}$+0.110$ (2\%)\\$-0.110$ (2\%) \end{tabular} & 1.25\\
\\
CT18 & 5.0003(3) & \begin{tabular}[c]{@{}c@{}}$+3.6151$ (72\%)\\$-1.9623$ (39\%)\end{tabular} & 5.127(3) & \begin{tabular}[c]{@{}c@{}}$+0.636$ (12\%)\\$-1.045$ (20\%)\end{tabular} & \begin{tabular}[c]{@{}c@{}}$+0.299$ (6\%)\\$-0.245$ (5\%) \end{tabular} & 1.03\\
\midrule\midrule
\end{tabular}
}
\caption{\textit{Integrated fiducial cross sections at LO and NLO in QCD for the $pp \to t\bar{t}t\bar{t}$ process in the $4\ell$ channel at the LHC with $\sqrt{s} = $ 13.6 TeV. Results are presented for MSHT20, NNPDF3.1 and CT18 PDF sets. They are evaluated using $\mu_0=2m_t$ and $\mu_0=E_T/4$. Also given are theoretical uncertainties coming from the scale variation and from PDFs. MC integration errors are displayed in parentheses.} In the last column the ${\cal K}$-factor is shown.}
\label{tab:lo_nlo} 
\end{table}
\begin{table}[t!]
\centering
\begin{tabular}{ccccc}
\midrule\midrule
\textsc{Scale} & \textsc{Order} & PDF & $\sigma$ [ab] & $\mathcal{K} = \sigma^{\textrm{NLO}}/\sigma^{\textrm{LO}}$ \\
\midrule\midrule
$\mu_0 = 2m_t$ & LO &  \verb|NNPDF3.1_lo_as_0130| &  $3.7389(2)^{+72\%}_{-39\%}$ &  1.30\\
\\
 & LO &  \verb|NNPDF3.1_lo_as_0118| &  $2.8835(2)^{+65\%}_{-37\%}$ &  1.68\\
\\
 & LO &  \verb|NNPDF3.1_nlo_as_0118| &  $3.1068(2)^{+68\%}_{-38\%}$ &  1.56\\
\\
 & NLO &  \verb|NNPDF3.1_nlo_as_0118| &  $4.846(2)^{+13\%}_{-21\%}$ &  -\\
\midrule\midrule
$\mu_0 = E_T/4$ & LO &  \verb|NNPDF3.1_lo_as_0130| &  $4.0930(3)^{+73\%}_{-39\%}$ &  1.25 \\
\\
 & LO &  \verb|NNPDF3.1_lo_as_0118| &  $3.1356(3)^{+66\%}_{-37\%}$ &  1.64\\
\\
 & LO &  \verb|NNPDF3.1_nlo_as_0118| &  $3.3633(2)^{+69\%}_{-38\%}$ &  1.52\\
\\
 & NLO &  \verb|NNPDF3.1_nlo_as_0118| &  $5.126(3)^{+12\%}_{-21\%}$ &  -\\
\midrule\midrule
\end{tabular}
\caption{\textit{Integrated fiducial cross sections at LO and NLO in QCD for the $pp \to t\bar{t}t\bar{t}$ process in the $4\ell$ channel at the LHC with $\sqrt{s} = $ 13.6 TeV. The LO and NLO NNPDF3.1 PDF sets are employed for LO predictions with different values of $\alpha_s (m_Z)$ using $\mu_R = \mu_F = \mu_0$ where $\mu_0=2m_t$ and $\mu_0=E_T/4$. Also given are NLO predictions and theoretical uncertainties coming from the scale variation. MC integration errors are displayed in parentheses. In the last column the ${\cal K}$-factor is also shown.}}
\label{tab:lo_nlo_nnpdf} 
\end{table}
\begin{figure}[t!]
        \centering        
        \includegraphics[width=0.8\textwidth]{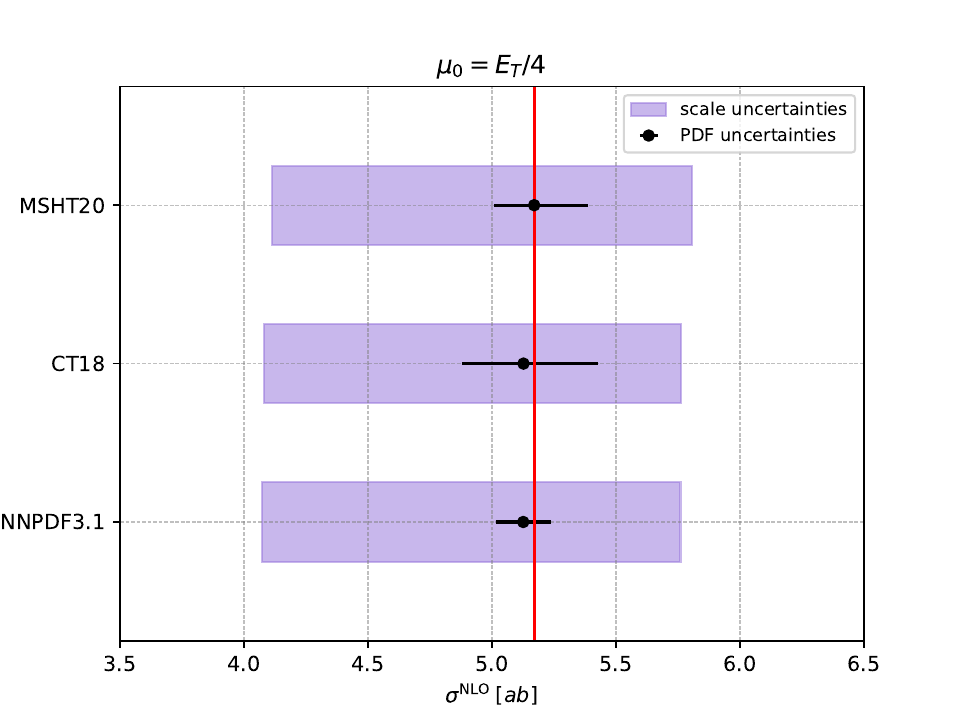}  \caption{\textit{Integrated fiducial cross sections at NLO in QCD for the $pp \to t\bar{t}t\bar{t}$  process in the $4\ell$ channel at the LHC with $\sqrt{s} = $ 13.6 TeV. Scale and PDF uncertainties are depicted for the dynamical scale setting $\mu_R = \mu_F = \mu_0 = E_T /4$. Results are evaluated using the following NLO PDF sets: MSHT20, CT18 and NNPDF3.1. The red vertical line represents the central prediction for our default MSHT20 PDF set.}}
         \label{fig:pdf_unc}
\end{figure}
\begin{figure}[t!]
\centering   
        \includegraphics[width=0.49\textwidth]{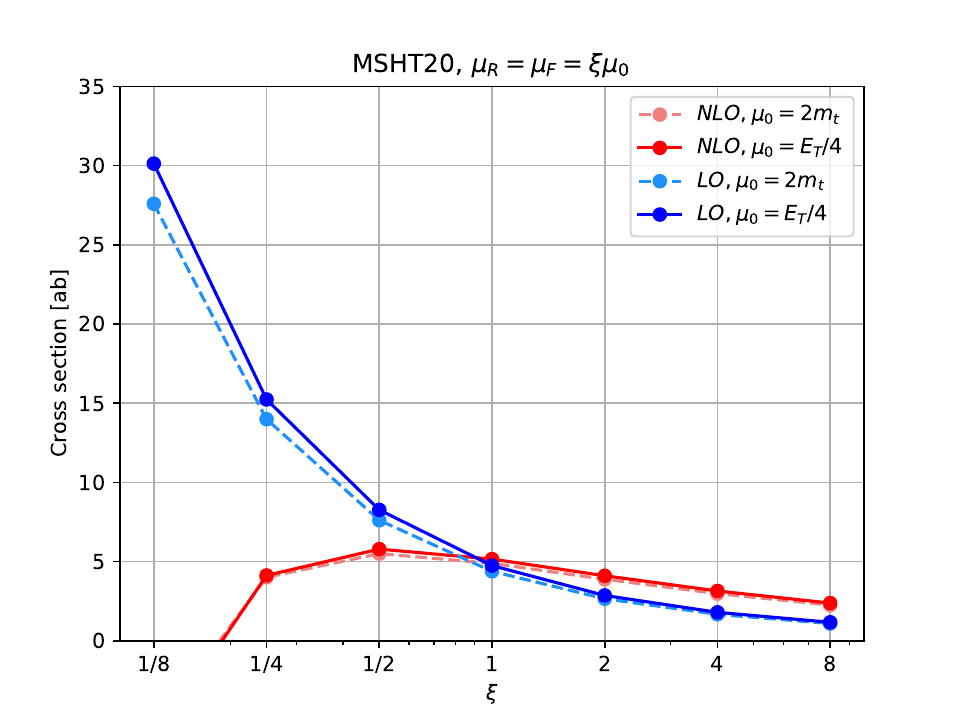}\\
                \includegraphics[width=0.49\textwidth] {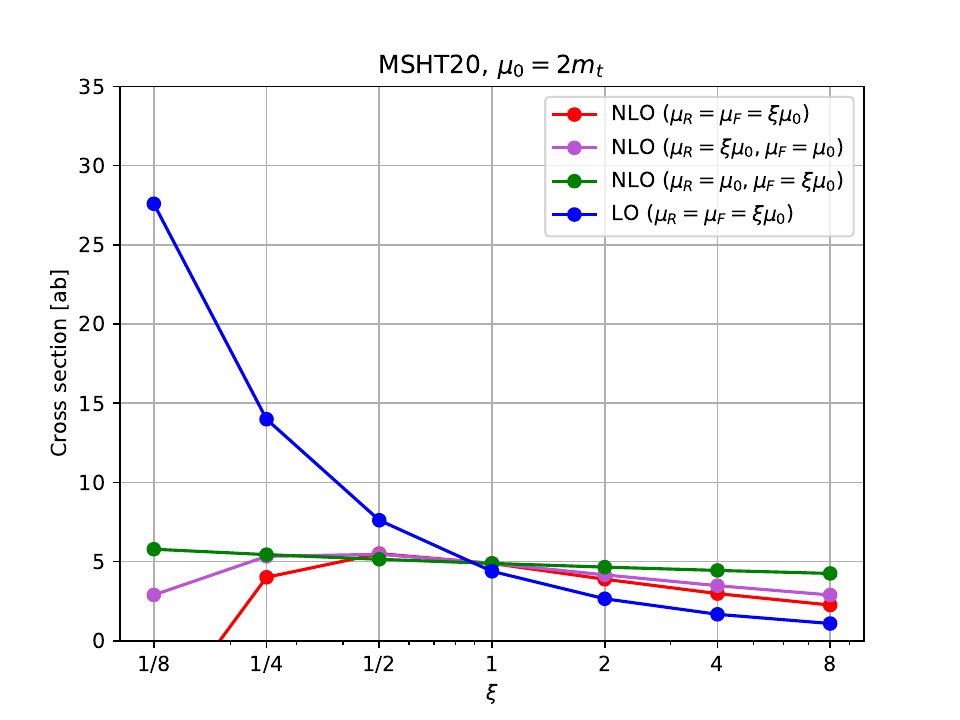}    
        \includegraphics[width=0.49\textwidth]{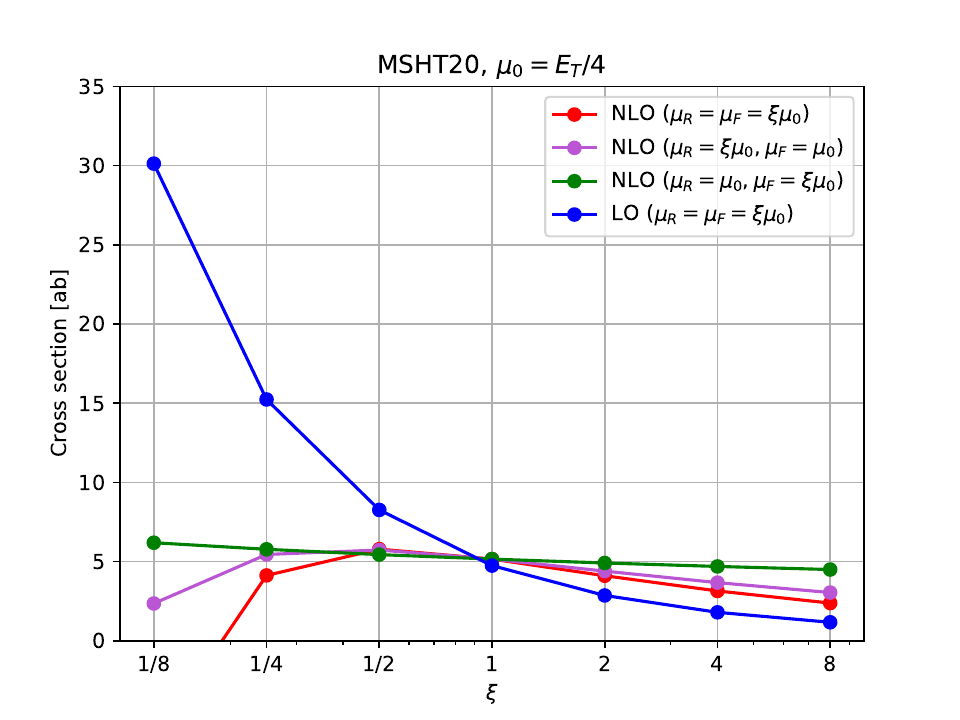} 
    \caption{\label{fig:ksi}\it Scale dependence of the integrated fiducial cross section at LO and NLO for  the $pp \to t\bar{t}t\bar{t}$ process in the $4\ell$ channel at the LHC with $\sqrt{s} = $ 13.6 TeV. The LO and the NLO MSHT20 PDF sets are employed. Renormalisation and factorisation scales are set to the common value $\mu_R=\mu_F=\mu_0$ where $\mu_0=2m_t$ and $\mu_0=E_T/4$.
    Both scales are varied simultaneously by a factor of $\xi$ in the following range  $\xi \in \left\{0.125, \dots, 8 \right\}$. For each case of $\mu_0$ also shown is the variation of $\mu_R$ with fixed $\mu_F$ and the variation of $\mu_F$ with fixed $\mu_R$.}
\end{figure}
\begin{figure}[t!]  
        \centering
        \includegraphics[width=0.49\textwidth]{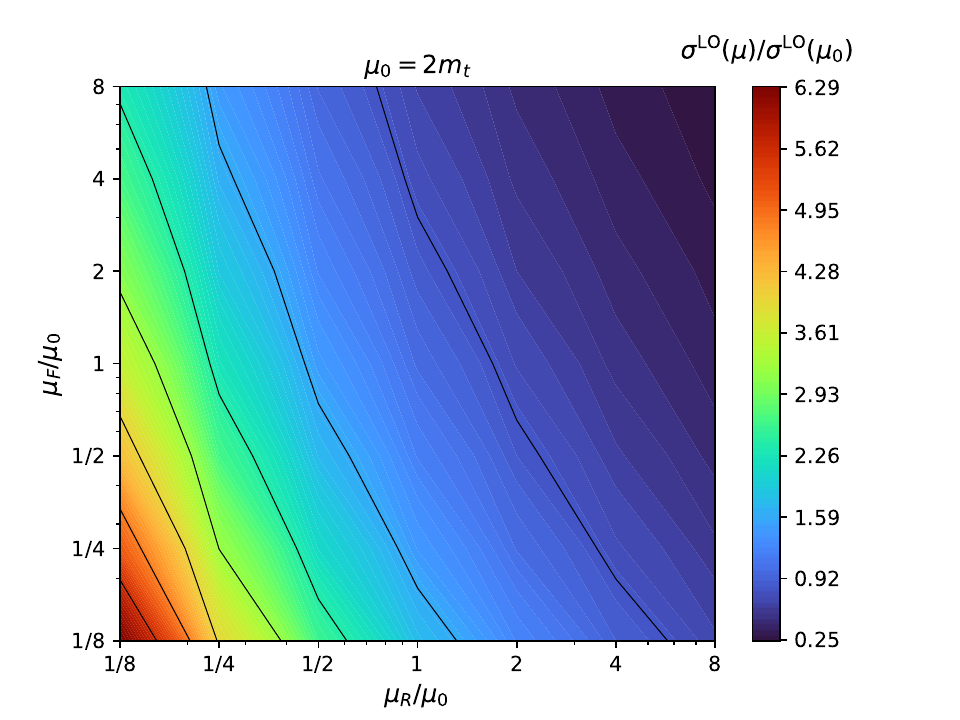}
        \includegraphics[width=0.49\textwidth]{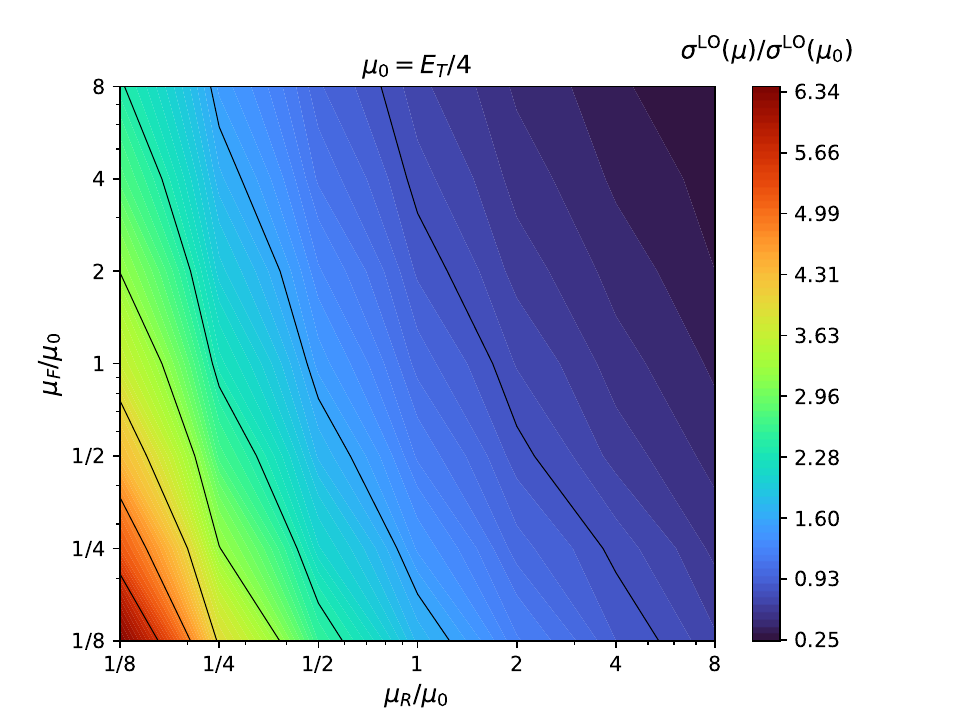}
        \includegraphics[width=0.49\textwidth]{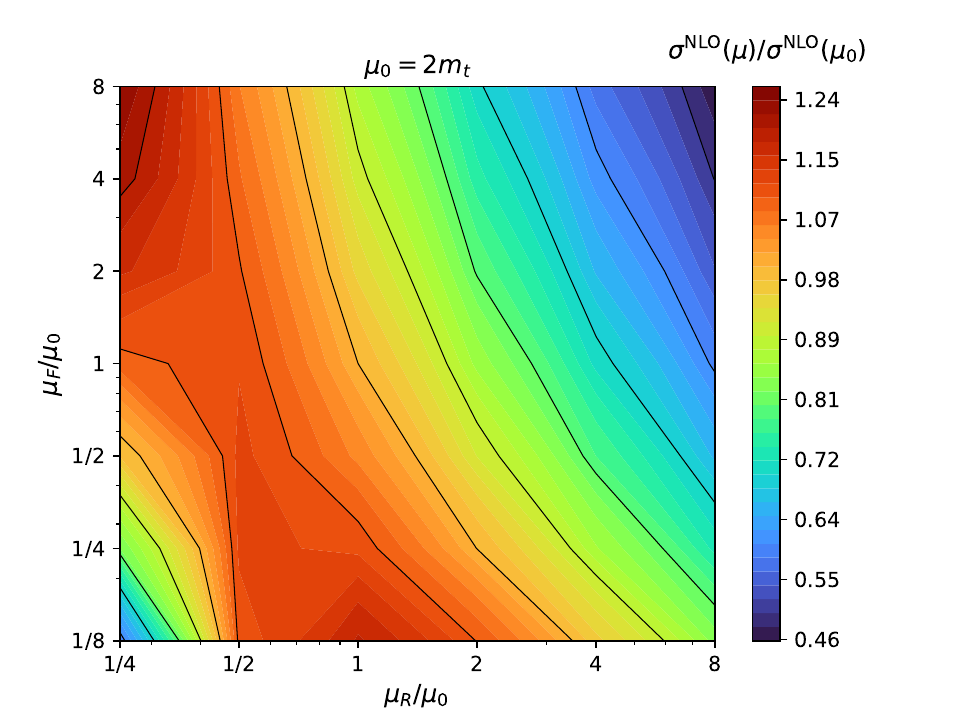} 
        \includegraphics[width=0.49\textwidth]{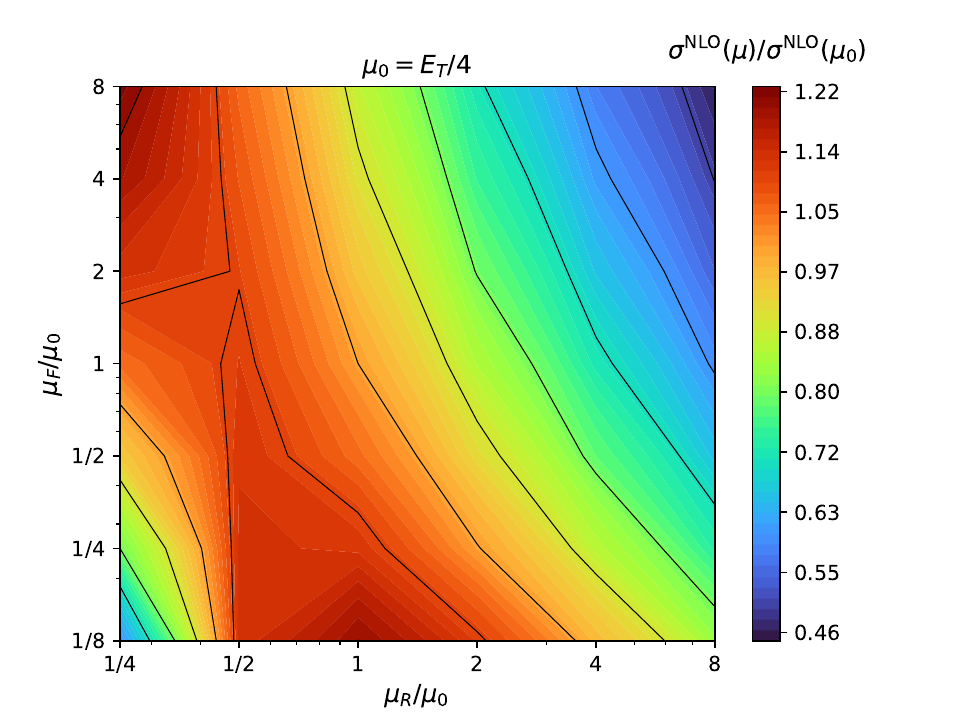}
    \caption{\textit{Contour plots for the scale dependence in the ($\mu_R, \mu_F$) plane at LO (upper plots) and NLO (lower plots) for our default PDF set MSHT20 and two scale choices $\mu_0 = 2m_t \; \textrm{and} \; \mu_0 = E_T/4$. Results are presented for the $pp \to t\bar{t}t\bar{t}$  process in the $4\ell$ channel at the LHC with $\sqrt{s} = $ 13.6 TeV.}}
    \label{fig:2D_plots}
\end{figure}

Within a given PDF set we can not assess the differences that enter into the parameterisation of the various PDF sets. To investigate this issue,  following the recommendation of the PDF4LHC group for the usage of PDFs suitable for applications at the LHC that are connected to SM processes, we additionally present our results for NNPDF3.1 and CT18 PDF sets. The LO and NLO integrated fiducial cross sections for the three PDF sets are shown in Table \ref{tab:lo_nlo} for two different scale choices: $\mu_0=2m_t$ and $\mu_0=E_T/4$. The $\mathcal{K}$ factor, defined as ${\cal K}=\sigma^{\rm NLO}/\sigma^{\rm LO}$, is also provided in the last column. The $68\%$ C.L. PDF uncertainties for NNPDF3.1 are up to $2\%$ only, whereas for the CT18 PDF set they are of the order of  $6\%$. We observe substantial variations in the values of the LO cross section for the three LO PDF sets that are provided for $\alpha_s(m_Z) = 0.130$.  These differences, ranging from $5\%$ to $15\%$, result 
in a variety of the  $\mathcal{K}$ factor: ${\cal K}=(1.25-1.30)$ for NNPDF3.1 and ${\cal K}=(1.03-1.04)$ for CT18. As expected, using NLO PDF sets instead for the LO calculations would correspond to lower values of $\sigma^{\rm LO}$ and, consequently, higher ${\cal K}$ factors. 
Similar effects would be observed when using LO PDF sets with $\alpha_s(m_Z)=0.118$.  In order to substantiate these statements in Table  \ref{tab:lo_nlo_nnpdf} we present LO integrated fiducial cross sections as obtained with the following PDF sets provided by the NNPDF collaboration: the LO NNPDF3.1 PDF set with  $\alpha_s(m_Z) = 0.130$ and $\alpha_s(m_Z) = 0.118$ as well as the NLO NNPDF3.1 PDF set. For consistency and readability, the NLO cross section is also displayed. Again, the results are given for two scale settings $\mu_0=2m_t$ and $\mu_0=E_T/4$. We can observe that the LO results are simply not adequate to describe the $pp\to t\bar{t}t\bar{t}$ process in the $4\ell$ channel.  They are sensitive not only to scale variation, but also to the choice of PDF set and the value of  $\alpha_s(m_Z)$ used.  This large dependence is reflected in the obtained ${\cal K}$-factor, which for the process at hand ranges from ${\cal K}=1.25$ to ${\cal K}=1.68$ within a given PDF set and from ${\cal K}=1.03$ to ${\cal K}=1.68$ once the three PDF sets, which are recommended by the PDF4LHC working group, are employed.  On the other hand, at the NLO QCD level and independent of the scale setting, the differences among $\sigma^{\rm NLO}$ for the three PDF sets (MSHT20, NNPDF3.1 and CT18) are of the order of $1\%$ only. Generally, as explained by the various PDF collaborations there are large differences between the LO and NLO PDF sets, both for central values and uncertainties. In particular, the LO uncertainties are significant due to the poor quality of the fit. This causes the shift in quark PDFs that can be substantial, while the gluon PDF at small $x$ is completely different between LO and NLO, see e.g. Ref. \cite{NNPDF:2017mvq}. This is due to the fact that the singular small-$x$ behaviour of the quark to gluon splittings only starts at NLO. In addition, the PDF global analyses involve data from deep-inelastic scattering (DIS) experiments, fixed-target Drell-Yan  (DY) data, and collider measurements from the Tevatron and LHC. However, gluon initiated DIS and DY processes, e.g. electroweak boson production  processes from ATLAS, CMS and LHCb that are very important for the determination of the gluon PDF, vanish at LO.

The graphical representation of the NLO results for the three PDF sets: MSHT20, CT18 and NNPDF3.1 for the $pp\to t\bar{t}t\bar{t}$ process in the $4\ell$ top-quark decay channel is provided in Figure \ref{fig:pdf_unc}. Since the results at the integrated fiducial cross-section level are similar for $\mu_0=2m_t$ and $\mu_0=E_T/4$, for simplicity we display the results only for the dynamic scale choice. For comparison, we also show the central prediction for our default MSHT20 PDF set, which is represented by a red vertical line. Figure \ref{fig:pdf_unc} shows a clear representation of the theoretical-error budget for our NLO QCD predictions for this process. We can observe that even though the scale uncertainties are asymmetric for the three PDF sets, they remain consistent in magnitude and are of the order of $20\%$. In each case, they also dominate the size of the final theoretical error. 

The graphical representation of the result for the scale dependence for both scale settings: $\mu_0=2m_t$ and $\mu_0= E_T /4$ is shown in Figure \ref{fig:ksi}. The behaviour of LO and NLO cross sections for the default  MSHT20 PDF sets is shown while varying the renormalisation and factorisation scales simultaneously $\mu_R = \mu_F = \xi \mu_0$ by a factor of $\xi$ in the range $\xi \in  \left\{0.125, ..., 8\right\}$.  For each case of $\mu_0$ also shown is the variation of $\mu_R$ with fixed $\mu_F$ and the variation of $\mu_F$ with fixed $\mu_R$. As already discussed, the LO scale dependence is large illustrating the well known fact that the LO prediction can only provide a rough estimate of the cross section for the process at hand. A significant reduction in the scale uncertainty can be seen when NLO QCD corrections are incorporated. We can further observe that the primary source of NLO scale uncertainty is related to the variation of $\mu_R$ (purple line). Changes in the factorisation scale $\mu_F$ (green line) have a relatively small impact on the integrated fiducial cross section. Lastly, we remark that scale choices with $ \xi < 0.25 $ might yield negative values for the integrated NLO fiducial cross section and thus should be avoided since they correspond to unphysical results.

The dependence of the LO and NLO cross sections on $\mu_R$ and $\mu_F$, which are varied independently around a central value of the scale, is presented in Figure \ref{fig:2D_plots}. Thus, this time we plot the distribution of the LO and NLO cross sections in the $\mu_R - \mu_F$ plane. In addition to the previous three special cases shown in Figure \ref{fig:ksi}: $\mu_R=\mu_F=\xi\mu_0$, $\mu_R=\mu_0,\,\mu_F=\xi\mu_0$ and $\mu_R=\xi\mu_0,\,\mu_F=\mu_0$, also other cases are depicted here. These contour plots provide complementary information to the previous scale dependence plots. We display the ratio of the integrated fiducial cross section $\sigma^{\rm (N)LO}(\mu_{R,\, F}=\xi\mu_0)$ to the one calculated using the central value of the scale $\sigma^{\rm (N)LO}(\mu_{R,\, F}=\mu_0)$. It should be noted that the region with $\mu_R < \mu_0/4$ is not included in the NLO plots due to the negative values of the integrated cross section, as already explained above. Analysing the entire plotted range of $\xi$, as we move toward lower values of $\mu_R$ and $\mu_F$, we can see a significant increase, exceeding a factor of $6$, in the $\sigma^{\rm LO} (\mu_{R,\,F}=\xi\mu_0) / \sigma^{\rm LO} (\mu_{R,\,F}=\mu_0)$ ratio. On the other hand, larger scale values yield smaller cross section ratios. At NLO, the scale variations are less pronounced, reaching an upper limit of $1.22-1.24$ and a lower limit of $0.46$ for the $\sigma^{\rm NLO}(\mu_{R,\, F}=\xi \mu_0) / \sigma^{\rm NLO} (\mu_{R,\, F} =\mu_0) $ ratio for both scale settings. Thus, when the whole range of $\xi$ is considered the uncertainties are substantial, even at the NLO level. Again we can observe that the scale variation is driven by the changes in $\mu_R$. 

We can conclude this part by saying that scanning all potential $(\mu_R, \mu_F)$ pairs in the range $0.5 \le \mu_R/\mu_0\le 2$ and $0.5 \le \mu_F/\mu_0 \le 2 $, without applying the additional requirement of $0.5 \le \mu_R/\mu_F \le 2$, would not change the size of the NLO scale uncertainties.  Indeed, for the two additional extreme cases $(\mu_R/\mu_0,\mu_F/\mu_0)=\left\{(2,0.5), (0.5,2)\right\}$, that are usually omitted, the magnitude of the scale uncertainties is in the range of $8\%-10\%$ only,  independently of the scale setting. Thus, the size of the NLO QCD theoretical error for the $pp \to t\bar{t}t\bar{t}$ process in the $4\ell$ channel would remain the same even if these two cases were included. 
\begin{table}[t!]
\centering
\label{tab:lodec}
\resizebox{\columnwidth}{!}{
\begin{tabular}{ccccc}
\midrule\midrule
Decay treatment & $\sigma_i^{\rm NLO}$ [ab] & $+\delta_{scale}$ [ab] & $-\delta_{scale}$ [ab] & $\sigma_i^{\rm NLO}/\sigma^{\rm NLO}_{\rm exp}-1$\\
\midrule\midrule
\multicolumn{5}{c}{\centering $\mu_R = \mu_F = \mu_0 = 2m_t$}\\
\midrule\midrule
${\rm full}$ & 5.462(3) & $+0.156$ (3\%) & $-0.853$ (16\%) & $+11.6\%$ \\
\\
${\rm LO_{dec}}$ & 5.295(3) & $+1.123$ (21\%) & $-1.224$ (23\%) & $+8.2\%$\\
\\
${\rm exp}$ & 4.895(2) & $+0.624$ (13\%)  & $-1.002$ (20\%) & $-$\\
\midrule\midrule
\multicolumn{5}{c}{\centering $\mu_R = \mu_F = \mu_0 = E_T/4$}\\
\midrule\midrule
${\rm full}$ & 5.735(3) & $+0.139\;(2\%)$ & $-0.882\;(15\%) $ & $+10.9\%$\\
\\
${\rm LO_{dec}}$ & 5.646(3) & $+1.225\;(22\%)$ & $-1.317\;(23\%)$ & $+9.2\%$\\
\\
${\rm exp}$ & 5.170(3) & $+0.638\;(12\%)$ & $-1.056\;(20\%)$ & $-$ \\
\midrule
\midrule
\end{tabular}
}
\caption{\textit{Integrated fiducial cross sections at NLO in QCD for  the $pp \to t\bar{t}t\bar{t}$  process in the $4\ell$ channel at the LHC with $\sqrt{s} = $ 13.6 TeV for three different scenarios: $\sigma^{\rm NLO}_{\rm full}$, $\sigma^{\rm NLO}_{\rm LO_{dec}}$ and  $\sigma^{\rm NLO}_{\rm exp}$. All results are given for our default setup with the MSHT20 PDF set and for two different scale choices: $\mu_0=2m_t$ and $\mu_0=E_T/4$. Also given are theoretical uncertainties coming from the scale variation.  MC integration errors are displayed in parentheses. In the last column the ratio to the $\sigma^{\rm NLO}_{\rm exp}$ result is 
provided.}}
\label{tab:scenarios} 
\end{table}

In the next step, we compare our NLO QCD predictions at the integrated fiducial cross-section level, labelled as  $\sigma^{\rm NLO}_{\rm exp}$, to $\sigma^{\rm NLO}_{\rm LO_{dec}}$ and $\sigma^{\rm NLO}_{\rm full}$. As described earlier, $\sigma^{\rm NLO}_{\rm full}$ corresponds to the full NWA calculation that takes into account QCD corrections in   $t\bar{t}t\bar{t}$ production and top-quark decays, while  $\Gamma_t^{\rm NLO}$ is treated as a fixed parameter. On the other hand, the result for  $\sigma^{\rm NLO}_{\rm LO_{dec}}$ utilises NLO QCD corrections in the production stage only. In this case, top-quark decays are described with the LO accuracy using $\Gamma_t^{\rm LO}$. The NLO QCD results for the three scenarios are summarised in Table \ref{tab:scenarios}. A few comments are in order. Firstly, we can notice that the difference between the result in the full NWA and the expanded one is in the range of $11\%-12\%$, thus, well within the corresponding NLO uncertainties, which are at the $15\%-16\%$ and $20\%$ level,  for $\sigma^{\rm NLO}_{\rm full}$ and $\sigma^{\rm NLO}_{\rm exp}$ respectively. Secondly, the expansion of $\Gamma_t^{\rm NLO}$ in terms of $\alpha_s$ results in the increased size of the scale dependence. Furthermore, we can compare  the  $\sigma^{\rm NLO}_{\rm LO_{dec}}$ result with the $\sigma^{\rm NLO}_{\rm exp}$ one. The omission of  NLO QCD corrections in top-quark decays increases not only the value of the integrated fiducial cross section by $8\%-9\%$, but also the size of the theoretical uncertainties due to the scale dependence from $20\%$ to $23\%$. Had we compared $\sigma^{\rm NLO}_{\rm LO_{dec}}$ with the full $\sigma^{\rm NLO}_{\rm full}$ result rather than with the expanded one, the difference in the absolute value of the integrated fiducial cross section would be much smaller. Indeed, since various values for the top-quark width are used in both results, the difference is at the level of $2\%-3\%$ only depending on the scale setting.   We note that NLO QCD corrections to the top-quark width are negative and of the order of $9\%$. The $\alpha_s$ expansion  of $\Gamma_t$ through NLO in QCD with our input parameters and for $\alpha_s(\mu_R=m_t)=0.1076705$ has the following form
\begin{equation}
\Gamma_t= (1.4806842 - 1.1803223 \,\alpha_s ) \, {\rm GeV}\,.
\end{equation}
On the other hand, the change in the size of the scale dependence would be more substantial as we would observe the increase from $15\%-16\%$ to $23\%$. 

We conclude this section by stating that the higher-order QCD effects in top-quark decays for the $pp \to t\bar{t}t\bar{t}$ process in the $4\ell$ channel are substantial, and amount to half of the theoretical uncertainties due to scale variation. Moreover, their inclusion reduces the overall size of theoretical uncertainties.  Finally, as expected, the NLO QCD result calculated with the $1/\Gamma_t$ term not expanded has the smallest theoretical uncertainties, which are in the range of $15\%-16\%$.

%
\section{Differential fiducial cross-section distributions}
\label{sec:diff}
%

While interesting, the NLO QCD corrections to the integrated fiducial cross section do not give us a complete picture of the impact of the higher-order effects on the $pp \to t\bar{t}t \bar{t}$ process in the $4\ell$ channel. To understand how NLO QCD corrections affect specific observables and/or phase-space regions, it is necessary to analyse the results at the differential fiducial cross-section level.  To this end, we start with the comparison between LO and NLO QCD computations for several observables. Additionally, we can inspect the size of theoretical uncertainties in different parts of the phase space.

It should be mentioned here that measuring the differential cross-section distributions shown below will be rather difficult even at the High-Luminosity phase of the LHC. The reason behind this is the expected number of events for this process. Indeed, considering  only $\ell^\pm=e^\pm, \mu^\pm$, the expected number of events is $N=(15-20)$  per experiment, assuming an integrated luminosity of $(3-4) ~{\rm ab}^{-1}$.  On the other hand,  when $\ell^\pm=e^\pm, \mu^\pm,\tau^\pm$ we can estimate $N=(157-209)$ events  for both ATLAS and CMS. We note that the numbers given here correspond to our definition of the fiducial phase-space region. These numbers will change when more inclusive event selection cuts are applied. Nevertheless, proper modelling of differential cross-section distributions is important for example for the many preparatory studies and various analyses that are currently being conducted for this process. In addition,  the analysis at the differential cross-section level for the SM $pp \to t\bar{t}t\bar{t}t + X$  process is a necessary step towards a correct interpretation of possible signals of new physics that may arise in the $4\ell$ channel.
\begin{figure}[!t]
        \includegraphics[width=0.5\linewidth]{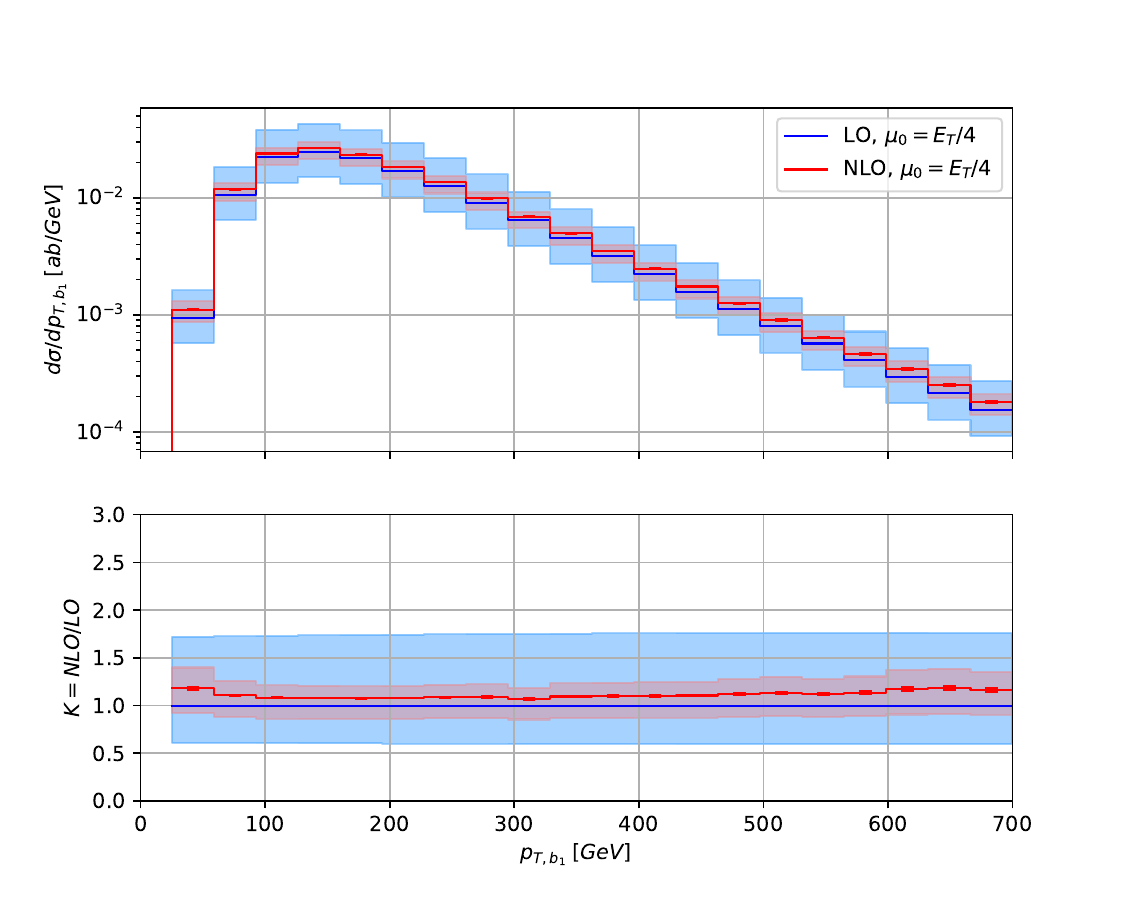}
        \includegraphics[width=0.5\linewidth]{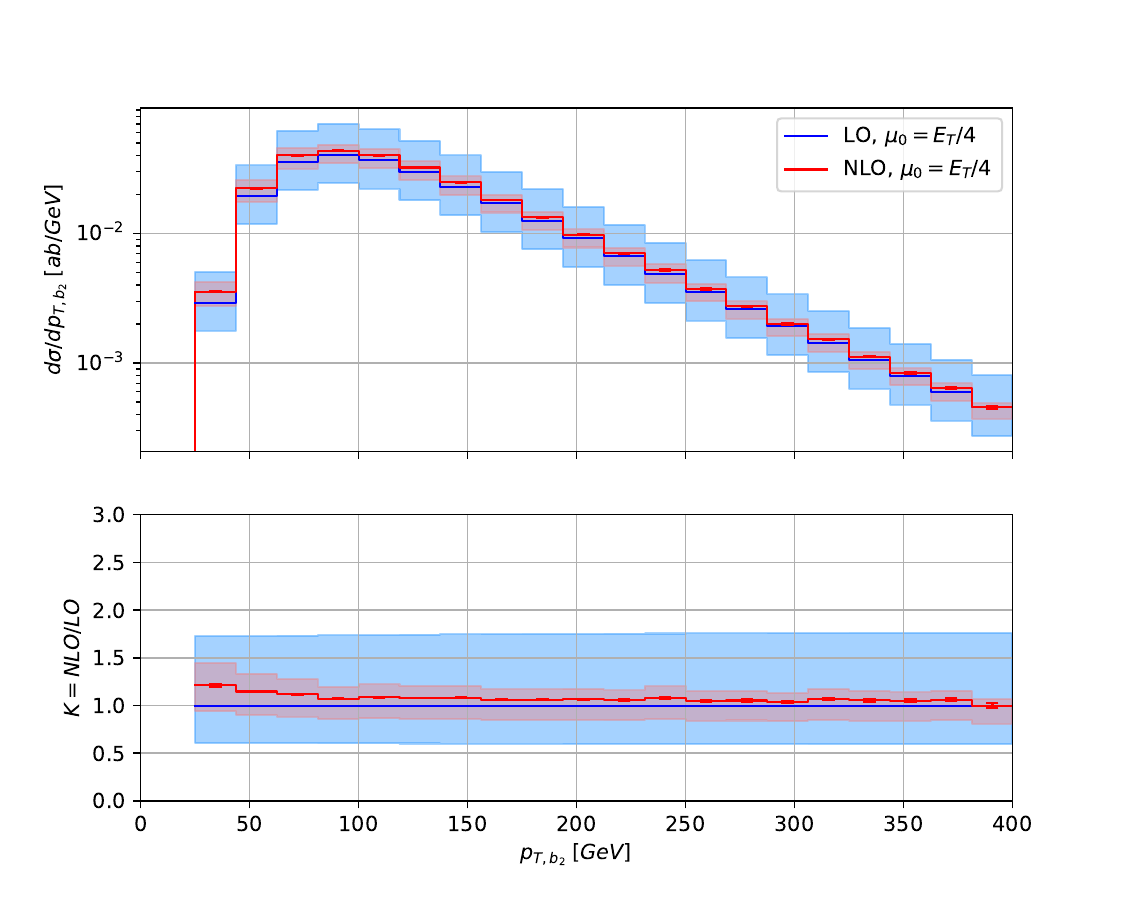}
        \includegraphics[width=0.5\textwidth]{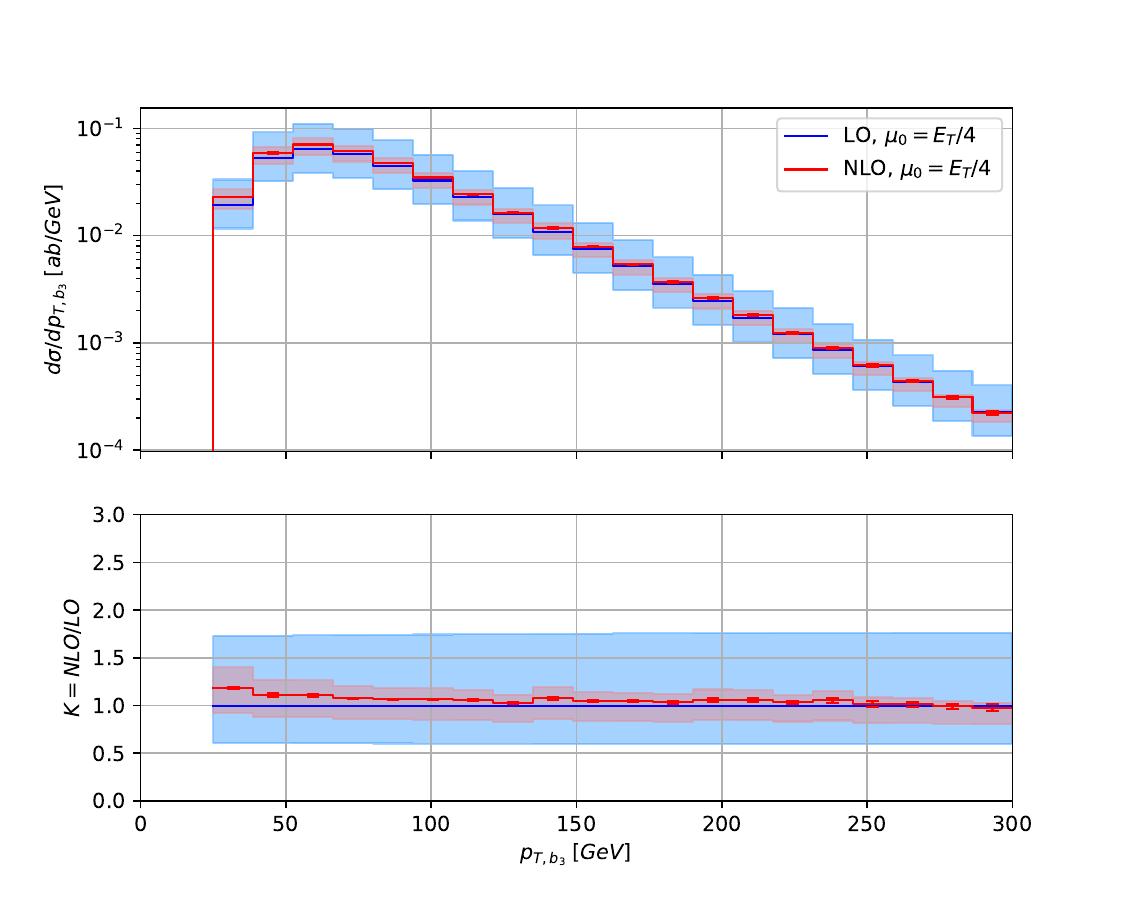} 
        \includegraphics[width=0.5\textwidth]{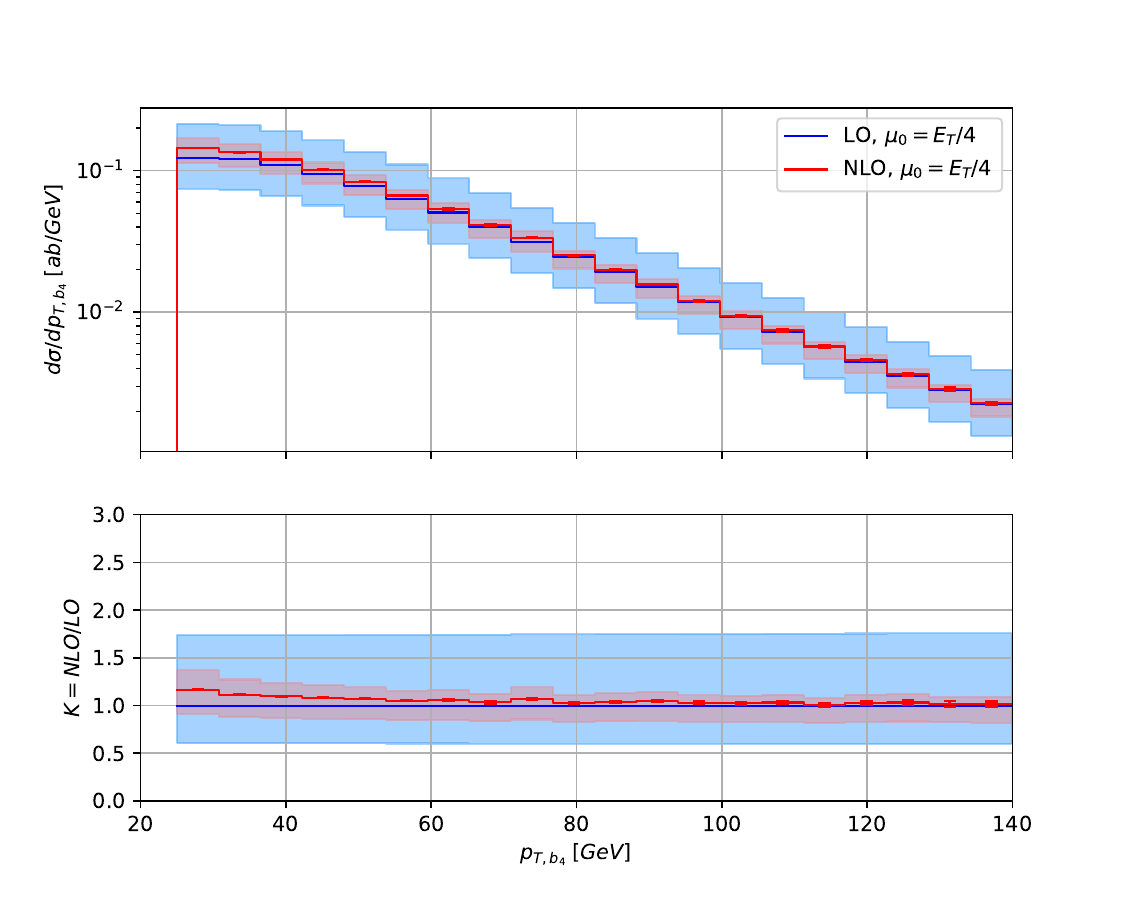}
\caption{\textit{Differential cross-section distributions for  the $pp \to t\bar{t}t\bar{t}$  process in the $4\ell$ channel at the LHC with $\sqrt{s} = 13.6$ TeV. The transverse momenta of the four $b$-jets, that are ordered according to their $p_T$, are displayed for the dynamical scale setting $\mu_R=\mu_F=\mu_0=E_T/4$ and the (N)LO MSHT20 PDF set.  The blue (red) curve corresponds to the LO (NLO) result. Also shown are the corresponding uncertainty bands resulting from scale variations. The lower panels present the differential ${\cal K}$-factor together with its uncertainty band and the relative scale uncertainties of the LO cross section.  Monte Carlo integration errors are displayed in both panels. }}
    \label{fig:ptb}
\end{figure}
\begin{figure}[!t]  
        \includegraphics[width=0.5\linewidth]{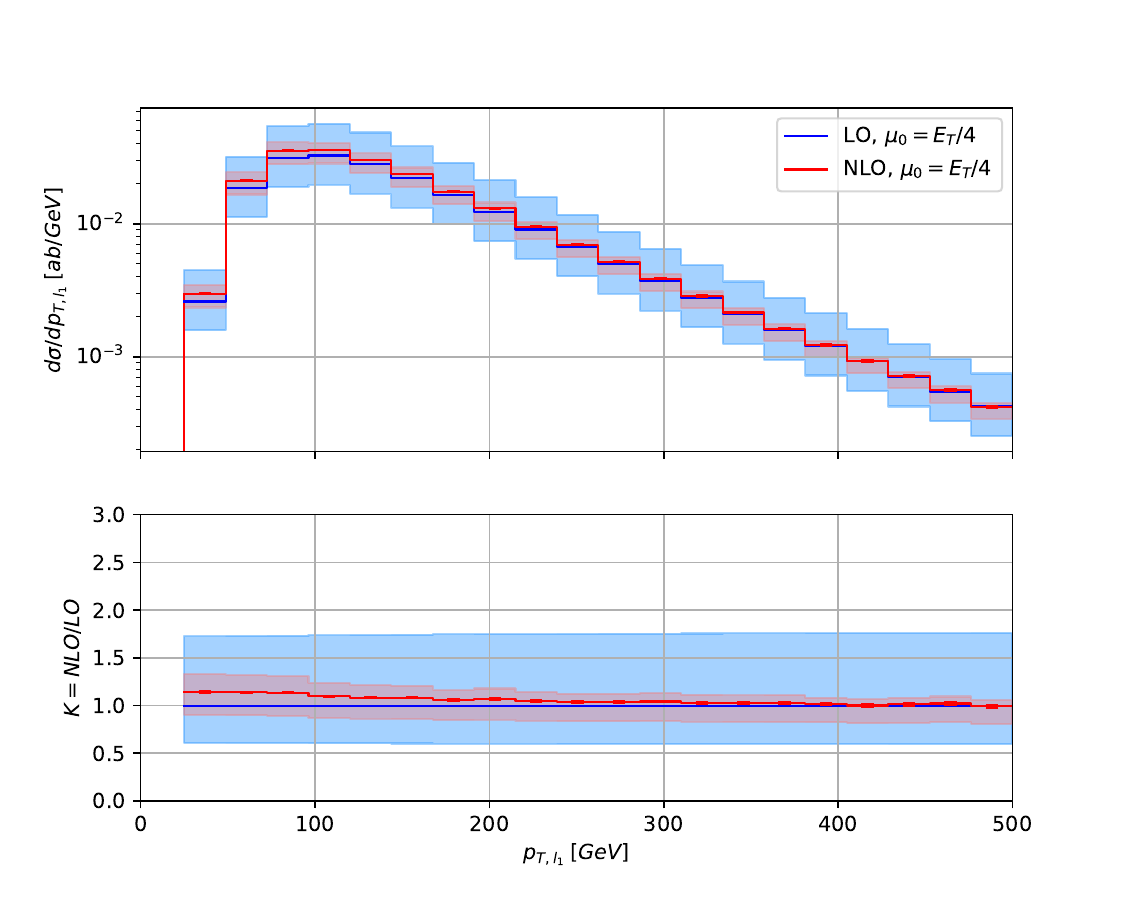}
        \includegraphics[width=0.5\linewidth]{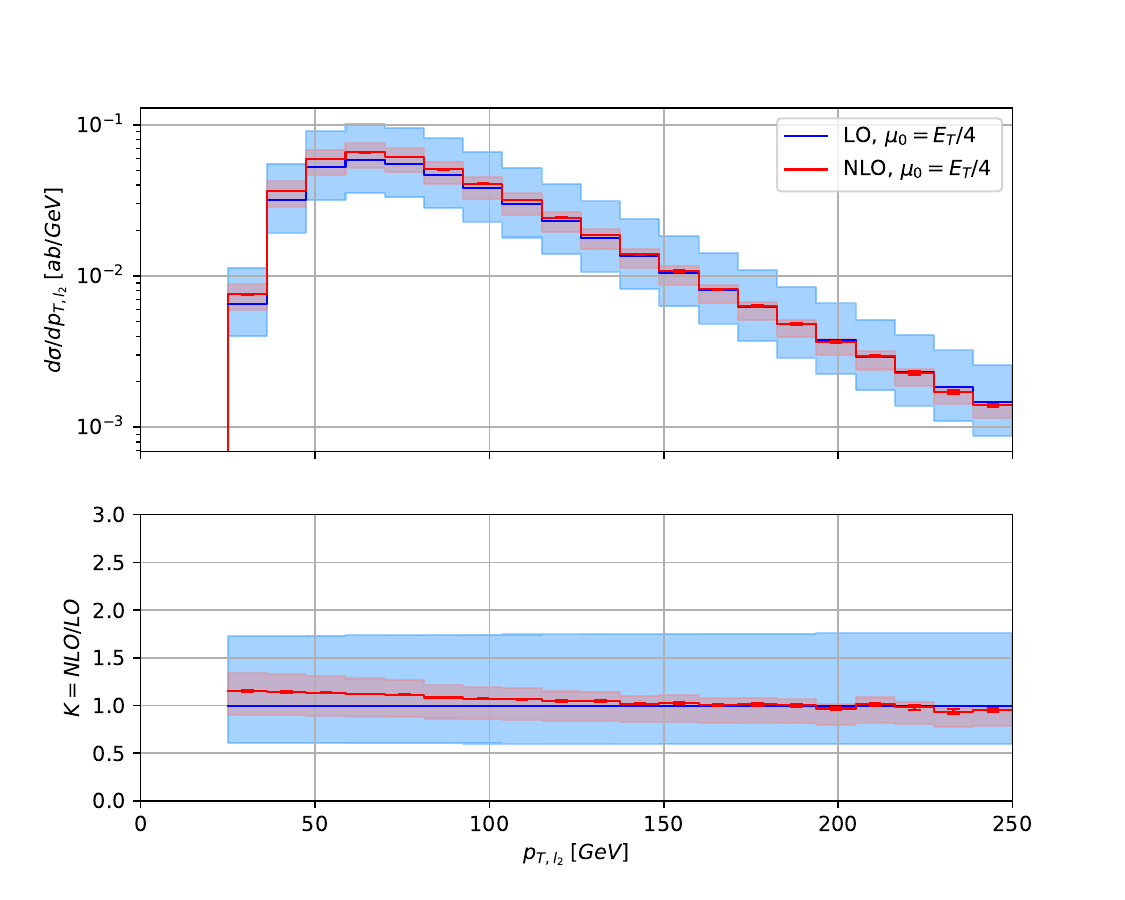}
        \includegraphics[width=0.5\textwidth]{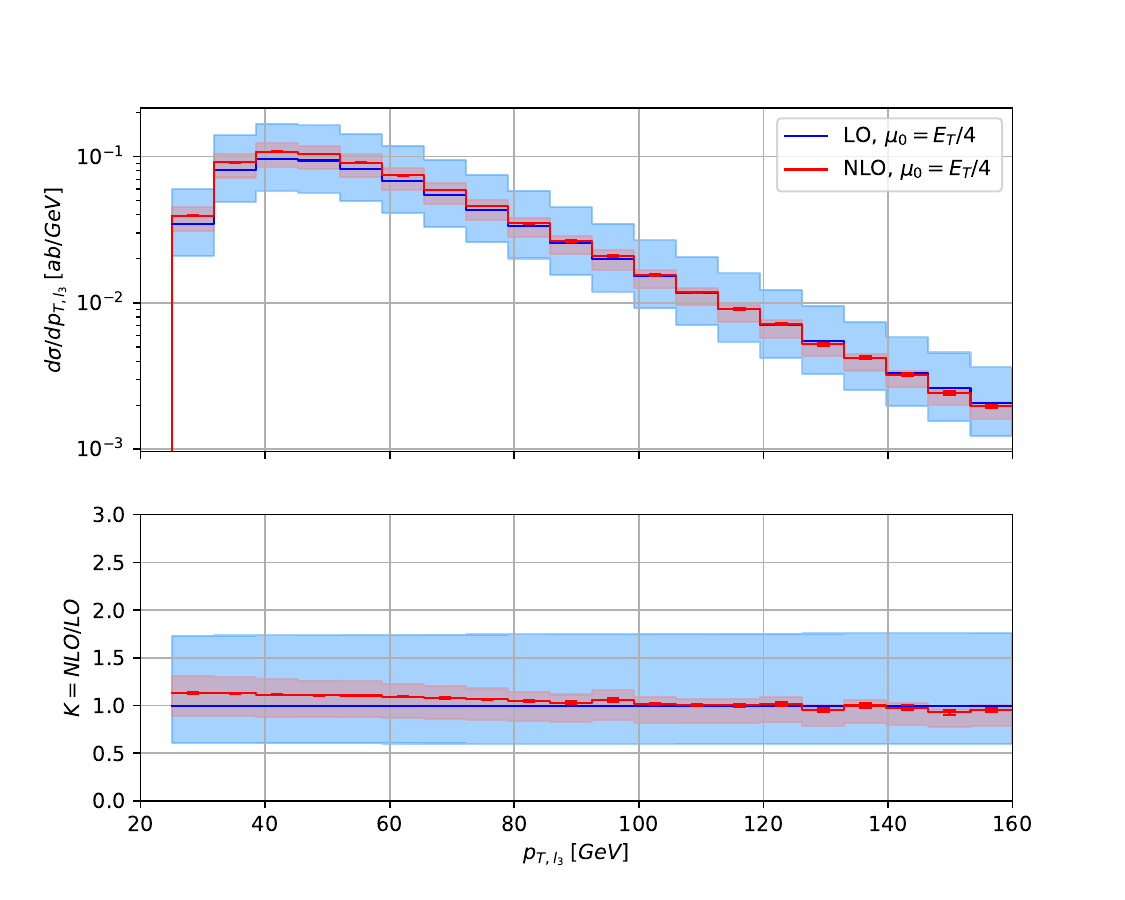} 
        \includegraphics[width=0.5\textwidth]{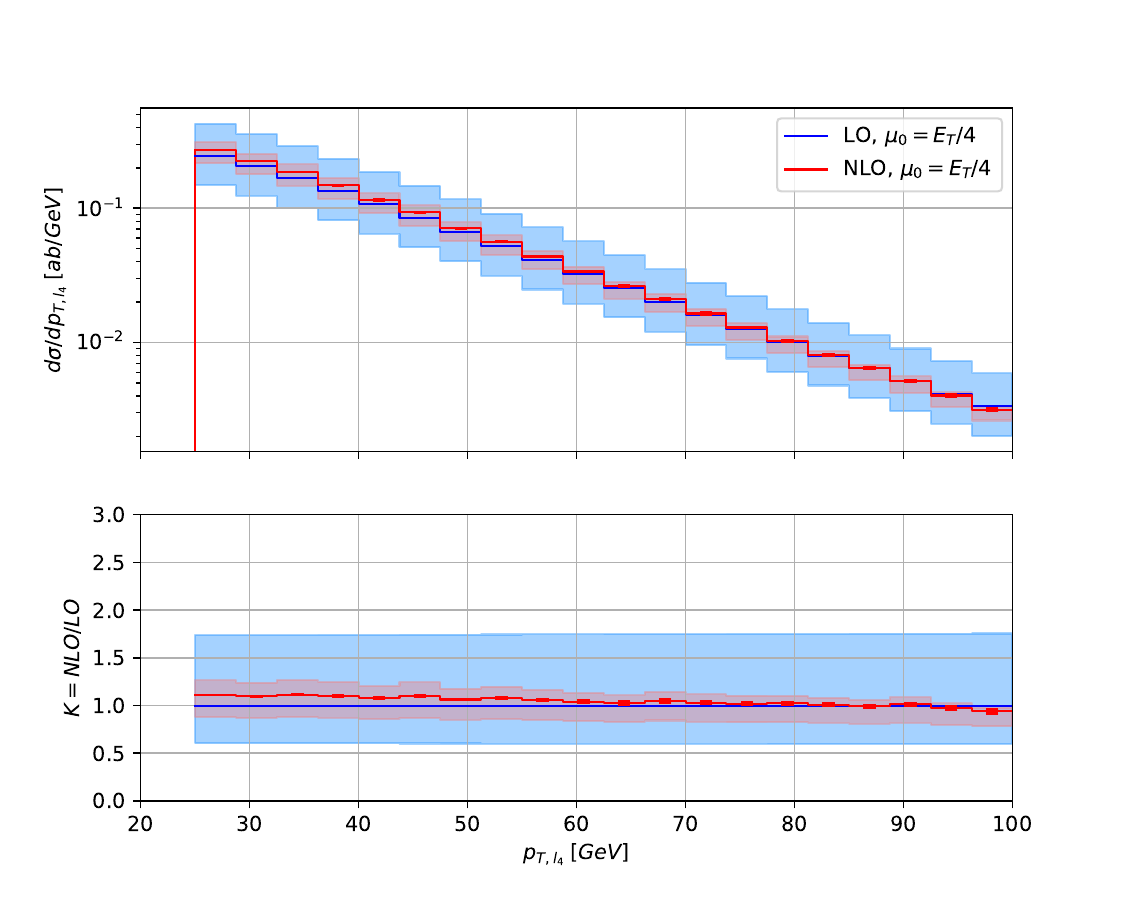}
\caption{\textit{Differential cross-section distributions for the $pp \to t\bar{t}t\bar{t}$ process in the $4\ell$ channel at the LHC with $\sqrt{s} = 13.6$ TeV. The transverse momenta of the charged leptons, that are ordered according to their $p_T$, are displayed for the dynamical scale setting $\mu_R=\mu_F=\mu_0=E_T/4$ and the (N)LO MSHT20 PDF set. The blue (red) curve corresponds to the LO (NLO) result. Also shown are the corresponding uncertainty bands resulting from scale variations. The lower panels display the differential ${\cal K}$-factor together with its uncertainty band and the relative scale uncertainties of the LO cross section. Monte Carlo integration errors are displayed in both panels.}}
    \label{fig:ptl}
\end{figure}
\begin{figure}[!t]  
        \includegraphics[width=0.5\linewidth]{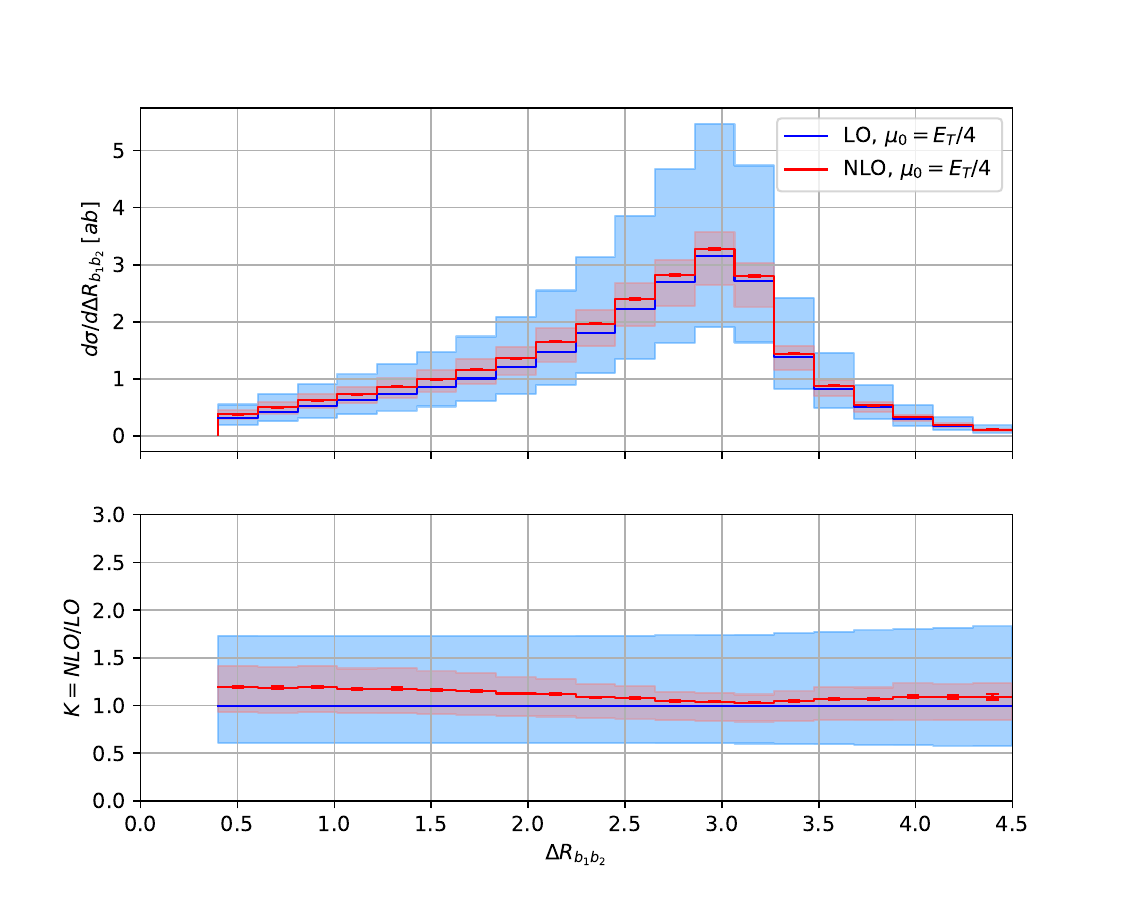}
        \includegraphics[width=0.5\linewidth]{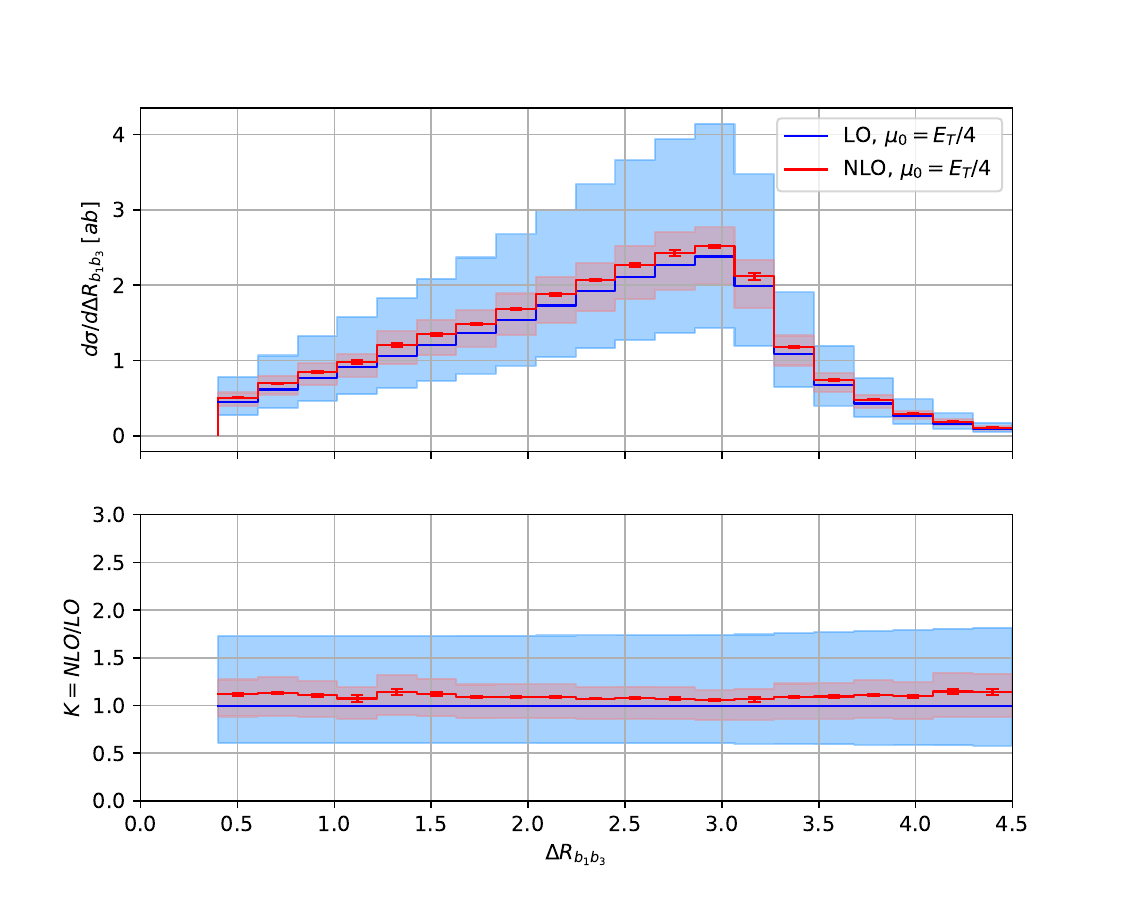}
        \includegraphics[width=0.5\textwidth]{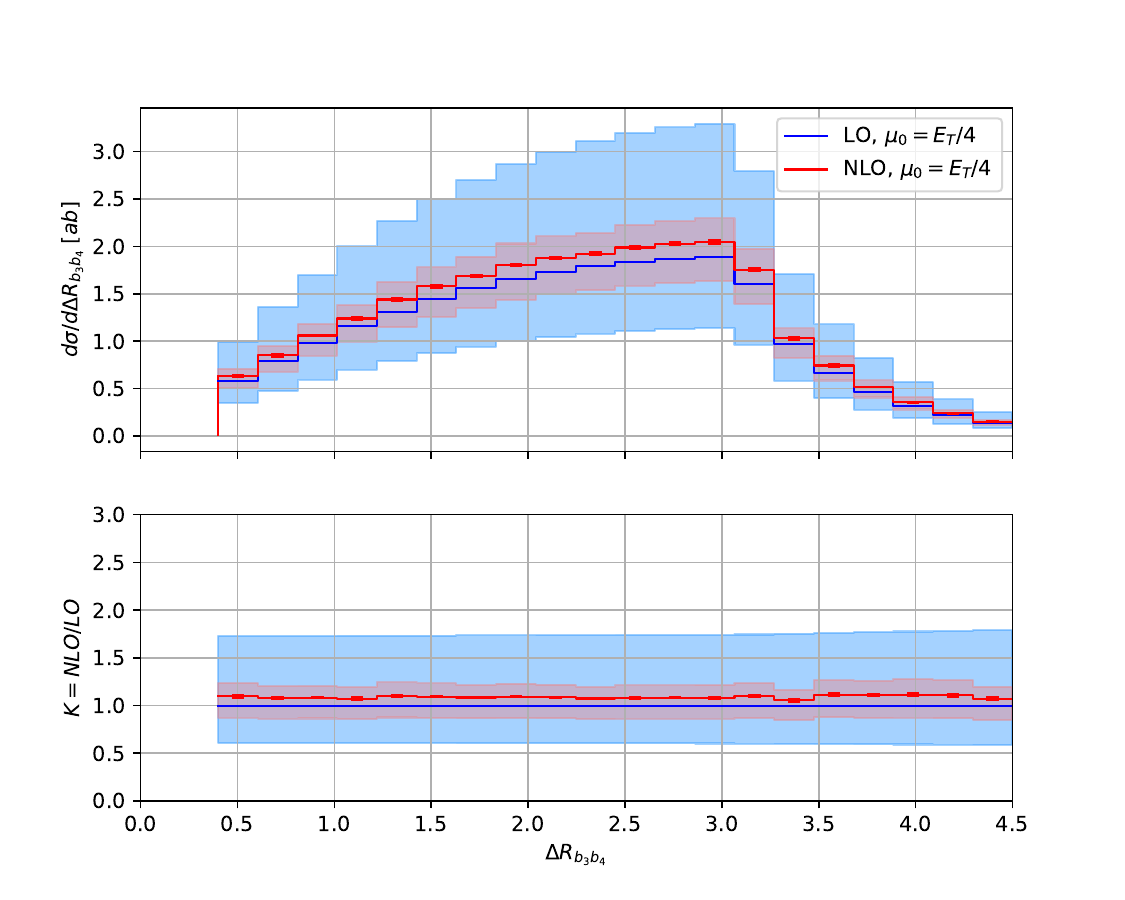} 
        \includegraphics[width=0.5\linewidth]{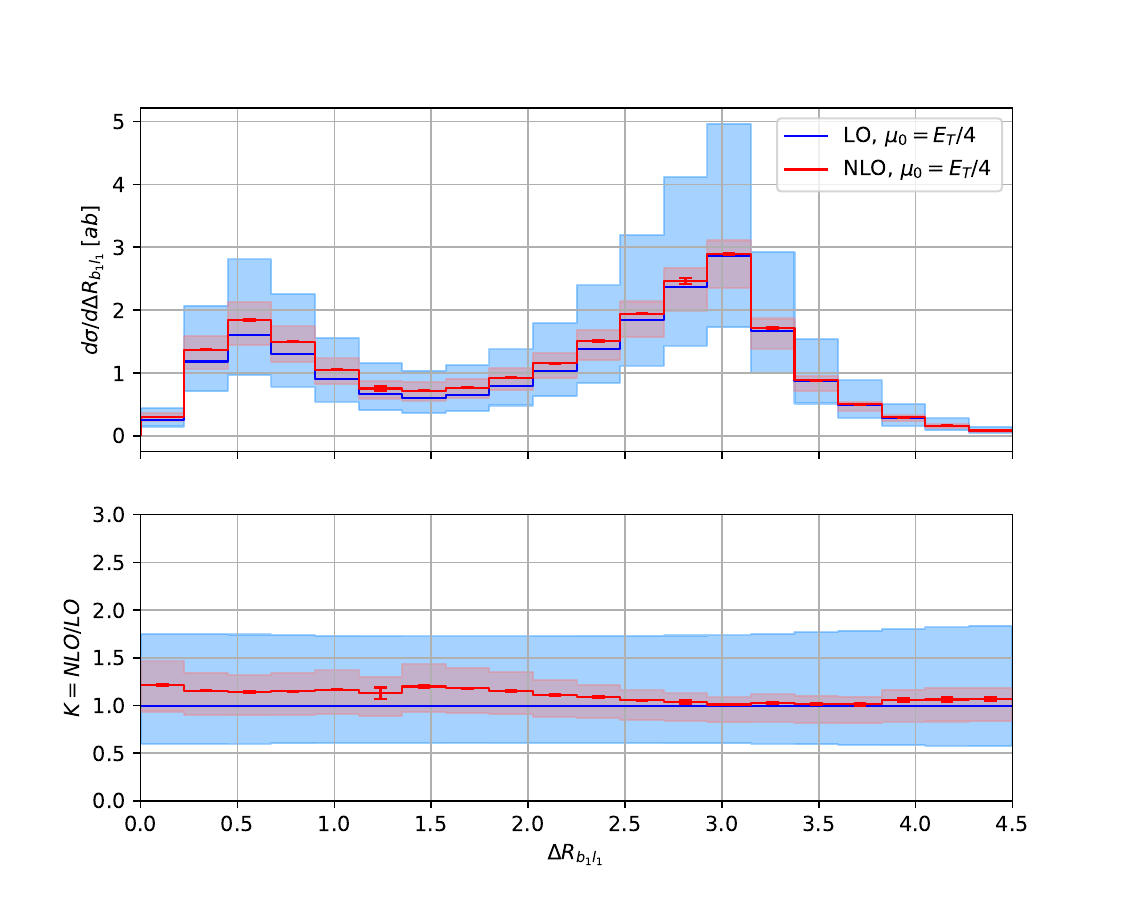}
\caption{\textit{Differential cross-section distributions for the $pp \to t\bar{t}t\bar{t}$ process in the $4\ell$ channel at the LHC with $\sqrt{s} =  13.6$ TeV. The $\Delta R_{bb}$ separations between $b$-jets are displayed for the dynamical scale setting $\mu_R=\mu_F=\mu_0=E_T/4$ and the (N)LO MSHT20 PDF set. The blue (red) curve corresponds to the LO (NLO) result. Also shown are the corresponding uncertainty bands resulting from scale variations. The lower panels display the differential ${\cal K}$-factor together with its uncertainty band and the relative scale uncertainties of the LO cross section. Monte Carlo integration errors are displayed in both panels.}}
    \label{fig:drbb}
\end{figure}
\begin{figure}[!t]  
        \includegraphics[width=0.5\linewidth]{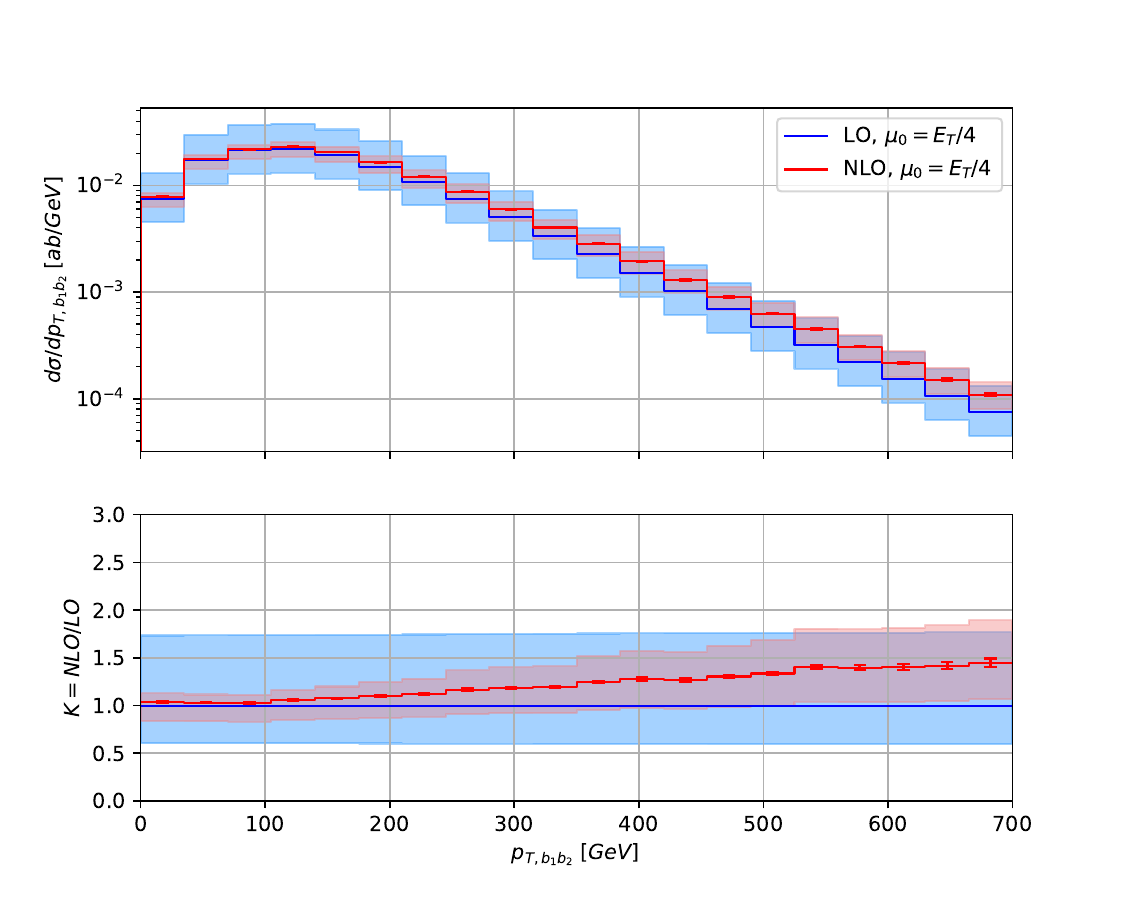}
        \includegraphics[width=0.5\linewidth]{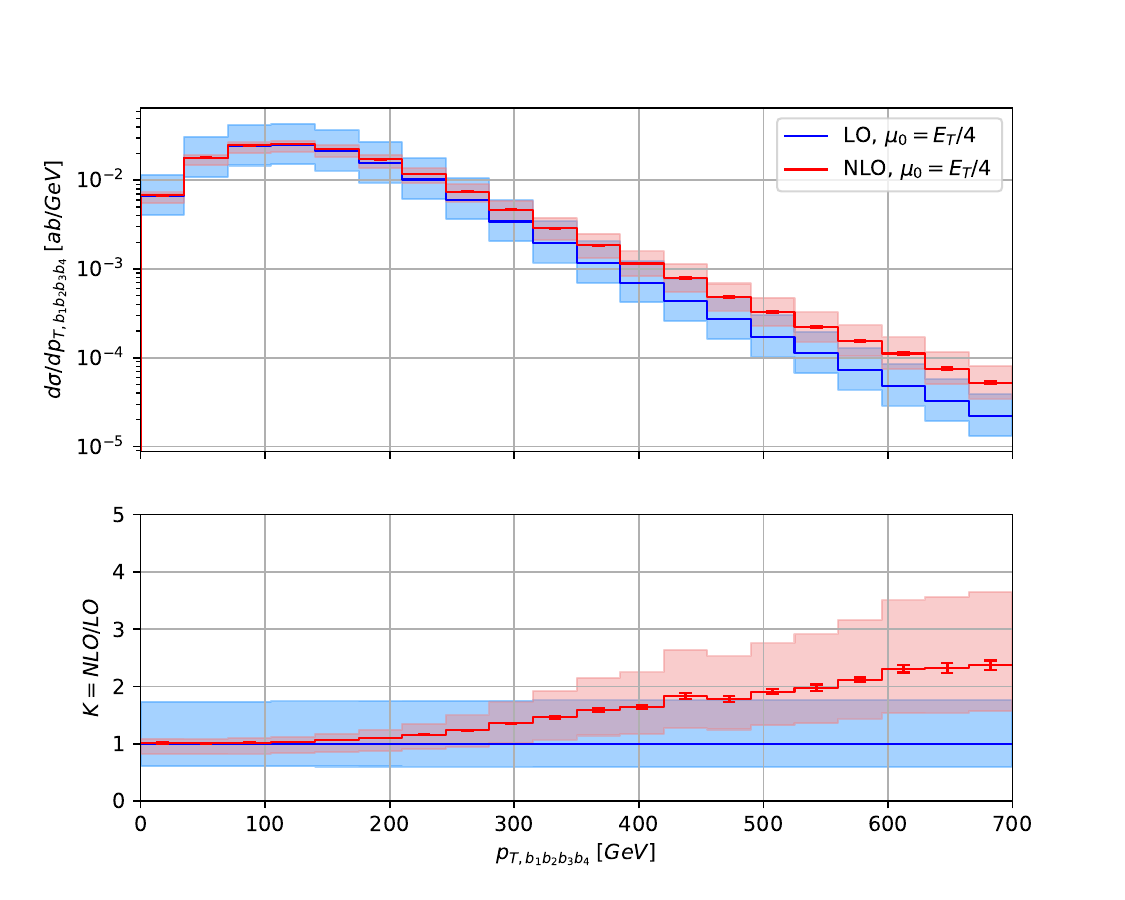}
\caption{\textit{Differential cross-section distributions  for the $pp \to t\bar{t}t\bar{t}$ process in the $4\ell$ channel  at the LHC with $\sqrt{s} = $ 13.6 TeV. The transverse momentum of the $b_1b_2$ system and  the system comprising all $b$-jets are displayed for the dynamical scale setting $\mu_R=\mu_F=\mu_0=E_T/4$ and the (N)LO MSHT20 PDF set. The blue (red) curve corresponds to the LO (NLO) result. Also shown are the corresponding uncertainty bands resulting from scale variations. The lower panels display the differential ${\cal K}$-factor together with its uncertainty band and the relative scale uncertainties of the LO cross section. Monte Carlo integration errors are displayed in both panels.}}
    \label{fig:ptbb}
\end{figure}

We present our findings only for the dynamical scale choice $\mu_R=\mu_F=\mu_0 = E_T/4$ and for the default (N)LO MSHT20 PDF set.  It is a well known fact that the fixed scale choice is not adequate at the differential cross-section level. For example, for some values of $\mu_R = \mu_F=\xi \mu_0$, where $\mu_0= 2 m_t$, the NLO results in the high-energy tails of dimensionful distributions become negative.  Furthermore, in  similar phase-space regions the LO and NLO scale variation bands no longer overlap. Finally, it can even happen that the size of the scale variation at the NLO level actually exceeds the one of the LO prediction. All these effects have already been observed and discussed in detail in our previous studies for the $pp \to t\bar{t}+X$ process, where $X= j, \gamma, Z, W^\pm, H, jj,\gamma\gamma, b\bar{b}$, see e.g. Refs. \cite{Bevilacqua:2015qha,Bevilacqua:2016jfk,Bevilacqua:2018woc,Bevilacqua:2019cvp,Bevilacqua:2020pzy,Bevilacqua:2021cit,Stremmer:2021bnk,Bevilacqua:2022ozv,Bevilacqua:2022nrm,Stremmer:2023kcd}. Similar effects with a fixed scale setting have been examined for the $pp \to t\bar{t}$ process at NLO in QCD  in e.g.  Ref.  \cite{Denner:2012yc}. 

In Figure \ref{fig:ptb} - Figure \ref{fig:ptbb} the upper panels show the absolute LO (blue curve) and NLO (red curve) predictions together with their corresponding scale uncertainty bands. The lower panels display the differential ${\cal K}$-factor together with their scale uncertainty bands as well as the relative scale uncertainties of the LO cross sections. Specifically, the results presented in the lower panels are normalised to the corresponding LO cross sections according to 
\begin{equation}
  {\cal K}^{\rm NLO}(\xi\mu_0)= \frac{d\sigma^{\rm NLO}(\xi\mu_0)dX}{d\sigma^{\rm LO}(\mu_0)/dX}\,, 
   \quad\quad\quad \quad 
   {\rm and}
   \quad\quad\quad \quad 
   {\cal K}^{\rm LO}(\xi\mu_0)=\frac{d\sigma^{\rm LO}(\xi\mu_0)dX}{d\sigma^{\rm LO}(\mu_0)/dX}\,, 
\end{equation}
where $\mu_0$ is the central value of the scale and $X$ represents the plotted observable.  The error bands are determined similarly to the case of the integrated cross-section case, i.e. with the help of the  7-point scale variation that is utilised bin-by-bin. 

We start our discussion with the transverse momentum of the $b$ jets that are presented in Figure \ref{fig:ptb}. The four $b$-jets  are ordered according to their $p_T$, thus,  $p_{T,\,b_1}$ $(p_{T, \,b_4})$ corresponds to the transverse momentum of the hardest (softest) $b$-jet. We notice that the size of the LO scale uncertainties is similar to that at the integrated cross-section level. The size of the ${\cal K}$-factor, on the other hand, changes from ${\cal K}=0.98$ to ${\cal K}=1.23$  depending on the $b$-jet and the phase-space region. Thus, distortions in the shape of the $b$-jet spectra up to $25\%$ can be observed after NLO QCD corrections are taken into account. We can further notice a substantial reduction in the size of theoretical uncertainties when moving from LO to NLO. The latter uncertainties are at the $20\%$ level. 

A fairly similar picture emerges for the transverse momenta of the four charged leptons that are depicted in Figure \ref{fig:ptl}. In this case, we receive NLO QCD corrections in the range of $(-7\%,+15\%)$, while NLO QCD uncertainties are consistently at the $20\%$ level, thus, again almost $4$ times smaller than at the LO level. 

Next, we shift our focus to certain dimensionless observables, highlighting the most noteworthy among those we have examined. In Figure \ref{fig:drbb} we plot the angular separation between the first two hardest $b$-jets $(\Delta R_{b_1b_2})$ and the first and the third hardest $b$-jet $(\Delta R_{b_1b_3})$. Also displayed is the angular separation between the two softest $b$-jets $(\Delta R_{b_3b_4})$ as well as the hardest $b$-jet and the hardest lepton $(\Delta R_{b_1l_1})$. In the case of  $\Delta R_{b_1b_2}$ and $\Delta R_{b_1b_3}$ the $b$-jets are mostly produced in back-to-back configurations. For the $\Delta R_{b_3b_4}$ observable, on the other hand, a significant number of events can also be found for small values of $\Delta R_{b_3b_4}$. A distinct feature can be noticed for the $\Delta R_{b_1 \ell_1}$ differential cross-section distribution. In particular, beyond the typical back-to-back configuration, a notable additional peak near  $\Delta R_{b_1 \ell_1} \approx 0.5$ is present. This configuration corresponds to the decay of highly energetic top quarks, resulting in small angular separations between their decay products. In all four cases, the NLO QCD corrections are in the range of  $(3\%-22\%)$. Also in the case of dimensionless observables, NLO uncertainties are rather constant and at the level of $20\%$. This should be compared to the corresponding LO uncertainties that are up to even $83\%$.

\begin{figure}
        \includegraphics[width=0.5\linewidth]{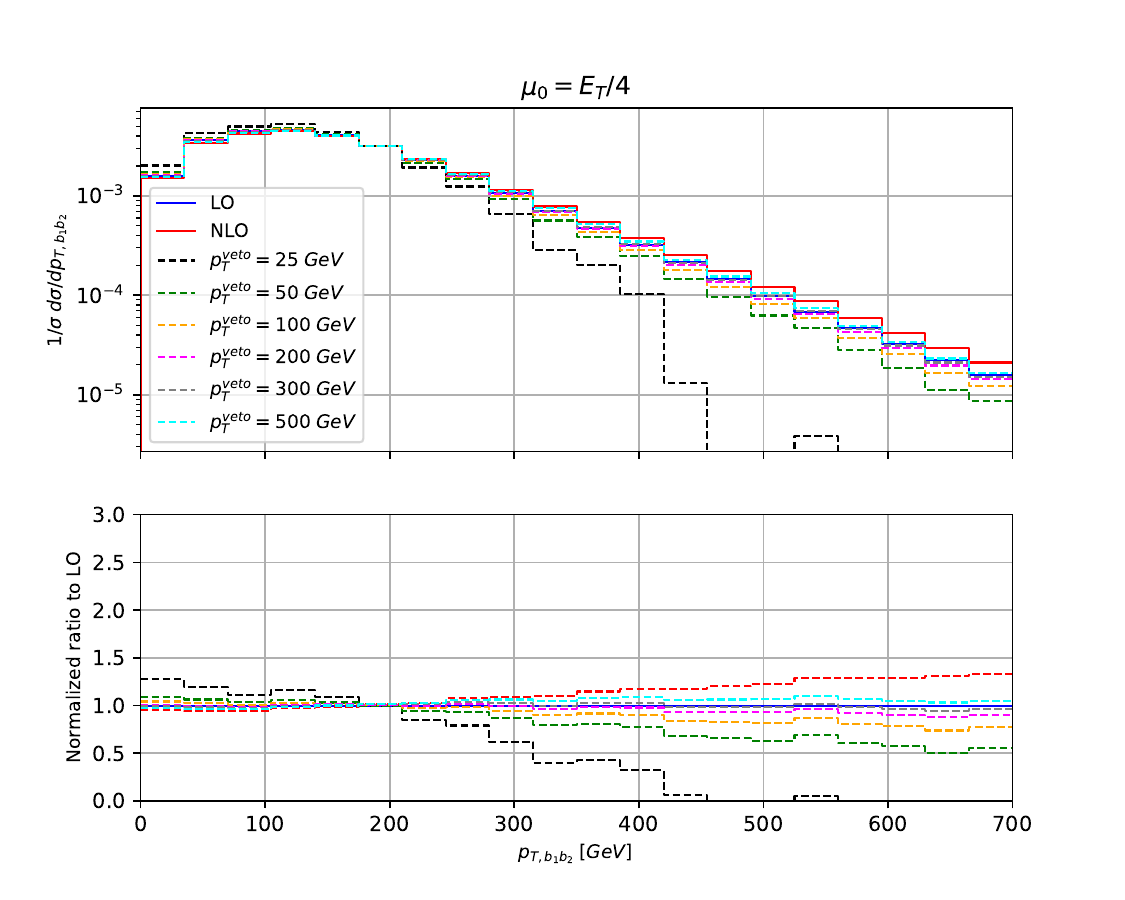}
        \includegraphics[width=0.5\linewidth]{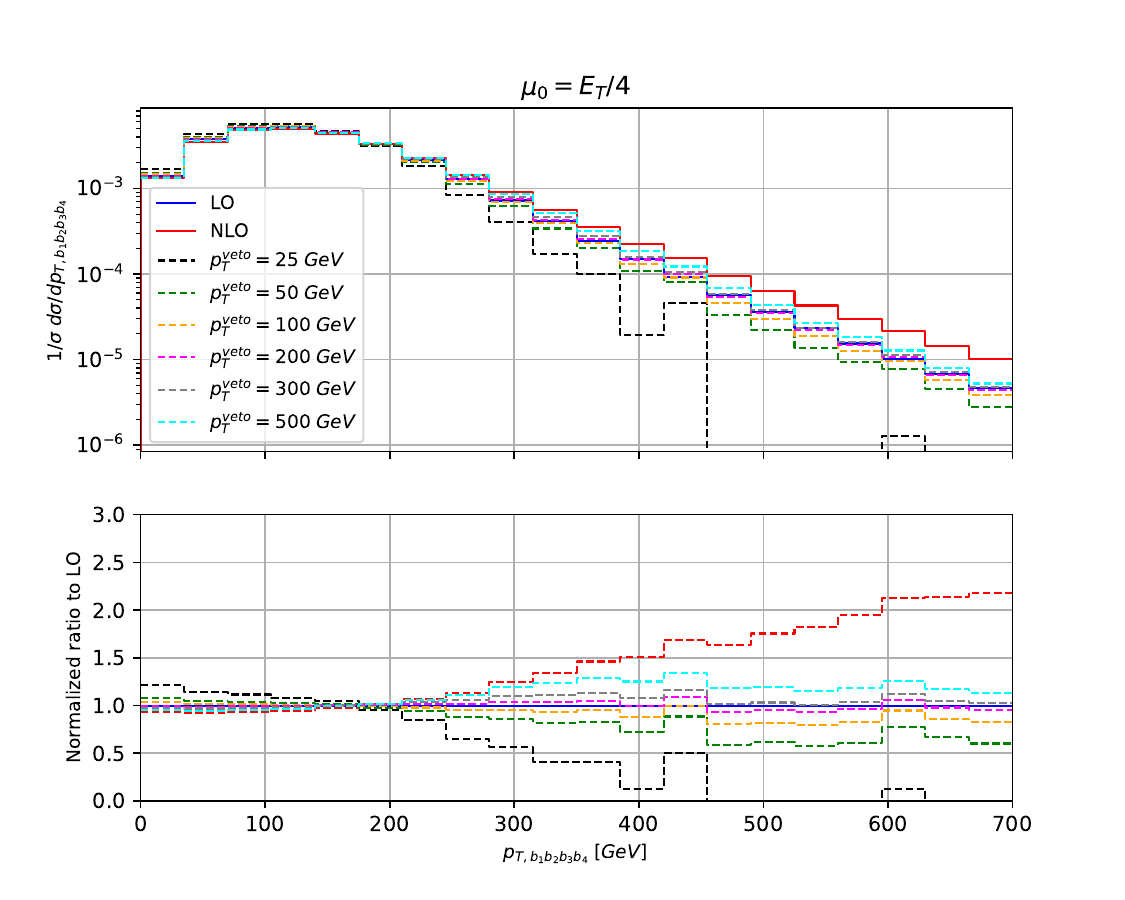}
\caption{\textit{Normalised differential cross-section distributions for the $pp \to t\bar{t}t\bar{t}$ process in the $4\ell$ channel  at the LHC with $\sqrt{s} = $ 13.6 TeV. The transverse momentum of the $b_1b_2$ system and  the system comprising all $b$-jets are displayed for the dynamical scale setting $\mu_R=\mu_F=\mu_0=E_T/4$ and the (N)LO MSHT20 PDF set. The blue (red) curve corresponds to the LO (NLO) result. Also shown are the NLO results with the $p_{T}^{veto}$ cut of $p_T^{veto}=(25,50,100,200,300,500)$ GeV. The lower panels present the ratio to the normalised LO cross section.}}
    \label{fig:ptbb_normalised}
\end{figure}

The observables discussed so far share a common characteristic, namely their NLO QCD corrections are rather stable and of the order of $20\%$. Moreover, the corresponding NLO uncertainties are of a similar magnitude. Finally, the NLO uncertainty bands consistently coincide with the corresponding LO ones. However, there are certain types of observables for which NLO QCD corrections can become very large, particularly, in some phase-space regions and the NLO predictions might no longer fall within the LO uncertainty bands. For illustrative purposes, in Figure \ref{fig:ptbb}, we show differential cross-section distributions for the transverse momentum of the $b_1b_2$ and $b_1b_2b_3b_4$ systems. At LO, in the rest frame of the (anti-)top quark the maximum transverse momentum of the resulting $b$-jet is given by  $p_{T,\,b}=(m_t^2-m_W^2)/2m_t \approx 67.5$ GeV. In order to produce $b$-jets with $p_{T,\,b} > 67.5$  GeV they need to be boosted via the $p_{T}$ of the parent particle. Consequently, at LO the $b\bar{b}b\bar{b}$  system with high $p_T$ is kinematically strongly disfavoured. At NLO, on the other hand,  the $t\bar{t}t\bar{t}$ system can acquire large transverse momentum by recoiling against the extra light jet, resulting in a significant enhancement of events at large values of $p_{T,\,b_1b_2b_3b_4}$. Indeed, we can observe that the differential ${\cal K}$-factor starts to rapidly increase for $p_{T, \,b_1b_2b_3b_4} \gtrsim  270$ GeV and NLO QCD predictions can be as much as $2.4$ times larger than the LO ones in the tails of the $p_{T, \,b_1b_2b_3b_4}$ distribution. We can also notice that the NLO theoretical uncertainties in those phase-space regions substantially increase and are of the order of $50\%$. Finally, the NLO results are not within the corresponding LO uncertainties that are at the level of  $76\%$. Similar but less pronounced effects can also be observed for the $p_{T,\,b_1b_2}$ observable in the same plotted range. In this case, the NLO QCD corrections increase maximally to $45\%$, while NLO uncertainties to $30\%$. To illustrate the strong dependence of these two observables on extra jet radiation, we can analyse them again, this time also applying a veto on the additional light- or $b$-jet.  The difference between the NLO cross section without and with this jet veto is given by a LO calculation, since in the latter case the existence of the five well-separated jets is demanded. While the four $b$-jets still need to have $p_{T, \,b}> 25$ GeV, the additional jet, if resolved and passing all the cuts that are also required for the $b$-jet, should additionally have $p_{T, \,j} < p_T^{veto}$, with $p_T^{veto}=(25,50,100,200,300,500)$ GeV, where $j=g,u,d,c,s,b$. Jet vetos significantly affect the normalisation of differential cross-section distributions, consequently, we study only their normalised versions.  In Figure \ref{fig:ptbb_normalised} we show the normalised LO and NLO QCD differential cross-section distributions for $p_{T,\, b_1b_2b_3b_4}$ and $p_{T, \,b_1b_2}$. Also depicted are the NLO QCD results with the jet veto of $p_T^{veto}=(25,50,100,200,300,500)$ GeV. For various values of the $p_T^{veto}$ cut large differences in the differential factor ${\cal K}$ can be observed. The substantial higher-order QCD corrections that we could previously see in the tails of the two distributions have now been reduced to just a few per cent. For $p_T^{veto}=25$ GeV we obtain negative values of the NLO cross section for 
$p_{T} \geq 470$ GeV for both observables.  This pathological behaviour of fixed-order calculations can be cured by using shower effects or analytic resummation. It should be noted, however, that the jet veto might substantially increase the size of the NLO uncertainty bands. For example, for $p_T^{veto}=50$ GeV the NLO uncertainties increase by a factor of $3-7$ depending on the observable. The observed large scale  dependence for the more exclusive NLO results for $p_{T,\, b_1 b_2}$ and $p_{T, \, b_1 b_2 b_3 b_4}$ gives cause for concern, as jet vetoes are widely used in experimental analyses at the LHC. For example, vetoing the additional jet activity is crucial for suppressing various SM background processes or in various searches for new physics effects. Indeed, in many analyses carried out by the ATLAS and CMS collaborations the most common jet veto scheme comprises imposing a maximum transverse momentum cut on anti-$k_T$ jets. Therefore, in such cases, in order to reduce theoretical uncertainties and to improve the convergence of perturbative predictions, approaches similar to those described in e.g. Ref. \cite{Berger:2010xi,Stewart:2010tn,Becher:2012qa,Banfi:2012yh,Banfi:2012jm,Michel:2018hui,Campbell:2023cha} would have to be used. 
\begin{figure}[!t]
        \includegraphics[width=0.5\linewidth]{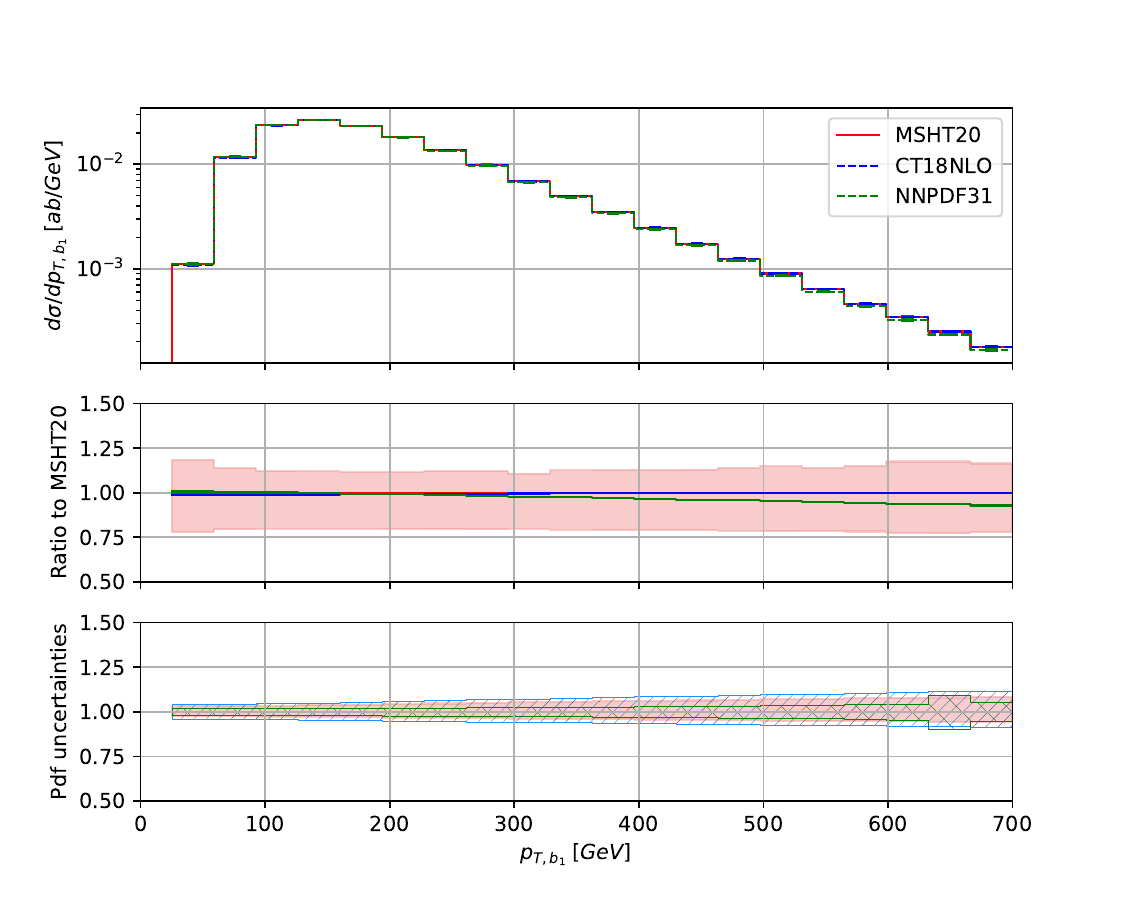}
        \includegraphics[width=0.5\linewidth]{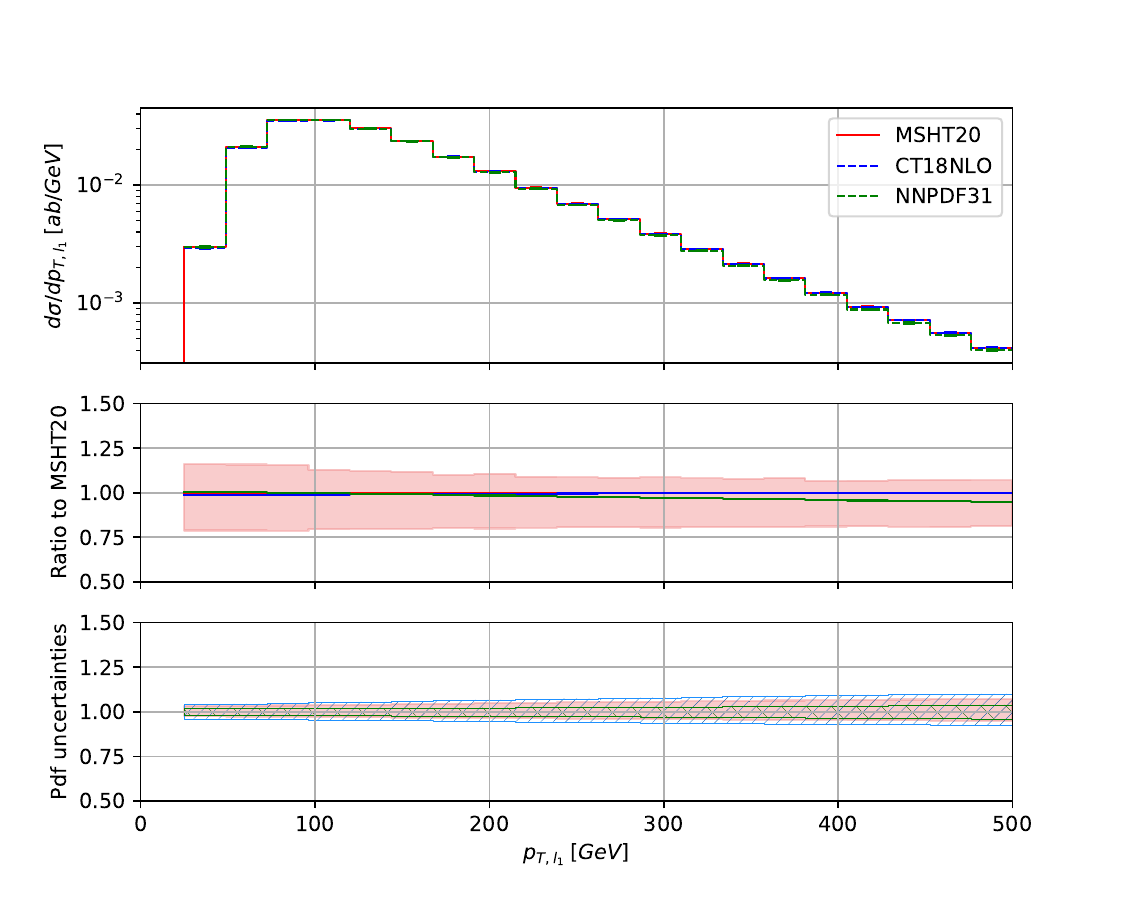}
        \includegraphics[width=0.5\textwidth]{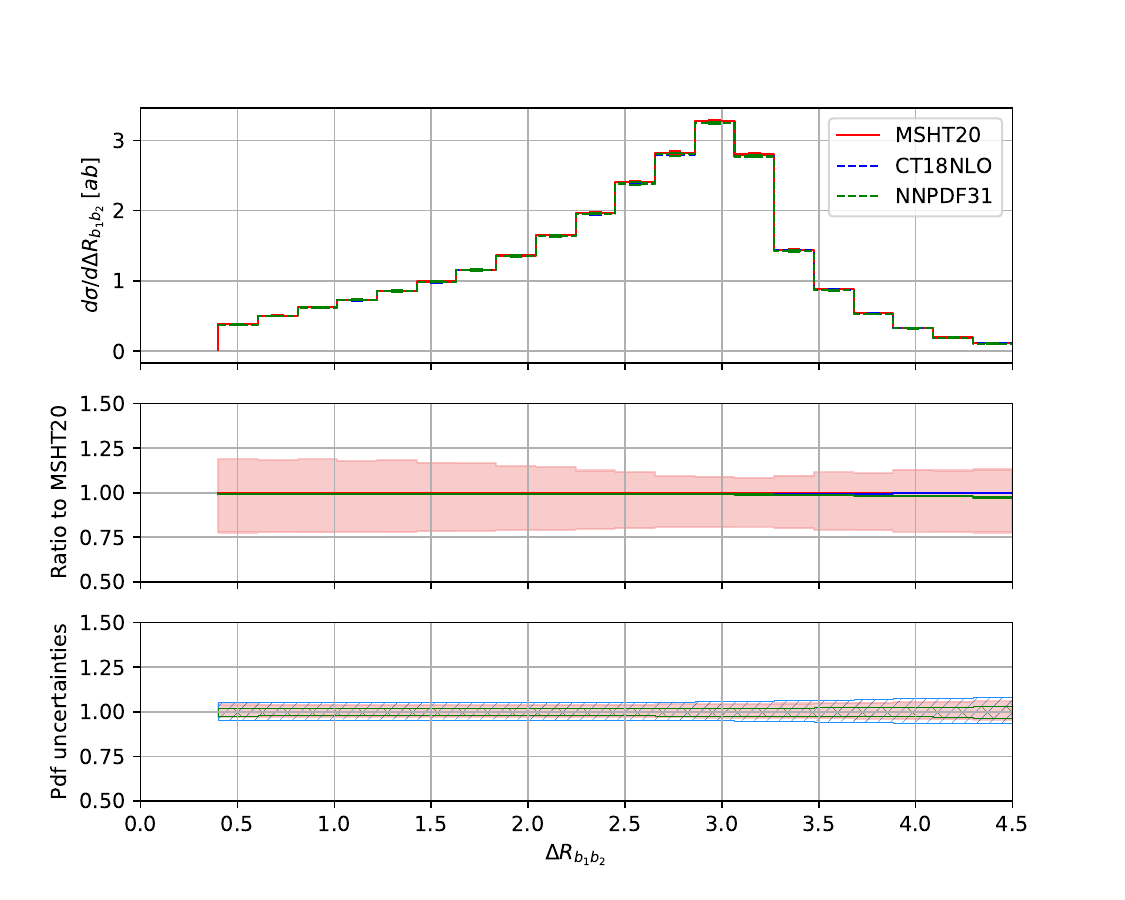} 
        \includegraphics[width=0.5\textwidth]{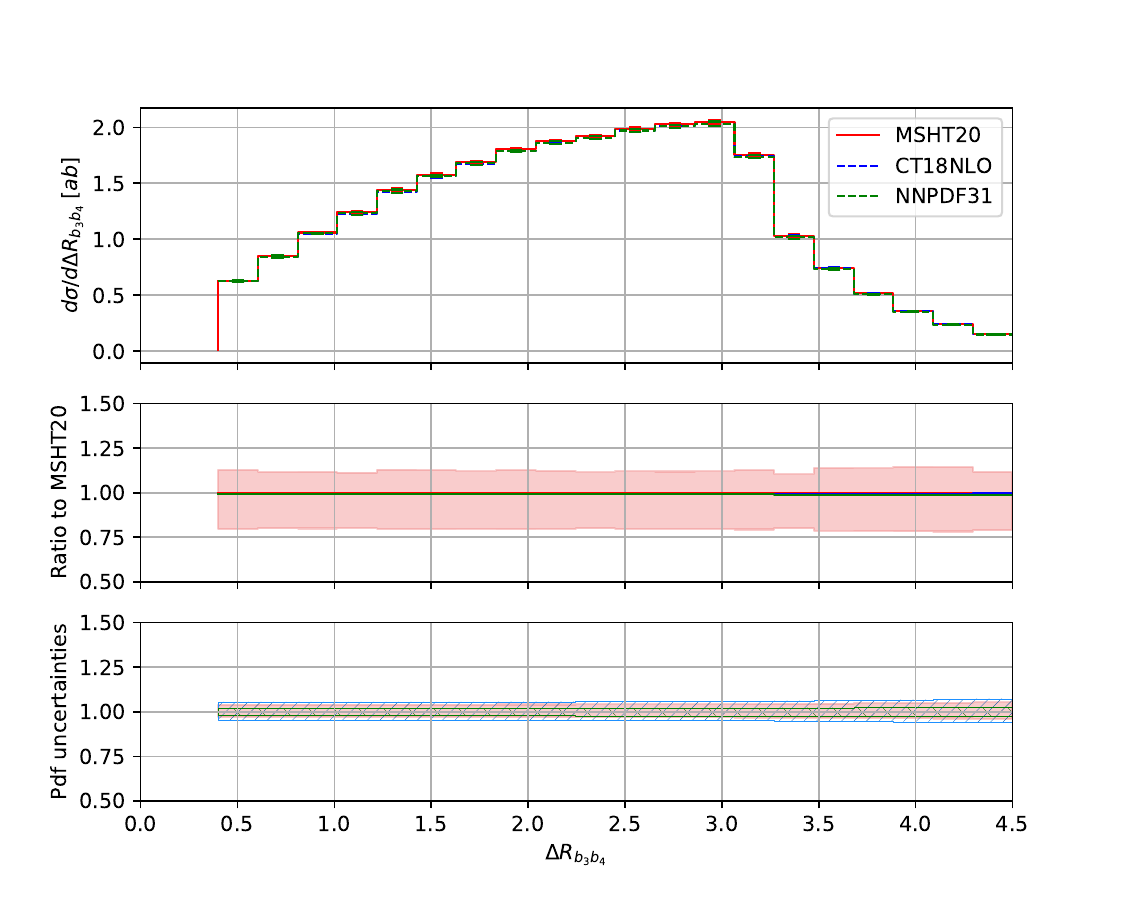}
    \caption{\textit{Differential cross-section distributions for the $pp \to t\bar{t}t\bar{t}$ process in the $4\ell$ channel at the LHC with $\sqrt{s} = $ 13.6 TeV. The transverse momentum of the hardest $b$-jet and charged lepton are presented together with the $\Delta R_{bb}$ separation between the hardest and softest $b$-jets for $\mu_R =\mu_F = \mu_0 = E_T /4$. The upper panels show the absolute NLO QCD predictions for the three PDF sets MSHT20 (solid red curve), CT18 (dashed blue curve) and NNPDF3.1 (dashed green curve). The middle panels display the ratio to the results with the MSHT20 PDF set as well as their scale dependence. The lower panels present the relative size of internal PDF uncertainties calculated separately for each PDF set. Monte Carlo integration errors are displayed in upper panels only.}}
    \label{fig:diff_pdfs}
\end{figure}
\begin{figure}[!t]  
        \includegraphics[width=0.5\linewidth]{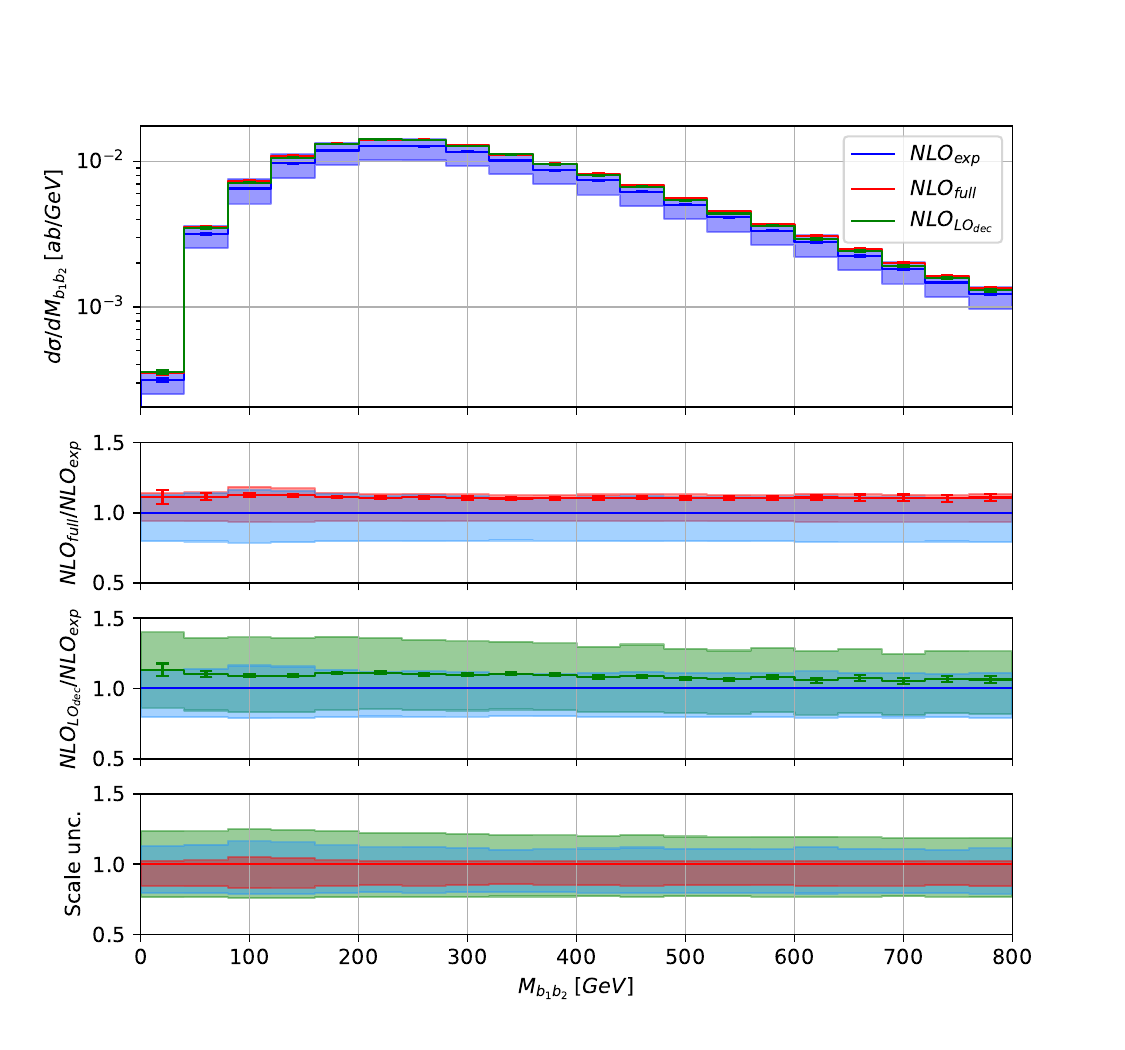}
        \includegraphics[width=0.5\linewidth]{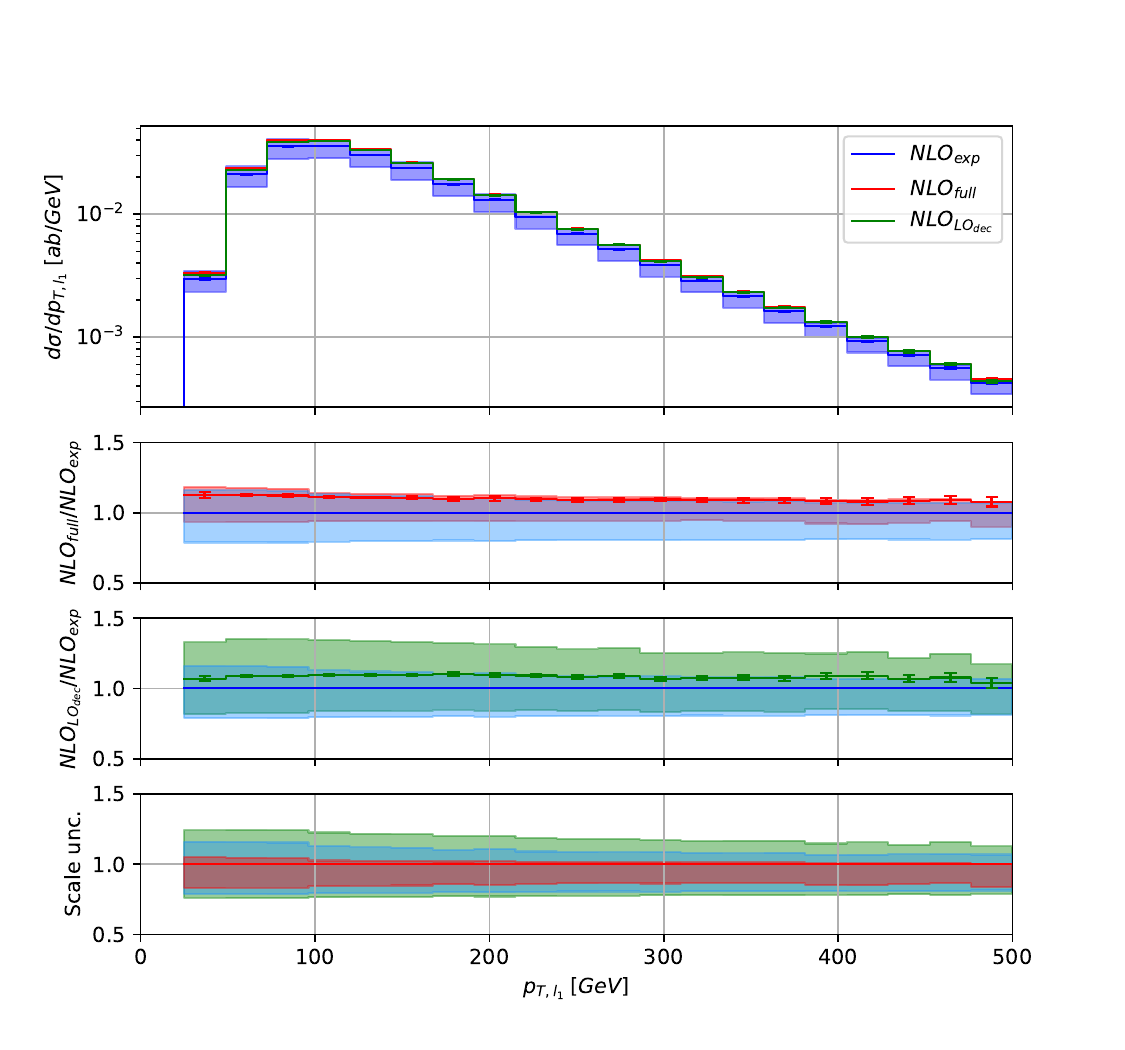}
        \includegraphics[width=0.5\textwidth]{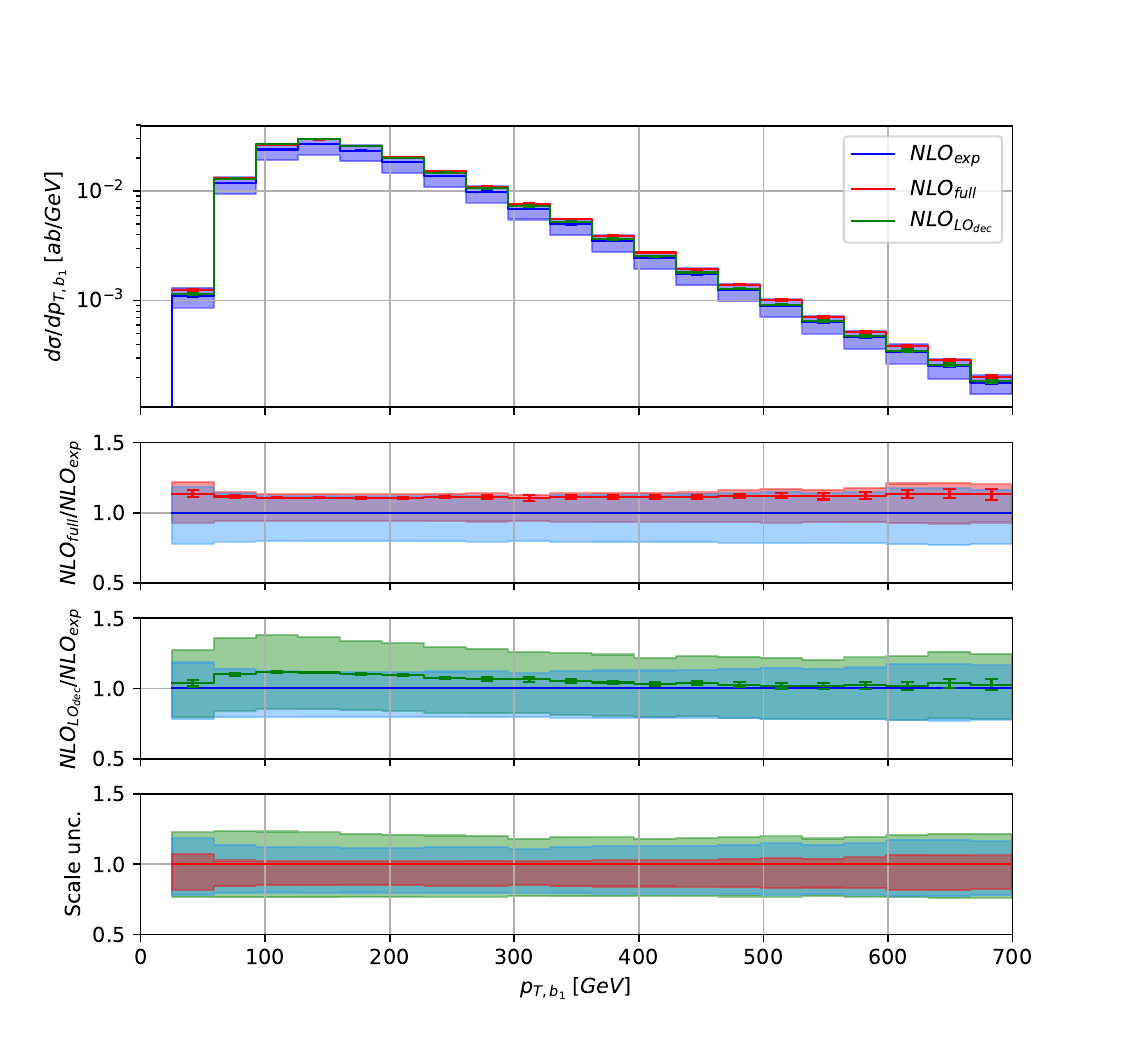} 
        \includegraphics[width=0.5\linewidth]{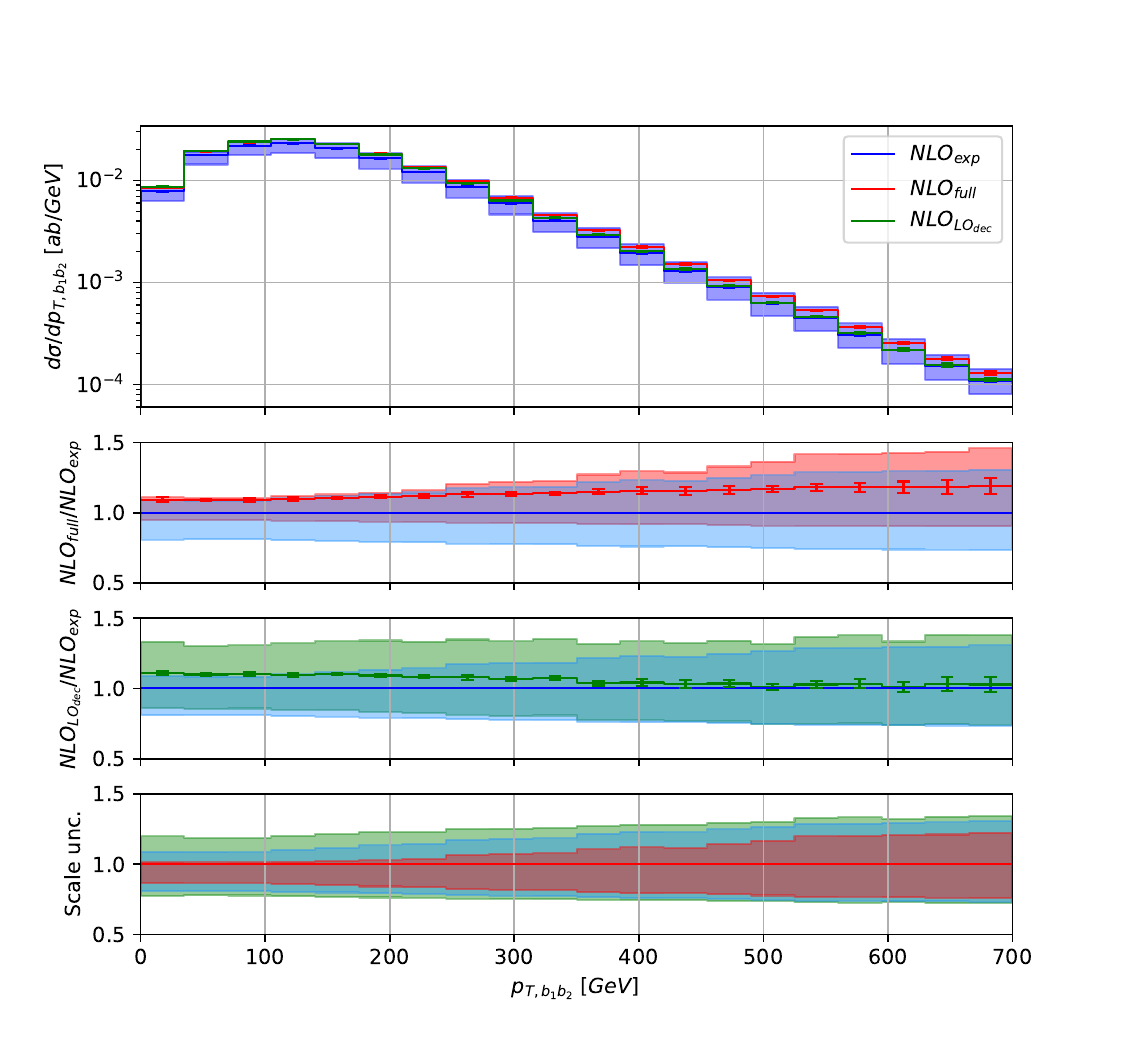}
\caption{\textit{Differential cross-section distributions  for the $pp \to t\bar{t}t\bar{t}$ process in the $4\ell$ channel at the LHC with $\sqrt{s} = 13.6$  TeV. The upper panels show the absolute NLO QCD predictions for $d\sigma^{\rm NLO}_{\rm exp}/dX$ (blue curve), $d\sigma^{\rm NLO}_{\rm full}/dX$ (red curve)  and $d\sigma^{\rm NLO}_{ \rm LO_{dec}}/dX$ (green curve), where $X=M_{b_1b_2},\, p_{T, \,\ell_1},\, p_{T, \,b_1}$ and $p_{T, \,b_1b_2}$. Results are given for $\mu_R = \mu_F =\mu_0  = E_T /4$ and the NLO MSHT20 PDF set. Also provided are theoretical uncertainties as obtained from the scale dependence. The two middle panels display the ratios of $d\sigma^{\rm NLO}_{\rm full}$ and $d\sigma^{\rm NLO}_{\rm LO_{dec}}$  to $d\sigma^{\rm NLO}_{\rm exp}$ together with their relative NLO scale uncertainties. The lowest panels compare the relative size of NLO scale uncertainties for the three approaches, normalised to their corresponding NLO results. Monte Carlo integration errors are displayed in all panels except the last one.}}
    \label{fig:diff_exp}
\end{figure}
\begin{figure}[!t]  
        \includegraphics[width=0.5\textwidth]{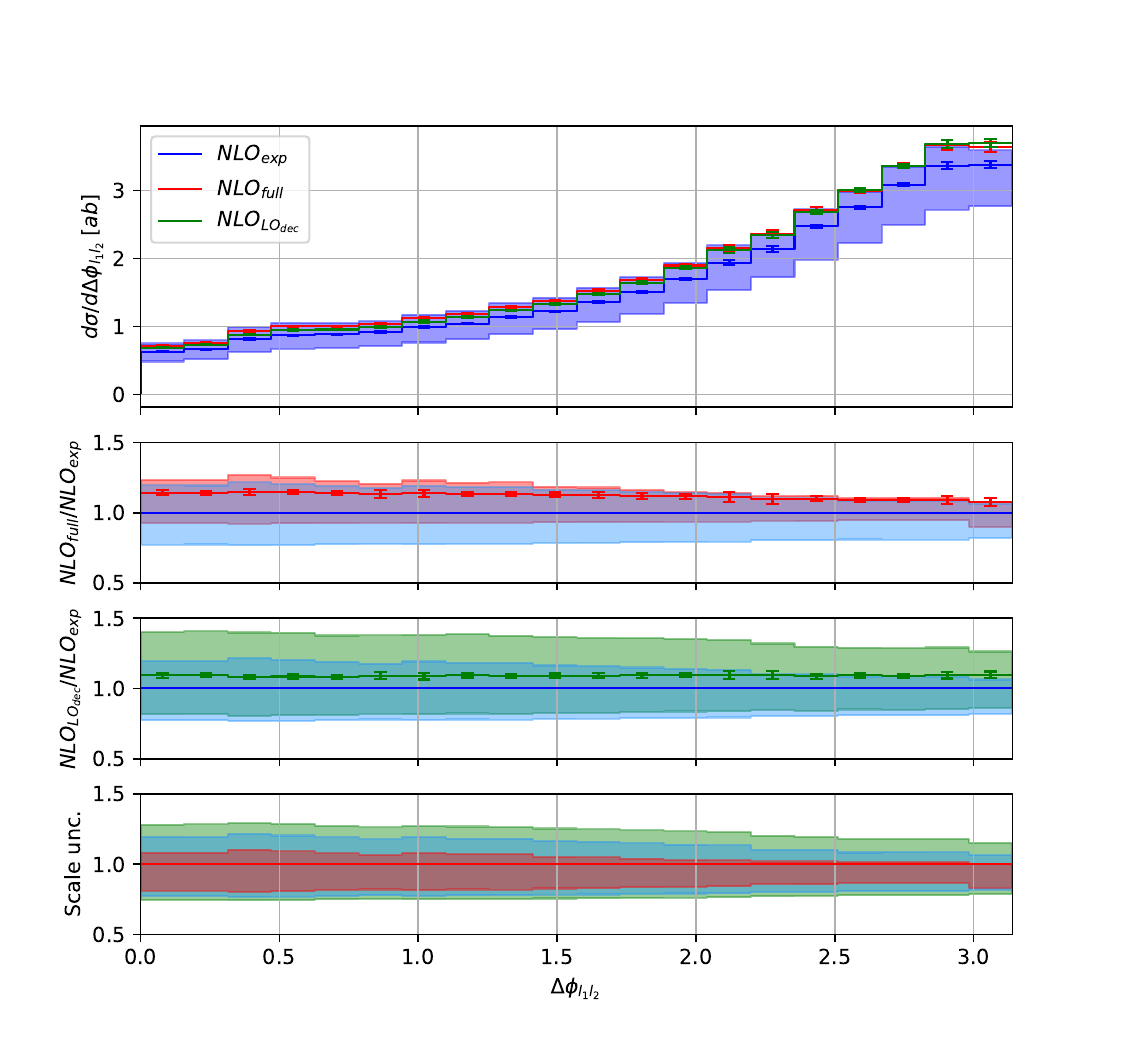} 
        \includegraphics[width=0.5\linewidth]{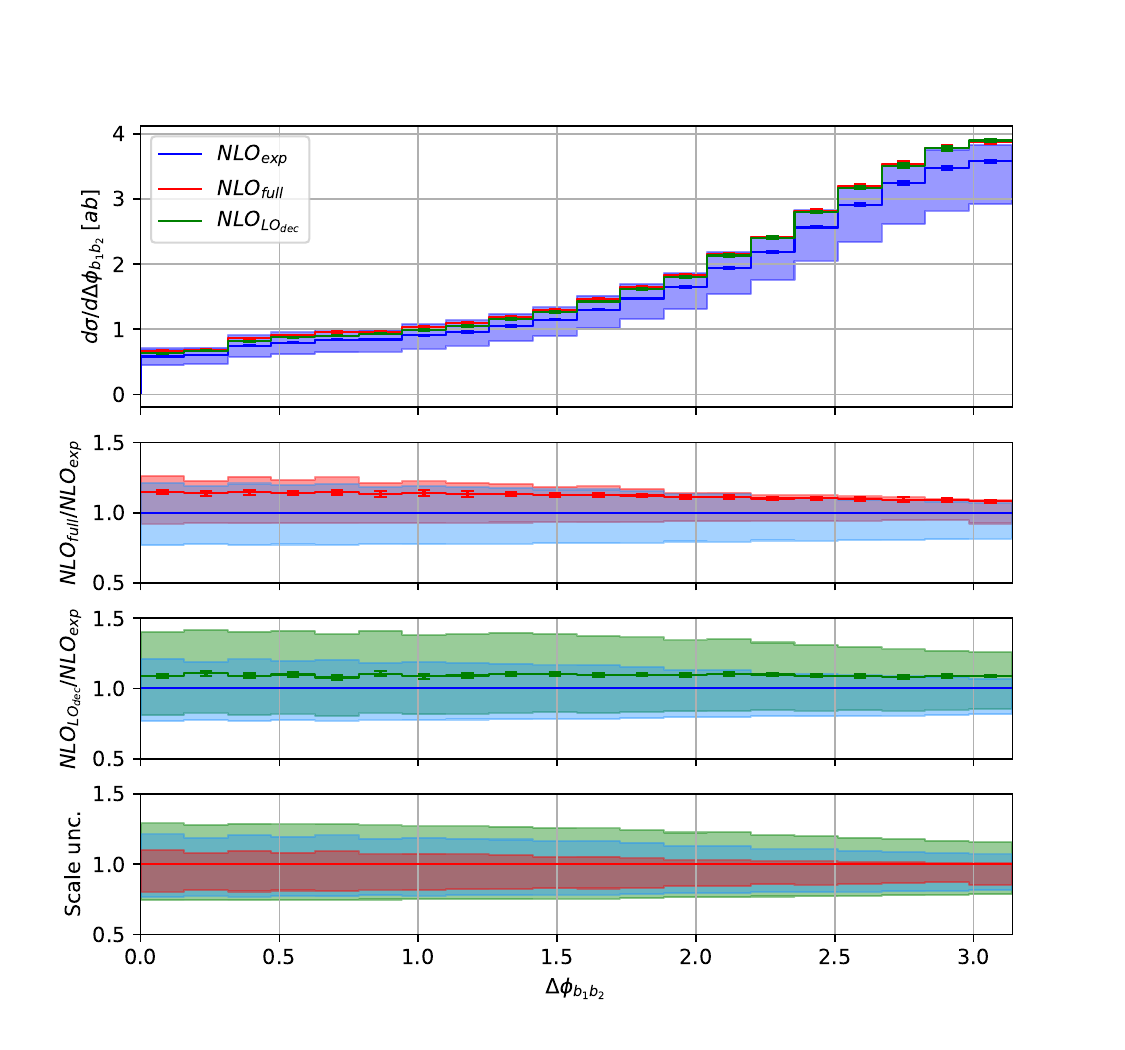}
\caption{\textit{Differential cross-section distributions  for the $pp \to t\bar{t}t\bar{t}$ process in the $4\ell$ channel at the LHC with $\sqrt{s} = 13.6$  TeV. The upper panels show the absolute NLO QCD predictions for $d\sigma^{\rm NLO}_{\rm exp}/dX$ (blue curve), $d\sigma^{\rm NLO}_{\rm full}/dX$ (red curve)  and $d\sigma^{\rm NLO}_{ \rm LO_{dec}}/dX$ (green curve), where $X=\Delta \phi_{\ell_1\ell_2}$ and $\Delta \phi_{b_1b_2}$. Results are given for $\mu_R = \mu_F =\mu_0  = E_T /4$ and the NLO MSHT20 PDF set. Also provided are theoretical uncertainties as obtained from the scale dependence. The two middle panels display the ratios of $d\sigma^{\rm NLO}_{\rm full}$ and $d\sigma^{\rm NLO}_{\rm LO_{dec}}$  to $d\sigma^{\rm NLO}_{\rm exp}$ together with their relative NLO scale uncertainties. The lowest panels compare the relative size of NLO scale uncertainties for the three approaches, normalised to their corresponding NLO results. Monte Carlo integration errors are displayed in all panels except the last one.}}
    \label{fig:diff_exp_angular}
\end{figure}

In the next step, we turn our attention to the PDF uncertainties. We have already shown that at the integrated fiducial cross-section level, such uncertainties are negligible when compared to the theoretical uncertainties coming from scale variation. In the following, we would like to check whether this is still the case at the differential cross-section level. We again present NLO QCD results for the same three PDF sets: MSHT20, CT18, and NNPDF3.1. We focus on investigating the consistency of these predictions and the size of the corresponding PDF uncertainties in different regions of the fiducial phase space. In Figure \ref{fig:diff_pdfs} we present afresh the following observables $p_{T,\, b_1}$, $p_{T,\, \ell_1}$, $\Delta R_{b_1 b_2}$ and $\Delta R_{b_3b_4}$. This time, however, the upper panels display the absolute differential cross-section distributions at NLO in QCD for the three PDF sets: MSHT20
(solid red curve), CT18 (dashed blue curve) and NNPDF3.1 (dashed green curve). The middle panels depict the ratio to the NLO result with the MSHT20 PDF set as well as its scale dependence. Finally, the lower panels present the relative size of the internal PDF uncertainties separately for each PDF set. In this case, all curves are normalised to their corresponding central predictions for  $\mu_0=E_T/4$. 

We can confirm the stability of our predictions with respect to the different PDF sets employed. For example, for the dimensionful observables, the differences between the NLO results are up to $5\%$,  $7\%$ for the $p_{T,\,\ell_1}$, $p_{T,\, b_1}$ distribution, respectively. The results with the NNPDF3.1 PDF set show the most significant differences with respect to the distributions obtained with the default MSHT20  PDF set. Nevertheless, any deviations observed are well within the range of the scale uncertainties, represented by the red uncertainty bands, that are in the range of $16\%-23\%$. By examining the lower panels of the $p_{T,\,\ell_1}$ and $p_{T,\, b_1}$ distributions, we can conclude that the internal PDF uncertainties tend to increase in the high-$p_T$ regions, where they account for a substantial fraction of the theoretical error. Indeed, for all three PDF sets these uncertainties are of the order of $10\%$. In the case of dimensionless observables the differences among the NLO QCD results for MSHT20, CT18 and NNPDF3.1 are negligible, while the internal PDF uncertainties are well below  $10\%$. In summary, after examining various observables, we conclude that different PDF sets yield consistent results for the central value of the scale. However, the internal PDF uncertainties may be more significant in the tails of some dimensionful observables. In particular, they can reach half the value of the theoretical error due to scale variation.

In the end, we examine the previously defined three approaches $d\sigma^{\rm NLO}_{\rm exp}/dX$, $d\sigma^{\rm NLO}_{\rm full}/dX$ and $d\sigma^{\rm NLO}_{ \rm LO_{dec}}/dX$ as a function of the observable $X$, where $X=M_{b_1b_2},\, p_{T, \,\ell_1},\, p_{T, \, b_1}$ and $p_{T, \,b_1b_2}$. These results are shown in Figure \ref{fig:diff_exp}.  The upper panels show the absolute NLO QCD predictions for the three approaches together with their corresponding theoretical uncertainties as obtained from the scale dependence.  The two middle panels display the ratios  of  $d\sigma^{\rm NLO}_{\rm full}/dX$ and $d\sigma^{\rm NLO}_{\rm LO_{dec}}/dX$ to the $d\sigma^{\rm NLO}_{\rm exp}/dX$ result, together with their relative uncertainties. The lowest panels compare again the relative size of NLO scale uncertainties for the three approaches, this time, however,  the findings are normalised to their corresponding NLO results for $\mu_0=E_T/4$. For the three observables studied $M_{b_1b_2},\, p_{T, \,\ell_1}$ and $p_{T, \, b_1}$, our observations and conclusions are very similar.  The differences between the full results and the expanded ones are of the order of $10\%$, which should be compared to the corresponding NLO uncertainties that are in the range of  $15\%-25\%$.  On the other hand, NLO QCD corrections in top-quark decays are consistently at the level of $10\%$. Larger effects could be observed for $p_{T, \, b_1b_2}$, especially in the tail of the distribution. In this case the differences between $d\sigma^{\rm NLO}_{\rm exp}/dp_{T, \, b_1b_2}$ and $d\sigma^{\rm NLO}_{\rm full}/dp_{T, \,b_1b_2}$ increase to $20\%$. However, the NLO uncertainties are also increasing and are now in the range of  $25\%-35\%$. Nevertheless, the size of higher-order QCD effects in top-quark decays remains similar to the case of the other three observables.  

In Figure \ref{fig:diff_exp_angular} we additionally display differential cross-section distributions as a function of the difference in azimuthal angle between the two hardest (charged) leptons, $\Delta \phi_{\ell_1\ell_2}$, and the two hardest $b$-jets,  $\Delta \phi_{b_1 b_2}$. The $\Delta \phi_{\ell_1\ell_2}$ observable is sensitive to signals of numerous beyond the SM scenarios, where among others new heavy states $(Y)$ might be produced in  the $pp \to t\bar{t} Y$ process with $Y \to t\bar{t}$ decays. Indeed, angular distributions of charged  leptons reflect spin correlations and can be employed to probe the ${\cal CP}$ numbers of $Y$. For $X=\Delta \phi_{\ell_1\ell_2}$ and $X=\Delta \phi_{b_1 b_2}$ it is interesting to see the difference between $d\sigma^{\rm NLO}_{\rm exp}/d X$ and $d\sigma^{\rm NLO}_{\rm LO_{dec}}/dX$. In both cases we can observe differences in the range of $8\%-11\%$. Furthermore, the differences between the full results and the expanded ones are up to  $15\%$, still well within the NLO uncertainties that are of the order of $25\%$. Again, the smallest NLO uncertainties can be seen for  the $d\sigma^{\rm NLO}_{\rm full}/dX$ result.

In short, the overall picture is rather similar to that already observed at the integrated fiducial cross-section level. Firstly, the magnitude of the NLO QCD uncertainties increases when higher-order QCD corrections to top-quark decays are neglected. The latter corrections are consistently at the $10\%$ level. Secondly, the differences between the expanded and unexpanded NLO results are within the obtained NLO uncertainties. However, the lack of the expansion of the $1/\Gamma_t$ factor in  $d\sigma^{\rm NLO}_{\rm full}/dX$ significantly reduces the size of the NLO uncertainties for all the observables that we have examined.  Thus, the best NLO QCD predictions for the $pp\to t\bar{t}t\bar{t}$ process in the $4\ell$ channel can be obtained by utilising the $d\sigma^{\rm NLO}_{\rm full}/dX$ results.

%
\section{Summary and Outlook}
\label{summary}
%
%

In this paper we have calculated NLO QCD corrections to $pp\to t\bar{t}t\bar{t}$ in the $4\ell$ top-quark decay channel for the LHC Run III energy of $\sqrt{s} = 13.6$ TeV. We have taken into account higher-order QCD effects in both the production and decays of the top quarks. The top-quark decays have been treated in the NWA, which maintains the spin correlations throughout the calculation. This is the first time that such a complete study for this process has been conducted at the NLO level in QCD. Indeed, up to now, top-quark decays in the $pp \to t\bar{t}t\bar{t}$ process have been treated only within parton shower frameworks. 

In the current study, the integrated fiducial cross sections and their theoretical uncertainties have been evaluated for two different scale settings: $\mu_0=2m_t$ and  $\mu_0=E_T/4$ as well as for the three PDF sets: MSHT20, CT18, and NNPDF3.1. For the default MSHT20 PDF set the  NLO QCD corrections are rather modest, of the order of $10\%$ only, regardless of the scale setting. For the other PDF sets, however, NLO QCD corrections might increase up to even $30\%$. The differences in the size of the ${\cal K}$-factor are driven by the underlining LO cross sections, and specifically by the value of $\alpha_s(m_Z)$ used in the LO PDF sets. The inclusion of higher-order QCD effects has a significant impact on the theoretical error from the $7$-point scale variation, which has decreased substantially from $74\%$ at LO to $20\%$ at NLO. Apart from the theoretical error due to the scale dependence,  for all three PDF sets we have also calculated the internal PDF uncertainties. These uncertainties are very small, of the order of $2\%-6\%$, depending on the PDF set used. In addition, the differences among the NLO results as evaluated with MSHT20, CT18, and NNPDF3.1  are of the order of $1\%$ only. Consequently, the PDF uncertainties are well below the theoretical uncertainties due to the scale dependence. The latter remains the dominant source of the theoretical error. 
Furthermore, we have shown that the inclusion of NLO QCD corrections in top-quark decays is necessary, as these higher-order effects are not negligible. Indeed, not only are they at the level of $10\%$, but they also impact the size of the  NLO theoretical error. Their omission increases the overall magnitude of this error. Finally, we have checked that the NLO QCD  result calculated with the $1/\Gamma_t$ term not expanded has the smallest theoretical uncertainties, which are at the level of  $15\%$ instead of $20\%$.

At the differential cross-section level, the size of the NLO QCD corrections depends on the studied observable and the examined region of the fiducial phase space. For the majority of the observables, and when employing the dynamical scale setting, $\mu_0=E_T/4$, the NLO QCD corrections are rather stable and moderate, of the order of $20\%$. Moreover, the corresponding NLO uncertainties are of similar magnitude. On top of that, the NLO uncertainty bands consistently coincide with the corresponding LO ones. However, there are certain types of observables for which not only the NLO QCD corrections can be very large, even up to $140\%$, but also the NLO uncertainty bands are outside of the LO ones. Finally, they comprise large NLO uncertainties. These are dimensionful observables that are very sensitive to additional jet radiation. We have been able to identify the following differential cross-section distributions for which this is the case:  $p_{T, \,b_1b_2b_3b_4}$, $p_{T, \,\ell_1 \ell_2 \ell_3 \ell_4}$, $p_T^{miss}$, $p_{T, \,b_1b_2}$ and $p_{T, \, \ell_1 \ell_2}$. 

In the next step, we have compared the following NLO predictions:  $d\sigma^{\rm NLO}_{\rm exp}/dX$, $d\sigma^{\rm NLO}_{\rm full}/dX$ and $d\sigma^{\rm NLO}_{ \rm LO_{dec}}/dX$ as a function of various observables $X$. By analysing these predictions we have concluded that the overall pattern is rather similar to that which has already been observed at the integrated fiducial cross-section level. Namely, the magnitude of the NLO uncertainties increases when NLO QCD corrections to top-quark decays are neglected. The higher-order corrections in top-quark decays are rather stable and of the order of  $10\%$. In addition, the difference between the expanded and unexpanded NLO results is within the obtained NLO scale uncertainties. Nonetheless, the lack of the expansion of the  $1/\Gamma_t$ term in $d\sigma^{\rm NLO}_{\rm full}/dX$, significantly reduces the size of the NLO uncertainties for all the observables that we have scrutinised. Consequently, the best NLO QCD predictions for the process at hand can be obtained by employing  $\sigma^{\rm NLO}_{\rm full}$ and $d\sigma^{\rm NLO}_{\rm full}/dX$.

In conclusion, we would like to mention that it would be beneficial to conduct a comparison of the results obtained in this work with the ones where top-quark decays are generated with the help of parton showers. Such a comparison could assess the extent to which parton shower effects can imitate higher-order effects in top-quark decays. We would also be able to verify the importance of NLO QCD corrections to angular observables that are sensitive to top-quark spin correlations. Furthermore, such a comparison could help to identify the phase-space regions that are indeed affected by resummed dominant soft-collinear logarithmic corrections from parton showers, which are absent from our fixed-order NLO QCD predictions. We plan to conduct such a comparison in the near future. 

In addition, it would be advantageous to examine the so-called  Matrix Element Corrections (MEC) to the \textsc{Pythia8} parton shower program \cite{Norrbin:2000uu} that can upgrade the accuracy of top-quarks decays to approximately NLO QCD in the shower evolution. Quite recently, a study has been carried out taking into account the MEC for the simplest case of $pp\to t\bar{t}$ production in di-lepton top-quark decay channel \cite{Frixione:2023hwz}. Another 
study has just been performed for the $pp\to t\bar{t}W^\pm$ process for the two same-sign leptons and jets $(2SS\ell)$ as well as three-lepton $(3\ell)$  final states \cite{Frederix:2024psm}. Unfortunately, such analyses are not yet available for the  $pp\to t\bar{t}t\bar{t}+X$ process. Although NLO plus parton shower predictions in the \textsc{Powheg Box} and \textsc{Mc$@$Nlo} framework for $pp \to t\bar{t}t\bar{t}+X$ production in the $\ell+jets$ channel are available in the literature
\cite{Jezo:2021smh}, the MEC to the top-quark decays are not included there. As this is a much more complicated process with two $t\bar{t}$ pairs and at least 12 final state particles, it would be interesting to test the validity of such approximate methods in such a complex environment. Another very important issue to look at when using the MEC is spin correlations. Having fixed-order predictions that describe top-quark decays and spin correlations with NLO QCD accuracy, we are in an excellent position to test both aspects.  Again, we plan to perform such a comparison in the near future.

The obtained NLO QCD results for the  $pp\to t\bar{t}t\bar{t}+X$ process are especially important for the High-Luminosity phase of the LHC (HL-LHC). The upgrade is currently in progress and ATLAS and CMS experiments are expected to start taking data from the beginning of 2029. The HL-LHC will allow us to study $pp\to t\bar{t}t\bar{t}$  production in the $4\ell$ channel in greater detail. We point out that our conclusions regarding the magnitude of NLO QCD corrections, the significance of higher-order effects in top-quark decays or the size of NLO uncertainties will remain the same if the center-of-mass energy of the LHC is increased from the current value to $14$ TeV. Finally, we note that, despite its relatively small cross section, a good theoretical control over the $pp\to t\bar{t}t\bar{t}$ process in the $4\ell$ channel is phenomenologically very relevant. Indeed, the NLO QCD uncertainties that we have estimated for this channel are at the level of $15\%-20\%$. For comparison, the projected uncertainties for the HL-LHC for the $4\ell$ channel with  $\ell^\pm=e^\pm, \mu^\pm$, combining ATLAS and CMS, are in the range of $\delta = 1/\sqrt{N} = 18\% - 16\%$. On the other hand, if $\ell^\pm=e^\pm, \mu^\pm, \tau^\pm$  the later uncertainties decrease to $8\%-7\%$. We note that these uncertainties are estimated solely on the basis of an integrated luminosity planned for the HL-LHC  that amounts to $(3-4)$ ${\rm ab}^{-1}$, see e.g. Ref. \cite{ZurbanoFernandez:2020cco}.  Thus, at the HL-LHC, among other things, we will gain sensitivity to higher-order effects in top-quark decays in the $pp\to t\bar{t}t\bar{t}+X$ process. However, even now, pending new results from HL-LHC, the NLO QCD results obtained in this paper are highly relevant for various preparatory studies.  Proper modelling of differential cross-section distributions, for example, is an invaluable contribution to the many Machine Learning studies that are currently being conducted for this process. We also believe that the NLO QCD analysis for the SM $pp\to t\bar{t}t\bar{t}+X$ (background) process at the LHC is a necessary step towards a correct interpretation of possible signals of new physics that may arise in the $4\ell$ channel.

Finally, although the focus of this paper has been on providing NLO QCD results for the $pp \to t\bar{t}t\bar{t}$ process in the $4\ell$ channel in the NWA, and the calculation of higher-order corrections to the full $2 \to   12$  process is a huge task at the moment, we can still estimate the impact of non-resonant and off-shell effects at the integrated and differential fiducial cross-section level with the help of the LO predictions. In fact, at the integrated cross-section level the impact of these omitted contributions is of the order of $1\%$ only, and therefore well within the NWA uncertainty defined by ${\cal O}(\Gamma_t/m_t)\approx 0.9\%$, see e.g. Ref.  \cite{Fadin:1993kt}. On the other hand, at the differential cross-section level, neglecting triple-, double-, single- and non-resonant top-quark contributions as well as Breit-Wigner top-quark propagators has a significant impact. In general, full off-shell effects are noticeable for dimensionful observables in the following phase-space regions: in high $p_T$ tails and in the vicinity of kinematical endpoints. Since the first type of observables is beyond the reach of the LHC, only the second type should be thoroughly investigated. Indeed, we have checked that differential cross-sectional distributions as a function of the minimum invariant mass of the $b$-jet and the positively charged lepton, $M(\ell^+ b)$, the reconstructed invariant mass of the top quark, $M(t)$ as well as the stransverse mass of the top quark and $W$ gauge boson, $M_{T2}(t)$ and $M_{T2}(W)$, respectively, are among those observables that are very sensitive to the full off-shell effects in the $pp \to t\bar{t}t\bar{t}$ process. We refer the reader to Ref. \cite{Bevilacqua:2022twl} for detailed definitions of these observables.  In such cases, it would be beneficial to merge the NLO QCD predictions as calculated in the NWA with the LO full off-shell results. To avoid double counting, the quadruple-resonant top-quark contributions must be properly subtracted from the full off-shell result. Such studies are left for the future. 

\acknowledgments{

This work was supported by the German Research Foundation (Deutsche Forschungsgemeinschaft - DFG) under grant 396021762 - TRR 257: \textit{Particle Physics Phenomenology after the Higgs Discovery}, and grant 400140256 - GRK 2497: \textit{The Physics of the Heaviest Particles at the LHC.}

Furthermore, we acknowledge support by the Federal Ministry of Education and Research (Bundesministerium f\"ur Bildung und Forschung - BMBF)  under grant BMBF-Projekt 05H21PACCA: \textit{Run 3 of CMS at the LHC: Theoretical studies of physics at particle accelerators}.

The authors gratefully acknowledge the computing time provided to them at the NHR Center NHR4CES at RWTH Aachen University (project number \texttt{p0020216}). This is funded by the Federal Ministry of Education and Research, and the state governments participating on the basis of the resolutions of the GWK for national high performance computing at universities.}


\bibliographystyle{JHEP}

\begingroup\raggedright\endgroup

\end{document}